\documentclass[a4paper,12pt]{article}
\usepackage{graphicx}
\usepackage{mathtools}
\usepackage{amssymb}
\usepackage{amsfonts}
\usepackage[utf8]{inputenc}
\usepackage[footnotesize]{caption}
\usepackage[font=scriptsize]{subcaption}
\usepackage{color}
\usepackage{braket}
\usepackage{cite}
\usepackage{hyperref}
\usepackage{url}
\usepackage{multirow}
\usepackage{relsize}
\usepackage{fullpage}
\usepackage{makecell}
\usepackage{blkarray}
\usepackage{fullpage}

\newcommand{\Antonio}[1]{{\color{red}#1}}
\newcommand{\newc}{\newcommand}
\newc{\be}{\begin{equation}}
\newc{\ee}{\end{equation}}
\newc{\bea}{\begin{eqnarray}}
\newc{\eea}{\end{eqnarray}}
\newc{\simlt}{~\mbox{\smaller\(\lesssim\)}~}
\newc{\simgt}{~\mbox{\smaller\(\gtrsim\)}~}

\newcommand{\pmatr}[1]{\begin{pmatrix} #1 \end{pmatrix}}

\begin{document}

\begin{titlepage}
\begin{center}
{\bf\Large
\boldmath{
%How to make any $Z'$ model flavourful
%Flavourful $Z'$ models with a vector-like family
%Flavourful $Z'$ models for $R_K$ and $R_{K^*}$ from mixing with a non-universal vector-like family
%How to make any $Z'$ model non-universal for $R_{K^{(*)}}$
%$SU(5)$ with a vector sector for the muon anomalies and the Yukawa relations
%Muon anomalies and the $SU(5)$ Yukawa relation with a vector sector
Muon anomalies and the $SU(5)$ Yukawa relations
%$SU(5)$ with vector sector for $R_{K^{(*)}}$ and $Y_e\neq Y_d^T$
%\vspace{1mm}
%\\  with applications to $R_K$ and $R_{K^*}$ 
}
} \\[12mm]
A. E. C\'arcamo Hern\'andez$^{\dagger}$ 
\footnote{E-mail: \texttt{antonio.carcamo@usm.cl}}
and
Stephen~F.~King$^{\star}$%
\footnote{E-mail: \texttt{king@soton.ac.uk}}
\\[-2mm]
\end{center}
\vspace*{0.50cm}
\centerline{$^{\dagger}$ \it
Universidad T\'{e}cnica Federico Santa Mar\'{\i}a and Centro Cient\'{\i}fico-Tecnol\'{o}gico de Valpara\'{\i}so,}
\centerline{\it
Casilla 110-V, Valpara\'{\i}so, Chile}
\vspace*{0.50cm}
\centerline{$^{\star}$ \it
School of Physics and Astronomy, University of Southampton,}
\centerline{\it
SO17 1BJ Southampton, United Kingdom }
\vspace*{1.20cm}

\begin{abstract}
{\noindent
We show that, within the framework of $SU(5)$ Grand Unified Theories (GUTs), multiple vector-like families at the GUT scale which transform under
a gauged $U(1)'$ (under which the three chiral families are neutral) can result in a single vector-like family at low energies which 
can induce non-universal and flavourful $Z'$ couplings,
which can account for the B physics anomalies in $R_{K^{(*)}}$. In such theories, we show that the same muon couplings which explain $R_{K^{(*)}}$
also correct the Yukawa relation $Y_e=Y_d^T$ in the muon sector without the need for higher Higgs representations.
To illustrate the mechanism, we construct a concrete a model based on $SU(5)\times A_4 \times Z_3\times Z_7$ with two vector-like families at the GUT scale,
and two right-handed neutrinos,
leading to a successful fit to quark and lepton (including neutrino) masses, mixing angles and CP phases,
 where the constraints from lepton flavour violation require $Y_e$ to be diagonal. } 
\end{abstract}
\end{titlepage}

\section{Introduction}

Most $Z'$ models \cite{Langacker:2008yv} have universal couplings to the three families of quarks and leptons.
The reason for this is both theoretical and phenomenological. Firstly many theoretical models naturally 
predict universal $Z'$ couplings. Secondly, 
from a phenomenological point of view, having universal couplings avoids
dangerous favour changing neutral currents (FCNCs) mediated by tree-level $Z'$ exchange.
The most sensitive processes 
involve the first two families, such as $K_0-\bar K_0$
mixing, $\mu - e$ conversion in muonic atoms, and so on, leading to stringent bounds on the $Z'$ mass and couplings
 \cite{Langacker:2008yv}.

Recently, the phenomenological motivation for considering non-universal $Z'$ models has increased due to 
mounting evidence for semi-leptonic $B$ decays which violate $\mu - e$ universality at rates which exceed those predicted by the SM
\cite{Descotes-Genon:2013wba,Altmannshofer:2013foa,Ghosh:2014awa}.
In particular, the LHCb Collaboration and other experiments have reported a number of anomalies in $B\rightarrow K^{(*)}l^+l^-$
decays such as the $R_K$ \cite{Aaij:2014ora} and $R_{K^*}$ \cite{Bifani} ratios of $\mu^+ \mu^-$ to $e^+ e^-$ final states, 
which are observed to be about $70\%$ of their expected values with a $4\sigma$ deviation from the SM,
and the $P'_5$ angular variable,
not to mention the $B\rightarrow \phi \mu^+ \mu^-$ mass distribution in $m_{\mu^+ \mu^-}$.

Following the recent measurement of $R_{K^*}$ \cite{Bifani}, a number of phenomenological analyses of these data, 
see e.g. \cite{Hiller:2017bzc,Ciuchini:2017mik,Geng:2017svp,Capdevila:2017bsm,Ghosh:2017ber,Bardhan:2017xcc}
favour a new physics operator of the $C^{NP}_{9\mu}=-C^{NP}_{10\mu}$ form \cite{Glashow:2014iga,DAmico:2017mtc},
\be
-\frac{1}{(31.5 \ {\rm TeV})^2} \; \bar b_L\gamma^{\mu} s_L \, \bar \mu_L \gamma_{\mu} \mu_L .
\label{G}
\ee
or of the $C^{NP}_{9\mu}$ form,
\be
 -\frac{1}{(31.5 \ {\rm TeV})^2} \; \bar b_L\gamma^{\mu} s_L \, \bar \mu \gamma_{\mu} \mu .
\label{G2}
\ee
or some linear combination of these two operators. Other solutions different than $C^{NP}_{9\mu}=-C^{NP}_{10\mu}$ allowing for a successful explanation of the $R_{K^*}$ anomalies are studied in detail in Ref. \cite{Descotes-Genon:2015uva}. However the solution $C^{NP}_{9\mu}=-C^{NP}_{10\mu}$ can provide a simultaneous explanation of the $R_{K^*}$ and $R_{D^*}$ anomalies \cite{Calibbi:2015kma}.

In a flavourful $Z'$ model, the new physics operator in Eq.\ref{G} will arise from tree-level $Z'$ exchange,
where the $Z'$ must dominantly couple to $\mu \mu $ over $e e $, 
and must also have the quark flavour changing coupling $b_L s_L$ which must dominate over $b_R s_R$.
The coefficient of the tree-level $Z'$ exchange operator is therefore of the form,
\be
\frac{C_{b_Ls_L} C_{\mu_L\mu_L}}{{M_Z'}^2}  \approx -\frac{1}{(31.5 \ {\rm TeV})^2}
\label{33}
\ee
In realistic models the product of the $Z'$ couplings $C_{b_Ls_L}C_{\mu_L\mu_L}$ is much smaller than unity since 
the constraint from the $B_s$ mass difference will imply that $\frac{|C_{b_Ls_L}|}{|C_{\mu_L\mu_L}|}\lesssim \frac{1}{50}$,
so if $ C_{\mu_L\mu_L}\lesssim 1$ then $C_{b_Ls_L}\lesssim 1/50$ which implies that $M_Z'\lesssim 5$ TeV, 
making the $Z'$ possibly observable at the LHC, depending on its coupling to light quarks. Studies of lepton-flavor violating B decays in generic $Z^{\prime}$ models before the $R_{K^*}$ measurement but compatible with it, are provided in Ref. \cite{Crivellin:2015era}. In addition, two and three Higgs doublet models with a non universal $U(1)^{\prime}$ gauge symmetry have been used as the first explanations for the $R_{K}$ and $R_{K^*}$ anomalies \cite{Crivellin:2015lwa}. An alternative explanation of the $R_{K}$ and $R_{K^*}$ anomalies in the framework of a two Higgs doublet model with two scalar singlets and non universal $U(1)^{\prime}$ gauge symmetry is provided in Ref. \cite{Bonilla:2017lsq}. Another explanation for the $R_{K}$ and $R_{K^*}$ anomalies 
is an extended inert doublet model having an extra non universal $U(1)'$ gauge symmetry, where the SM fermion mass hierarchy is generated from sequential loop supression~\cite{CarcamoHernandez:2019cbd,CarcamoHernandez:2019xkb}. Furthermore, the $R_{K}$ and $R_{K^*}$ anomalies can be explained in an aligned 2HDM with right-handed Majorana neutrinos mediating linear and inverse scale seesaw mechanisms to generate light active neutrino masses~\cite{DelleRose:2019ukt}. Apart from these explanations, the $R_{K}$ and $R_{K^*}$ anomalies can also be explained in models with extended $SU(3)_{C}\times SU(3)_{L}\times U(1)'$ symmetry, with nonminimal particle content, as done in Ref \cite{Descotes-Genon:2017ptp}. Finally, a vector leptoquark in the Standard Model representation $(3,1)_{2/3}$ arising from a Pati-Salam-like theory has been shown for the first time to provide a good fit to the $R_{K^*}$ anomalies \cite{Assad:2017iib}.

In a recent paper, we showed how to obtain a flavourful $Z'$ suitable for explaining $R_{K^*}$
by adding a fourth vector-like family with non-universal $U(1)'$ charges \cite{King:2017anf}.
The idea is that the $Z'$ couples universally to the three chiral families, which then mix with the non-universal fourth family
to induce effective non-universal couplings in the physical light mixed quarks and leptons. 
Such a mechanism has wide applicability, for example it was recently discussed in the context of 
F-theory models with non-universal gauginos \cite{Romao:2017qnu}.
Two explicit examples were discussed in \cite{King:2017anf}: an $SO(10)\rightarrow SU(5)\times U(1)_X$ 
model, where we identified $U(1)'\equiv U(1)_X$,
which however was subsequently shown to be not
consistent with both explaining $R_{K^*}$ and respecting the $B_s$ mass difference \cite{Antusch:2017tud},
and a fermiophobic model where the $U(1)'$ charges are not carried by the three chiral families, only by fourth vector-like family.
The fermiophobic looks more promising, since, with suitable couplings, it can overcome all the phenomenological 
flavour changing and collider constraints, and can in addition also provide an explanation for Dark Matter,
as recently discussed \cite{Falkowski:2018dsl}.

On the other hand, the existing pattern of Standard Model (SM) fermion masses is extended over a range of five orders of
magnitude in the quark sector and a much wider range of about 12 orders of magnitude, when neutrinos are included. Unlike in the quark sector where the mixing angles are very small, two of the three leptonic mixing angles, i.e., the atmospheric $\theta_{23}$ and the solar $\theta_{12}$ are large, while the reactor angle $\theta_{13}$ is comparatively  small. 
%While the mixing angles in the quark sector are very small, in the lepton sector two of the mixing angles are large, and one mixing angle is small. 
%This suggests a different kind of New Physics for the neutrino sector from the one present in the quark mass and mixing pattern. 
This suggests a different kind of underlying physics for the neutrino sector than what
should be responsible for the observed hierarchy of quark masses and mixing angles. That flavour puzzle of the SM indicates that New Physics has to be
advocated to explain the observed SM fermion mass and mixing pattern. %prevailing pattern of SM fermion masses and mixings. 
That SM ``flavor puzzle'' motivates to build models with additional scalars and fermions in their particle spectrum and with an extended gauge group, supplemented by discrete flavour symmetries, which are usually spontaneously broken, in order to generate the observed pattern of SM fermion masses and mixing angles. Recent reviews of discrete flavor groups can be found in Refs. \cite%
{Ishimori:2010au,Altarelli:2010gt,King:2013eh,King:2014nza,King:2017guk}. 
Several discrete groups such as $S_{3}$ 
\cite%
{Gerard:1982mm,Kubo:2003iw,Kubo:2003pd,Kobayashi:2003fh,Chen:2004rr,Mondragon:2007af,Mondragon:2008gm,Bhattacharyya:2010hp, Dong:2011vb,Dias:2012bh,Meloni:2012ci,Canales:2012dr,Canales:2013cga,Ma:2013zca,Kajiyama:2013sza,Hernandez:2013hea, Ma:2014qra,Hernandez:2014vta,Hernandez:2014lpa,Gupta:2014nba,Hernandez:2015dga,Hernandez:2015zeh,Hernandez:2016rbi,Hernandez:2015hrt,CarcamoHernandez:2016pdu,Arbelaez:2016mhg,Gomez-Izquierdo:2017rxi,Cruz:2017add,CarcamoHernandez:2018vdj}%
, $A_{4}$ \cite%
{Ma:2001dn,deMedeirosVarzielas:2005qg,He:2006dk,Chen:2009um,Burrows:2010wz,King:2011ab,Antusch:2011ic,Ahn:2012tv,Cooper:2012wf,Memenga:2013vc,King:2013hoa,King:2013iva,Ding:2013bpa,Felipe:2013vwa,Varzielas:2012ai, Ishimori:2012fg,King:2013hj,Antusch:2013wn,Hernandez:2013dta,Babu:2002dz,Altarelli:2005yx,Varzielas:2008jm,Gupta:2011ct,Morisi:2013eca, Altarelli:2005yp,Kadosh:2010rm,Kadosh:2013nra,delAguila:2010vg,Campos:2014lla,Vien:2014pta,King:2014iia,Joshipura:2015dsa,Hernandez:2015tna,Bjorkeroth:2015ora,Chattopadhyay:2017zvs,CarcamoHernandez:2017kra,CarcamoHernandez:2017cwi,CentellesChulia:2017koy,Bjorkeroth:2017tsz,Belyaev:2018vkl,King:2018fke,Bernigaud:2018qky,deAnda:2018ecu,CarcamoHernandez:2019pmy}%
, $S_{4}$ \cite%
{Patel:2010hr,Morisi:2011pm,Mohapatra:2012tb,BhupalDev:2012nm,Hagedorn:2012ut,Varzielas:2012pa,Ding:2013hpa,Ishimori:2010fs,Ding:2013eca,Hagedorn:2011un,Campos:2014zaa,Dong:2010zu,VanVien:2015xha,Dimou:2015cmw,King:2016yvg,Bjorkeroth:2017ybg,deAnda:2017yeb,deAnda:2018oik,CarcamoHernandez:2019eme}
, $D_{4}$ \cite%
{Frampton:1994rk,Grimus:2003kq,Grimus:2004rj,Frigerio:2004jg,Adulpravitchai:2008yp,Ishimori:2008gp,Hagedorn:2010mq,Meloni:2011cc,Vien:2013zra}%
, $Q_{6}$ \cite%
{Babu:2004tn,Kajiyama:2005rk,Kajiyama:2007pr,Kifune:2007fj,Babu:2009nn,
Kawashima:2009jv,Kaburaki:2010xc,Babu:2011mv,Araki:2011zg, Gomez-Izquierdo:2013uaa,Gomez-Izquierdo:2017med}%
, $T_{7}$ \cite%
{Luhn:2007sy,Hagedorn:2008bc,Cao:2010mp,Luhn:2012bc,Kajiyama:2013lja,Bonilla:2014xla,Vien:2014gza, Vien:2015koa,Hernandez:2015cra,Arbelaez:2015toa}%
, $T_{13}$ \cite{Ding:2011qt,Hartmann:2011dn,Hartmann:2011pq,Kajiyama:2010sb}%
, $T^{\prime }$ \cite%
{Aranda:2000tm,Sen:2007vx,Aranda:2007dp,Chen:2007afa,Frampton:2008bz,Eby:2011ph,Frampton:2013lva,Chen:2013wba,Vien:2018otl}%
, $\Delta (27)$ \cite{Branco:1983tn,deMedeirosVarzielas:2006fc,Ma:2007wu,Bazzocchi:2009qg,Varzielas:2012nn,Bhattacharyya:2012pi,Ferreira:2012ri,Ma:2013xqa,Nishi:2013jqa,Varzielas:2013sla,Aranda:2013gga,Ma:2014eka, Abbas:2014ewa,Abbas:2015zna,Varzielas:2015aua,Bjorkeroth:2015uou,Chen:2015jta,Vien:2016tmh,Hernandez:2016eod,CarcamoHernandez:2017owh,deMedeirosVarzielas:2017sdv,Bernal:2017xat,CarcamoHernandez:2018iel,deMedeirosVarzielas:2018vab,CarcamoHernandez:2018djj,CarcamoHernandez:2018hst}%
, $\Delta(54)$ \cite{Carballo-Perez:2016ooy}, $\Delta(96)$ \cite{King:2012in,King:2013vna,Ding:2014ssa}, $\Delta(6N^2)$ \cite{Ishimori:2014jwa,King:2014rwa,Ishimori:2014nxa} and $A_{5}$ \cite%
{Everett:2008et,Feruglio:2011qq,Cooper:2012bd,Varzielas:2013hga,Gehrlein:2014wda,Gehrlein:2015dxa,DiIura:2015kfa,Ballett:2015wia,Gehrlein:2015dza,Turner:2015uta,Li:2015jxa,Ding:2017hdv}
 have been implemented in extensions of the SM, to provide a nice description of the observed pattern of fermion masses and mixing angles. %explain the .

In this paper we focus on an $SU(5)\times U(1)'$ model with a vector-like fourth family
where the three chiral families do not couple to the $U(1)'$, but the fourth vector-like family has arbitrary $U(1)'$ charges 
for the different multiplets, which mix with the three families, thereby inducing effective non-universal couplings 
for the light physical mixed quarks and leptons.
The particular scheme we consider involves induced $Z'$ couplings to third family left-handed quark doublets and second family left-handed lepton doublets,
similar to the model discussed recently in \cite{Falkowski:2018dsl}.
However, in addition, we also allow induced $Z'$ couplings to the right-handed muon, in order to provide non-universality for both left-handed and right-handed muons, and hence give corrections to the physical muon Yukawa coupling.
We show that such an $SU(5)$ model with the vector sector
can account for the muon anomalies $R_{K^{(*)}}$ and correct the Yukawa relation $Y_e\neq Y_d^T$ without the need for higher Higgs representations.
The same applies to flavoured GUTs such as $SU(5)\times A_4$ with a vector sector. In addition, we study the implications of a $A_4$ flavoured $SU(5)\times U(1)'$ GUT theory with five generations of fermions, on SM fermion masses and mixings. To successfully describe the observed pattern of SM fermion masses and mixing angles, we supplement the $A_4$ family symmetry of that model by the $Z_3\times Z_7$ discrete group and we extend the particle content of our model by adding two right handed Majorana neutrinos and several $SU(5)$ singlet scalar fields. The discrete $A_4\times Z_3\times Z_7$ discrete group is needed in order to reproduce the specific patterns of mass matrices in the quark and lepton sectors, consistent with the low energy SM fermion flavor data. The two right handed Majorana neutrinos are required for the implementation of the type I seesaw mechanism at tree level to generate the masses for the light active neutrinos as pointed out for the first time in Refs. \cite{King:1999mb,King:2002nf}. In this framework, the active neutrinos acquire small masses scaled by the inverse of the large type-I seesaw mediators, thus providing a natural explanation for the smallness of neutrino masses.

The layout of the remainder of the paper is as follows. In section \ref{general} we describe a 2 Higgs doublet model with four generations of fermions, several scalar singlets, an extra $U(1)'$ gauge symmetry under which the SM fermions are neutral and the fourth generation of fermions is charged. In section \ref{SU5} we present the $SU(5)\times U(1)'$ GUT theory with five
%fourth and fifth 
 generations of fermions in the $\bar{\bf 5}$ and ${\bf 10}$ irreps of $SU(5)$. In section \ref{SU5A4} we outline the $A_4$ flavoured $SU(5)\times U(1)'$ GUT theory with five generations of fermions and we discuss its implications on SM fermion masses and mixings. Finally we conclude in section \ref{conclusion}. Appendix \ref{A4} provides a brief description of the $A_4$ discrete group.

\section{Standard Model with a vector sector}
\label{general}

\begin{table}
\centering
\footnotesize
\captionsetup{width=0.9\textwidth}
\begin{tabular}{| c c c c c  |}
\hline
\multirow{2}{*}{\rule{0pt}{4ex}Field}	& \multicolumn{4}{c |}{Representation/charge} \\
\cline{2-5}
\rule{0pt}{3ex}			& $SU(3)_c$ & $SU(2)_L$ & $U(1)_Y$ &$U(1)'$ \\ [0.75ex]
\hline \hline
\rule{0pt}{3ex}%
$Q_{Li}$ 		 & ${\bf 3}$ & ${\bf 2}$ & $1/6$ & $0$\\
$u_{Ri}$ 		 & ${\bf 3}$ & ${\bf 1}$ & $2/3$ & $0$\\
$d_{Ri}$ 		 & ${\bf 3}$ & ${\bf 1}$ & $-1/3$ & $0$\\
$L_{Li}$ 		 & ${\bf 1}$ & ${\bf 2}$ & $-1/2$ & $0$\\
$e_{Ri}$ 		 & ${\bf 1}$ & ${\bf 1}$ & $-1$ & $0$\\
$\nu_{Ri}$         & ${\bf 1}$ & ${\bf 1}$ & $0$ & $0$\\
\hline
\hline
\rule{0pt}{3ex}%
$H_u$ & ${\bf 1}$ & ${\bf 2}$ & $-1/2$ &$0$ \\
$H_d$ & ${\bf 1}$ & ${\bf 2}$ & $1/2$ &$0$ \\
\hline
\hline
\rule{0pt}{3ex}%
$Q_{L4}$,$\tilde{Q}_{R4}$   		 & ${\bf 3}$ & ${\bf 2}$ & $1/6$ & $q_{Q_4}$\\
$u_{R4}$,$\tilde{u}_{L4}$  		 & ${\bf 3}$ & ${\bf 1}$ & $2/3$ & $q_{u_4}$\\
$d_{R4}$,$\tilde{d}_{L4}$  		 & ${\bf 3}$ & ${\bf 1}$ & $-1/3$ & $q_{d_4}$\\
$L_{L4}$,$\tilde{L}_{R4}$  		 & ${\bf 1}$ & ${\bf 2}$ & $-1/2$ & $q_{L_4}$\\
$e_{R4}$,$\tilde{e}_{L4}$  		 & ${\bf 1}$ & ${\bf 1}$ & $-1$ & $q_{e_4}$\\
\hline
\hline
\rule{0pt}{3ex}%
$\phi_{Q,u,d,L,e}$ & ${\bf 1}$ & ${\bf 1}$ & $0$ &$q_{\phi_{Q,u,d,L,e}}$ \\
\hline
\end{tabular}
\caption{The general framework considered in this paper.
}
\label{tab:funfields1}
\end{table}
In this section we analyse the model defined in Table~\ref{tab:funfields1}.
The three chiral families and the Higgs doublets do not carry any $U(1)'$ charges.
We allow the vector-like family to carry arbitrary $U(1)'$ charges.
The scalars $\phi$ couple the vector-like family to the three chiral families.

\subsection{Higgs Yukawa couplings}

The Higgs Yukawa couplings of the first three chiral families $\psi_i$ are,
\be
{\cal L}^{Yuk}=y^u_{ij}H_u\overline{Q}_{Li}u_{Rj}+y^d_{ij}H_d\overline{Q}_{Li}d_{Rj}
+y^e_{ij}H_d\overline{L}_{Li}e_{Rj} +H.c.
\label{yuk}
\ee
where $i,j=1,\ldots ,3$.

In addition we allow the possibly of the fourth vector-like family Higgs Yukawa couplings,
\be
{\cal L}_4^{Yuk}=y^u_{4}H_u\overline{Q}_{L4}u_{R4}+y^d_{4}H_d\overline{Q}_{L4}d_{R4}
+y^e_{4}H_d\overline{L}_{L4}e_{R4} +H.c.
\label{yuk4}
\ee
although the existence of these couplings will depend on the choice of the $U(1)'$ charges for the vector-like family,
and some or all of these couplings could be zero.

\subsection{Heavy masses}

In this subsection we ignore the Higgs Yukawa couplings (which give electroweak scale masses)
and consider only the heavy mass Lagrangian (which gives multi-TeV masses).

The vector-like family can mix with the three chiral families via the $\phi$ scalars, and also can have explicit masses,
leading to the heavy Lagrangian,
\bea
{\cal L}^{heavy}&=&x^{Q}_{i}\phi_Q \overline{Q}_{Li}\tilde{Q}_{R4}
+x^{u}_{i}\phi_u \overline{\tilde{u}}_{L4} u_{Ri}
+x^{d}_{i}\phi_d \overline{\tilde{d}}_{L4} d_{Ri}
+x^{L}_{i}\phi_L \overline{L}_{Li}\tilde{L}_{R4}
+x^{e}_{i}\phi_e \overline{\tilde{e}}_{L4} e_{Ri}
\nonumber \\
&+& M^{Q}_{4}\overline{Q}_{L4}\tilde{Q}_{R4}
+M^{u}_{4}\overline{\tilde{u}}_{L4} u_{R4}
+M^{d}_{4}\overline{\tilde{d}}_{L4} d_{R4}
+M^{L}_{4} \overline{L}_{L4}\tilde{L}_{R4}
+M^{e}_{4} \overline{\tilde{e}}_{L4} e_{R4}
+H.c.
\eea
After the singlet fields $\phi$ develop vacuum expectation values (VEVs), the $U(1)'$ gauge symmetry is broken and yields a massive $Z'$
gauge boson whose mass is of order the largest VEV of the $\phi$ fields. 
Then may define
new mass parameters $M^Q_i=x^Q_i\langle \phi_Q \rangle$, and similarly for the other mass parameters,
to give,
\be
{\cal L}^{heavy}=
M^{Q}_{\alpha}\overline{Q}_{L\alpha}\tilde{Q}_{R4}
+M^{u}_{\alpha}\overline{\tilde{u}}_{L4} u_{R\alpha}
+M^{d}_{\alpha}\overline{\tilde{d}}_{L4} d_{R\alpha}
+M^{L}_{\alpha} \overline{L}_{L\alpha}\tilde{L}_{R4}
+M^{e}_{\alpha} \overline{\tilde{e}}_{L4} e_{R\alpha}
+H.c.
\ee
where $\alpha =1,\ldots ,4$ in a compact notation.

All these mass terms are heavy, of order a few TeV, and our first task is to identify the heavy mass states and integrate them out.
Actually only one linear combination of the four ``normal chirality'' states will get heavy, while the other three orthogonal linear combinations
will remain massless (ignoring the Higgs Yukawa couplings). 
We will identify the three physical massless families with the quarks and leptons of the Standard Model.

\subsection{Diagonalising the heavy masses}
\label{extrafamily}
We now focus on ${\cal L}^{heavy}$ (ignoring the Higgs Yukawa Lagrangian)
and show how the heavy masses may be diagonalised, denoting the fields in this basis by primes.
The goal is to identify the light states of the low energy effective Standard Model (SM) below the few TeV scale,
after the heavy states have been integrated out. 

In the primed basis, the fourth family is massive (before electroweak symmetry breaking),
\be
{\cal L}^{mass}=
\tilde{M}^{Q}_{4}\overline{Q'}_{L4}\tilde{Q}_{R4}
+\tilde{M}^{u}_{4}\overline{\tilde{u}}_{L4} u'_{R4}
+\tilde{M}^{d}_{4}\overline{\tilde{d}}_{L4} d'_{R4}
+\tilde{M}^{L}_{4} \overline{L'}_{L4}\tilde{L}_{R4}
+\tilde{M}^{e}_{4} \overline{\tilde{e}}_{L4} e'_{R4}
+H.c.
\label{vecmass}
\ee

The first three families in the primed basis have zero mass (before electroweak symmetry breaking), and 
are identified as the quarks and leptons of the SM.

The fields in the primed basis and the original basis are related by unitary $4\times 4$ mixing matrices,
\be
Q'_{L}=V_{Q_L}{Q}_{L}, \  u'_{R}=V_{u_R}{u}_{R},
\  d'_{R}=V_{d_R}{d}_{R},
\  L'_{L}=V_{L_L}{L}_{L}, 
\  e'_{R}=V_{e_R}{e}_{R}.
\label{primedbasis}
\ee

In our scheme we will consider only the non-zero mixing angles to be $\theta^{Q_L}_{34}$, in order to 
generate the $Z'$ coupling to the third family quark doublet including $b'_L$, and also 
$\theta^{L_L}_{24}$ and $\theta^{e_R}_{24}$ to generate the $Z'$ coupling to the second family lepton doublet including $\mu'_L$ and also $\mu'_R$,
in the primed basis. This is very similar to the model in \cite{Falkowski:2018dsl}, where the non-zero angles $\theta^{Q_L}_{34}$ and $\theta^{L_L}_{24}$ were
considered, and whose main focus was on the phenomenological viability of the model including Dark Matter. The model considered here includes in addition the non-zero angle $\theta^{e_R}_{24}$ which generates an additional $Z'$ coupling to $\mu'_R$, which is important for the main focus of the present paper,
namely the effect of the model on the $SU(5)$ Yukawa relations.

To summarise, in this paper we consider:
\be
	V_{Q_L}=V^{Q_L}_{34} = \pmatr{1&0&0&0\\0&1&0&0\\0&0&c^{Q_L}_{34}&s^{Q_L}_{34}
	\\ 0&0&-s^{Q_L}_{34}&c^{Q_L}_{34}}, \quad 
\label{34Q}
\ee

\be
	V_{L_L}=V^{L_L}_{24} = \pmatr{1&0&0&0\\0&c^{L_L}_{24}&0&s^{L_L}_{24}\\0&0&1&0\\
	0&-s^{L_L}_{24}&0&c^{L_L}_{24}}, \quad 
	\label{24L}
\ee

\be
	V_{e_R}=V^{e_R}_{24} = \pmatr{1&0&0&0\\0&c^{e_R}_{24}&0&s^{e_R}_{24}\\0&0&1&0\\
	0&-s^{e_R}_{24}&0&c^{e_R}_{24}},
	\label{24e}
\ee
denoting $c=\cos \theta$ and $s=\sin \theta$.

\subsection{The Lagrangian in the primed basis }
 
\subsubsection{Yukawa couplings in the primed basis}
\label{Yprimebasis}
In the original basis, the Yukawa couplings in Eq.\ref{yuk} may be written in terms of the three chiral families $\psi_i$ plus the same chirality 
fourth family $\psi_4$ in a $4\times 4$ matrix notation as,
\be
{\cal L}^{Yuk}=H_u\overline{Q}_{L }\tilde{y}^u u_{R }
+H_d\overline{Q}_{L}\tilde{y}^dd_{R}
+H_d\overline{L}_{L}\tilde{y}^ee_{R} +H.c.
\ee
where $\tilde{y}^u,\tilde{y}^d,\tilde{y}^e$ are $4\times 4$ matrices consisting of the original $3\times 3$ matrices, $y^u,y^d,y^e$,
but augmented by a fourth row and column, as follows: 
%consisting of all zeroes, apart from the (4,4) elements, for example,
\be
{\tilde y}^e =
	\pmatr{{y}^e_{11}&{y}^e_{12}&{y}^e_{13}&{y}^e_{14}\\{y}^e_{21}&{y}^e_{22}&{y}^e_{23}&{y}^e_{24}
	\\{y}^e_{31}&{y}^e_{32}&{y}^e_{33}&{y}^e_{34}\\
	{y}^e_{41}&{y}^e_{42}&{y}^e_{43}&{y}^e_{44}}.
		\label{yp1}
\ee

In the primed basis in Eq.\ref{primedbasis}, where only the fourth components of the fermions are very heavy,
the Yukawa couplings become,
\be
{\cal L}^{Yuk}=H_u\overline{Q'}_{L }\tilde{y}'^u u'_{R }
+H_d\overline{Q'}_{L } \tilde{y}'^d d'_{R }
+H_d\overline{L'}_{L } \tilde{y}'^e e'_{R }
+H.c.
\ee
where 
\be
\tilde{y}'^u =V_{Q_L}\tilde{y}^u V_{u_R}^{\dagger}, \ \   
\tilde{y}'^d =V_{Q_L}\tilde{y}^d V_{d_R}^{\dagger}, \ \
\tilde{y}'^e =V_{L_L}\tilde{y}^e V_{e_R}^{\dagger}
\label{ytp}
\ee

In the primed basis it is trivial to integrate out the heavy family by 
simply removing the fourth rows and colums of the primed
Yukawa matrices in Eq.\ref{ytp}, to leave the upper $3\times 3$ blocks, which describe the three massless families,
in the low energy effective theory involving the massless fermions $\psi'_i$,
\be
{\cal L}^{Yuk}_{light}=y'^u_{ij}H_u\overline{Q'}_{Li}u'_{Rj}+y'^d_{ij}H_d\overline{Q'}_{Li}d'_{Rj}
+y'^e_{ij}H_d\overline{L'}_{Li}e'_{Rj} +H.c.
\label{yukp}
\ee
where 
\be
{y}'^u_{ij} =(V_{Q_L}\tilde{y}^u V_{u_R}^{\dagger})_{ij}, \ \   
{y}'^d_{ij} =(V_{Q_L}\tilde{y}^d V_{d_R}^{\dagger})_{ij}, \ \
{y}'^e_{ij} =(V_{L_L}\tilde{y}^e V_{e_R}^{\dagger})_{ij}
\label{yp}
\ee
and $i,j=1,\ldots ,3$. The physical three family quark and lepton masses in the low energy effective theory should be calculated 
using the $3\times 3$ Yukawa matrices in Eq.\ref{yp}.

For example, from Eqs.\ref{24L}, \ref{24e}, \ref{yp1}, \ref{ytp}
we see that, if ${y}_{44}$ is large, then this mixing may enhance significantly ${y}'^e_{22}$
compared to its original value ${y}^e_{22}$,
%\be
\begin{eqnarray}
{\tilde y}'^e &=&\pmatr{1&0&0&0\\0&c^{L_L}_{24}&0&s^{L_L}_{24}\\0&0&1&0\\
	0&-s^{L_L}_{24}&0&c^{L_L}_{24}}\pmatr{{y}^e_{11}&{y}^e_{12}&{y}^e_{13}&{y}^e_{14}\\{y}^e_{21}&{y}^e_{22}&{y}^e_{23}&{y}^e_{24}
	\\{y}^e_{31}&{y}^e_{32}&{y}^e_{33}&{y}^e_{34}\\ {y}^e_{41}&{y}^e_{42}&{y}^e_{43}&{y}_{44}}%\\
	\pmatr{1&0&0&0\\0&c^{e_R}_{24}&0&-s^{e_R}_{24}\\0&0&1&0\\
	0&s^{e_R}_{24}&0&c^{e_R}_{24}}\nonumber\\
	&=&\pmatr{{y}^e_{11}&{y}^e_{12}&{y}^e_{13}&{y}'^e_{14}\\{y}^e_{21}&{y}'^e_{22}&{y}^e_{23}&{y}'^e_{24}
	\\{y}^e_{31}&{y}^e_{32}&{y}^e_{33}&{y}'^e_{34}\\
	{y}'^e_{41}&{y}'^e_{42}&{y}'^e_{43}&{y}'^e_{44}}
	\label{yp2}  
\end{eqnarray}

%\ee
where the 22 element of the $3\times 3$ light physical Yukawa matrix gets modified as follows,
\be
{y}'^e_{22}=c^{L_L}_{24}c^{e_R}_{24}{y}^e_{22}+c^{L_L}_{24}s^{e_R}_{24}{y}^e_{24}+s^{L_L}_{24}c^{e_R}_{24}{y}^e_{42}+s^{L_L}_{24}s^{e_R}_{24}{y}^e_{44}
\approx {y}^e_{22}+\theta^{e_R}_{24}{y}^e_{24}+\theta^{L_L}_{24}{y}^e_{42}+\theta^{L_L}_{24}\theta^{e_R}_{24}{y}^e_{44}
\label{e22}
\ee
where the approximation is for small angles.
This may be a rather large correction if ${y}^e_{44}\gg {y}^e_{22}$ or ${y}^e_{24}\gg {y}^e_{22}$ or ${y}^e_{42}\gg {y}^e_{22}$ even for small angle rotations.
Such an enhancement is not present for ${y}'^d_{22}$, due to the assumed zero angles $ \theta^{Q_L}_{24}=\theta^{d_R}_{24}=0$.
Therefore any relation between
${y}^e_{22}$ and ${y}^d_{22}$ will not be respected by the physical couplings ${y}'^e_{22}$ and ${y}'^d_{22}$, 
after the mixing with the vector-like family has been taken into account.

By a similar argument, turning on the mixing angles $ \theta^{L_L}_{14},\theta^{e_R}_{14}$ would lead to,
\be
{y}'^e_{11}
\approx {y}^e_{11}+\theta^{e_R}_{14}{y}^e_{14}+\theta^{L_L}_{14}{y}^e_{41}+\theta^{L_L}_{14}\theta^{e_R}_{14}{y}^e_{44},
\label{e11}
\ee
where these mixing angles $ \theta^{L_L}_{14},\theta^{e_R}_{14}$ could be much smaller than 
$\theta^{Q_L}_{24},\theta^{d_R}_{24}$ and still give a significant correction, since the $11$ element of the 
charged lepton matrix is more sensitive to such corrections than the $22$ element (since the electron mass is much smaller
than the muon mass).

\subsubsection{ $Z'$ gauge couplings in the primed basis}

There is a GIM mechanism in the electroweak sector leading to no flavour changing neutral currents (FCNCs). 
However in the physics of $Z'$ gauge bosons,
the $U(1)'$ charges depend on the family index $\alpha$. This leads to non-universality and possibly FCNCs due to 
$Z'$ gauge boson exchange, as we discuss.
After $U(1)'$ breaking, we have a massive $Z'$ gauge boson with diagonal gauge couplings
to the four families of quarks and leptons,
in the original basis,
\be
{\cal L}^{gauge}_{Z'}= g'Z'_{\mu}\left(
\overline{Q}_LD_Q\gamma^{\mu}{Q}_L
+ \overline{u}_RD_u\gamma^{\mu}u_R
+\overline{d}_RD_d\gamma^{\mu}d_R 
+\overline{L}_LD_L\gamma^{\mu}L_L
+\overline{e}_RD_e\gamma^{\mu}e_R
\right)
\label{gaugeZp}
\ee
where only the fourth family has non-zero charges,
\bea
&&D_Q={\rm diag}(0, 0, 0,q_{Q4}), \ 
D_u={\rm diag}(0, 0, 0,q_{u4}), \ 
D_d={\rm diag}(0,0,0,q_{d4}), \nonumber \\
&&D_L={\rm diag}(0,0,0,q_{L4}), \ 
D_e={\rm diag}(0,0,0,q_{e4}).
\label{Zpcharges1}
\eea

In the diagonal heavy mass (primed) basis, given by the unitary transformations in Eq.\ref{primedbasis},
the $Z'$ couplings to the four families of quarks and leptons in Eq.\ref{gaugeZp} becomes,
\be
{\cal L}^{gauge}_{Z'}= g'Z'_{\mu}\left(
\overline{Q}'_LD'_Q\gamma^{\mu}{Q}'_L
+ \overline{u}'_RD'_u\gamma^{\mu}u'_R
+\overline{d}'_RD'_d\gamma^{\mu}d'_R 
+\overline{L}'_LD'_L\gamma^{\mu}L'_L
+\overline{e}'_RD'_e\gamma^{\mu}e'_R
\right)
\label{gaugeZp1}
\ee
where 
\bea
&&D'_Q= V_{Q_L}D_QV_{Q_L}^{\dagger}, \ 
D'_u=V_{u_R}D_uV_{u_R}^{\dagger}, \ 
D'_d=V_{d_R}D_dV_{d_R}^{\dagger}, \nonumber \\
&&D'_L=V_{L_L}D_LV_{L_L}^{\dagger}, \ 
D'_e=V_{e_R}D_eV_{e_R}^{\dagger}.
\eea

In the low energy effective theory, after decoupling the fourth heavy family, Eq.\ref{gaugeZp1}
gives the $Z'$ couplings to the three massless families of quarks and leptons,
\be
{\cal L}^{gauge}_{Z'}= g'Z'_{\mu}\left(
\overline{Q}'_{L}\tilde{D}'_Q\gamma^{\mu}{Q}'_{L}
+ \overline{u}'_{R}\tilde{D}'_u\gamma^{\mu}u'_{R}
+\overline{d}'_{R}\tilde{D}'_d\gamma^{\mu}d'_{R}
+ \overline{L}'_{L}\tilde{D}'_L\gamma^{\mu}L'_{L}
+\overline{e}'_{R}\tilde{D}'_e\gamma^{\mu}e'_{R}
\right)
\label{gaugeZp2}
\ee
where the $3\times 3$ matrices $\tilde{D}'$ are given by,
\bea
&&(\tilde{D}'_Q)_{ij}= (V_{Q_L}D_QV_{Q_L}^{\dagger})_{ij}, \ 
(\tilde{D}'_u)_{ij}=(V_{u_R}D_uV_{u_R}^{\dagger})_{ij}, \ 
(\tilde{D}'_d)_{ij}=(V_{d_R}D_dV_{d_R}^{\dagger})_{ij}, \nonumber \\
&&(\tilde{D}'_L)_{ij}=(V_{L_L}D_LV_{L_L}^{\dagger})_{ij}, \ 
(\tilde{D}'_e)_{ij}=(V_{e_R}D_eV_{e_R}^{\dagger})_{ij},
\label{Dp}
\eea
where $i,j=1,\ldots ,3$.

Without the fourth family mixing all these $Z'$ couplings would be zero, since the three original
chiral families have zero $U(1)'$ charges. However with Eqs.\ref{34Q}, \ref{24L}, \ref{24e},
this mixing induces $Z'$ couplings to the third family left-handed quarks and to the muons,
as we discuss in the next subsection.

\subsection{Phenomenology}
\label{pheno}

The example we consider is one in which the quarks and leptons start out not coupling to the $Z'$ at all,
as in fermiophobic models. We show that such fermiophobic $Z'$ models  
may be converted to flavourful $Z'$ models via mixing with fourth and fifth vector-like family with $Z'$ couplings. We consider both fourth and fifth vector-like families of charged fermions to account for the $R_K$ and $R_{K^*}$ anomalies and at the same time to allow embedding the model in a $SU(5)$ GUT theory in such a way that the mixings between the heavy and light states will yield a realistic SM quark mass spectrum at low energies without adding a scalar field in the $\mathbf{45}$ irrep representation of $SU(5)$ as we will shown in detail in Section \ref{SU5A4}.         Without the inclusion of the fifth fermion family it will not be possible to embed our model in a $SU(5)$ GUT theory consistent with the low energy SM fermion flavor data and at the same time allowing for an explanation of the $R_K$ and $R_{K^*}$ anomalies, without invoking $\mathbf{45}$ irrep scalar of $SU(5)$. We start by considering the following scenario where the mixing matrices for the fermionic fields $Q_L$, $L_L$ and $e_R$ are:
\begin{eqnarray}
V_{Q_{L}} &=&V_{35}^{Q_{L}}=\left( 
\begin{array}{ccccc}
1 & 0 & 0 & 0 & 0 \\ 
0 & 1 & 0 & 0 & 0 \\ 
0 & 0 & c_{35}^{Q_{L}} & 0 & s_{35}^{Q_{L}} \\ 
0 & 0 & 0 & 1 & 0 \\ 
0 & 0 & -s_{35}^{Q_{L}} & 0 & c_{35}^{Q_{L}}%
\end{array}%
\right) , \\
V_{L_{L}} &=&V_{24}^{L}V_{14}^{L}V_{15}^{L}=\left( 
\begin{array}{ccccc}
c_{14}^{L}c_{15}^{L} & 0 & 0 & s_{14}^{L} & c_{14}^{L}s_{15}^{L} \\ 
-c_{15}^{L}s_{14}^{L}s_{24}^{L} & c_{24}^{L} & 0 & c_{14}^{L}s_{24}^{L} & 
-s_{14}^{L}s_{15}^{L}s_{24}^{L} \\ 
0 & 0 & 1 & 0 & 0 \\ 
-c_{15}^{L}c_{24}^{L}s_{14}^{L} & -s_{24}^{L} & 0 & c_{14}^{L}c_{24}^{L} & 
-c_{24}^{L}s_{14}^{L}s_{15}^{L} \\ 
-s_{15}^{L} & 0 & 0 & 0 & c_{15}^{L}%
\end{array}%
\right),\notag\\
% \left( 
% \begin{array}{ccccc}
% 1 & 0 & 0 & 0 & 0 \\ 
% 0 & c_{24}^{L} & 0 & s_{24}^{L} & 0 \\ 
% 0 & 0 & 1 & 0 & 0 \\ 
% 0 & -s_{24}^{L} & 0 & c_{24}^{L} & 0 \\ 
% 0 & 0 & 0 & 0 & 1%
% \end{array}%
% \right) \left( 
% \begin{array}{ccccc}
% c_{14}^{L} & 0 & 0 & s_{14}^{L} & 0 \\ 
% 0 & 1 & 0 & 0 & 0 \\ 
% 0 & 0 & 1 & 0 & 0 \\ 
% -s_{14}^{L} & 0 & 0 & c_{14}^{L} & 0 \\ 
% 0 & 0 & 0 & 0 & 1%
% \end{array}%
% \right) \left( 
% \begin{array}{ccccc}
% c_{15}^{L} & 0 & 0 & 0 & s_{15}^{L} \\ 
% 0 & 1 & 0 & 0 & 0 \\ 
% 0 & 0 & 1 & 0 & 0 \\ 
% 0 & 0 & 0 & 1 & 0 \\ 
% -s_{15}^{L} & 0 & 0 & 0 & c_{15}^{L}%
% \end{array}%
% \right) , \notag\\
V_{e_{R}} &=&V_{24}^{e_{R}}V_{14}^{e_{R}}V_{15}^{e_{R}}=\left( 
\begin{array}{ccccc}
c_{14}^{e_{R}}c_{15}^{e_{R}} & 0 & 0 & s_{14}^{e_{R}} & 
c_{14}^{e_{R}}s_{15}^{e_{R}} \\ 
-c_{15}^{e_{R}}s_{14}^{e_{R}}s_{24}^{e_{R}} & c_{24}^{e_{R}} & 0 & 
c_{14}^{e_{R}}s_{24}^{e_{R}} & -s_{14}^{e_{R}}s_{15}^{e_{R}}s_{24}^{e_{R}}
\\ 
0 & 0 & 1 & 0 & 0 \\ 
-c_{15}^{e_{R}}c_{24}^{e_{R}}s_{14}^{e_{R}} & -s_{24}^{e_{R}} & 0 & 
c_{14}^{e_{R}}c_{24}^{e_{R}} & -c_{24}^{e_{R}}s_{14}^{e_{R}}s_{15}^{e_{R}}
\\ 
-s_{15}^{e_{R}} & 0 & 0 & 0 & c_{15}^{e_{R}}%
\end{array}%
\right)\notag
% \left( 
% \begin{array}{ccccc}
% 1 & 0 & 0 & 0 & 0 \\ 
% 0 & c_{24}^{e_{R}} & 0 & s_{24}^{e_{R}} & 0 \\ 
% 0 & 0 & 1 & 0 & 0 \\ 
% 0 & -s_{24}^{e_{R}} & 0 & c_{24}^{e_{R}} & 0 \\ 
% 0 & 0 & 0 & 0 & 1%
% \end{array}%
% \right)\left( 
% \begin{array}{ccccc}
% c_{14}^{e_{R}} & 0 & 0 & s_{14}^{e_{R}} & 0 \\ 
% 0 & 1 & 0 & 0 & 0 \\ 
% 0 & 0 & 1 & 0 & 0 \\ 
% -s_{14}^{e_{R}} & 0 & 0 & c_{14}^{e_{R}} & 0 \\ 
% 0 & 0 & 0 & 0 & 1%
% \end{array}%
% \right) \left( 
% \begin{array}{ccccc}
% c_{15}^{e_{R}} & 0 & 0 & 0 & s_{15}^{e_{R}} \\ 
% 0 & 1 & 0 & 0 & 0 \\ 
% 0 & 0 & 1 & 0 & 0 \\ 
% 0 & 0 & 0 & 1 & 0 \\ 
% -s_{15}^{e_{R}} & 0 & 0 & 0 & c_{15}^{e_{R}}%
% \end{array}%
% \right) \notag
\label{Vfs}
\end{eqnarray}
In addition we consider that only the fourth and fifth families have nonvanishing charges:
\begin{equation}
D_Q={\rm diag}(0, 0, 0,q_{Q4},q_{Q5}), \ D_L={\rm diag}(0,0,0,q_{L4},q_{L5}), \ 
D_e={\rm diag}(0,0,0,q_{e4},q_{e5})
\label{Df}  
\end{equation}%}
Then, by replacing  in Eq. (\ref{Dp}) we find the following relations:
\begin{eqnarray}
	\tilde{D}'_Q&=&q_{Q5} \pmatr{0&0&0\\
	0&0&0\\
	0&0&(s^{Q}_{35})^2},  \\
	\tilde{D}'_L&=&\left( 
\begin{array}{ccc}
q_{L_{4}}\left( s_{14}^{L}\right) ^{2}+q_{L_{5}}\left( s_{15}^{L}\right)
^{2}\left( c_{14}^{L}\right) ^{2} & c_{14}^{L}s_{14}^{L}s_{24}^{L}\left[
q_{L_{4}}-q_{L_{5}}\left( s_{15}^{L}\right) ^{2}\right] & 0 \\ 
c_{14}^{L}s_{14}^{L}s_{24}^{L}\left[ q_{L_{4}}-q_{L_{5}}\left(
s_{15}^{L}\right) ^{2}\right] & q_{L_{4}}\left( s_{24}^{L}\right) ^{2}\left(
c_{14}^{L}\right) ^{2}+q_{L_{5}}\left( s_{15}^{L}\right) ^{2}\left(
s_{14}^{L}\right) ^{2}\left( s_{24}^{L}\right) ^{2} & 0 \\ 
0 & 0 & 0%
\end{array}%
\right)  \notag\\
	\tilde{D}'_e&=&\left( 
\begin{array}{ccc}
q_{e_{4}}\left( s_{14}^{e}\right) ^{2}+q_{e_{5}}\left( s_{15}^{e}\right)
^{2}\left( c_{14}^{e}\right) ^{2} & c_{14}^{e}s_{14}^{e}s_{24}^{e}\left[
q_{e_{4}}-q_{e_{5}}\left( s_{15}^{e}\right) ^{2}\right] & 0 \\ 
c_{14}^{e}s_{14}^{e}s_{24}^{e}\left[ q_{e_{4}}-q_{e_{5}}\left(
s_{15}^{e}\right) ^{2}\right] & q_{e_{4}}\left( s_{24}^{e}\right) ^{2}\left(
c_{14}^{e}\right) ^{2}+q_{e_{5}}\left( s_{15}^{e}\right) ^{2}\left(
s_{14}^{e}\right) ^{2}\left( s_{24}^{e}\right) ^{2} & 0 \\ 
0 & 0 & 0%
\end{array}%
\right)\notag 
\label{Dpexplicit3}
\end{eqnarray}
% \bea
% &&D_Q={\rm diag}(0, 0, 0,q_{Q4},q_{Q5}), \ 
% % D_u={\rm diag}(0, 0, 0,q_{u4}), \ 
% % D_d={\rm diag}(0,0,0,q_{d4}), \nonumber \\
% %&&
% D_L={\rm diag}(0,0,0,q_{L4},q_{L5}), \ 
% D_e={\rm diag}(0,0,0,q_{e4},q_{e5}).
% \label{Zpcharges1}
% \eea

%The mixing matrices given above correspond to the upper left $3\times 3$ blocks of the following matrices:
so that the $Z'$ couplings from Eq.\ref{gaugeZp2} become,
\begin{eqnarray}
\mathcal{L}_{Z^{\prime }}^{gauge} &=&g^{\prime }Z_{\lambda }^{\prime
}\left\{ q_{Q5}(s_{35}^{Q})^{2}\overline{Q}_{L3}^{\prime }\gamma ^{\lambda }{%
Q}_{L3}^{\prime }+\left[ q_{L_{4}}\left( s_{14}^{L}\right)
^{2}+q_{L_{5}}\left( s_{15}^{L}\right) ^{2}\left( c_{14}^{L}\right) ^{2}%
\right] \overline{L}_{L1}^{\prime }\gamma ^{\lambda }L_{L1}^{\prime
}\right\}   \notag \\
&&+g^{\prime }Z_{\lambda }^{\prime }\left[ q_{e_{4}}\left( s_{14}^{e}\right)
^{2}+q_{e_{5}}\left( s_{15}^{e}\right) ^{2}\left( c_{14}^{e}\right) ^{2}%
\right] \overline{e}_{R1}^{\prime }\gamma ^{\lambda }e_{R1}^{\prime }  \notag
\\
&&+g^{\prime }Z_{\lambda }^{\prime }\left[ q_{L_{4}}\left( s_{24}^{L}\right)
^{2}\left( c_{14}^{L}\right) ^{2}+q_{L_{5}}\left( s_{15}^{L}\right)
^{2}\left( s_{14}^{L}\right) ^{2}\left( s_{24}^{L}\right) ^{2}\right] 
\overline{L}_{L2}^{\prime }\gamma ^{\lambda }L_{L2}^{\prime }  \notag \\
&&+g^{\prime }Z_{\lambda }^{\prime }\left[ q_{e_{4}}\left( s_{24}^{e}\right)
^{2}\left( c_{14}^{e}\right) ^{2}+q_{e_{5}}\left( s_{15}^{e}\right)
^{2}\left( s_{14}^{e}\right) ^{2}\left( s_{24}^{e}\right) ^{2}\right] 
\overline{e}_{R2}^{\prime }\gamma ^{\lambda }e_{R2}^{\prime }  \notag \\
&&+g^{\prime }Z_{\lambda }^{\prime }c_{14}^{L}s_{14}^{L}s_{24}^{L}\left[
q_{L_{4}}-q_{L_{5}}\left( s_{15}^{L}\right) ^{2}\right] \left( \overline{L}%
_{L1}^{\prime }\gamma ^{\lambda }L_{L2}^{\prime }+\overline{L}_{L2}^{\prime
}\gamma ^{\lambda }L_{L1}^{\prime }\right)\notag\\
&&+g^{\prime }Z_{\lambda }^{\prime }c_{14}^{e}s_{14}^{e}s_{24}^{e}\left[
q_{e_{4}}-q_{e_{5}}\left( s_{15}^{e}\right) ^{2}\right] \left( \overline{e}%
_{R1}^{\prime }\gamma ^{\lambda }e_{R2}^{\prime }+\overline{e}_{R2}^{\prime
}\gamma ^{\lambda }e_{1R}^{\prime }\right)   
\label{gaugeZp7}
\end{eqnarray}
where the $Z'$ couples only to the third family left-handed quark doublets ${Q}'_{L3}=(t'_L,b'_L)$ 
and the muons
${L}'_{L2}=(\nu'_{\mu L},\mu'_L)$ and ${e}'_{R2}=\mu'_R$, where the primes indicate that these are the states before the 
Yukawa matrices are diagonalised.

Ignoring any charged lepton mixing amongst the three light families (to start with), this will lead the couplings,
\begin{eqnarray}
{\cal L}^{gauge}_{Z'}
&=& Z'_{\lambda}\left(
C_{b_Ls_L}\overline{b}_{L}\gamma^{\lambda}{s}_{L}
+ C_{\mu_L\mu_L}\overline{\mu}_{L}\gamma^{\lambda}\mu_{L}
+  C_{\mu_R\mu_R} \overline{\mu}_{R}\gamma^{\lambda}\mu_{R}+C_{e_Le_L}\overline{e}_{L}\gamma^{\lambda}e_{L}\right.\\
&+&C_{e_Re_R} \overline{e}_{R}\gamma^{\lambda}e_{R}+\left. C_{\mu_Le_L}\left(\overline{\mu}_{L}\gamma^{\lambda}e_{L}+\overline{e}_{L}\gamma^{\lambda}\mu_{L}\right)
+  C_{\mu_Re_R}\left)(\overline{\mu}_{R}\gamma^{\lambda}e_{R}+\overline{e}_{R}\gamma^{\lambda}\mu_{R}\right)\notag
 +\ldots \right)
\label{gaugeZp9}  
\end{eqnarray}
%where we have defined
with the different couplings of the $Z'$ gauge bosons with the charged leptonic fields appearing in Eq. \ref{gaugeZp9} are given by:\begin{eqnarray}
C_{b_{L}s_{L}} &\equiv &g^{\prime }q_{Q5}(s_{35}^{Q})^{2}(V_{dL}^{\prime
\dagger })_{32},\ \ \ \ C_{\mu _{L}\mu _{L}}\equiv g^{\prime }\left[
q_{L_{4}}\left( s_{24}^{L}\right) ^{2}\left( c_{14}^{L}\right)
^{2}+q_{L_{5}}\left( s_{15}^{L}\right) ^{2}\left( s_{14}^{L}\right)
^{2}\left( s_{24}^{L}\right) ^{2}\right]   \notag \\
C_{\mu _{R}\mu _{R}} &\equiv &g^{\prime }\left[ q_{e_{4}}\left(
s_{24}^{e}\right) ^{2}\left( c_{14}^{e}\right) ^{2}+q_{e_{5}}\left(
s_{15}^{e}\right) ^{2}\left( s_{14}^{e}\right) ^{2}\left( s_{24}^{e}\right)
^{2}\right] ,  \notag \\
C_{e_{L}e_{L}} &\equiv &g^{\prime }\left[ q_{L_{4}}\left( s_{14}^{L}\right)
^{2}+q_{L_{5}}\left( s_{15}^{L}\right) ^{2}\left( c_{14}^{L}\right) ^{2}%
\right] ,\ \ \ \ C_{e_{R}e_{R}}\equiv g^{\prime }\left[ q_{e_{4}}\left(
s_{14}^{e}\right) ^{2}+q_{e_{5}}\left( s_{15}^{e}\right) ^{2}\left(
c_{14}^{e}\right) ^{2}\right]   \notag \\
C_{\mu _{L}e_{L}} &\equiv &g^{\prime }\left\{ c_{14}^{L}s_{14}^{L}s_{24}^{L}
\left[ q_{L_{4}}-q_{L_{5}}\left( s_{15}^{L}\right) ^{2}\right] +\left[
q_{L_{4}}\left( s_{24}^{L}\right) ^{2}\left( c_{14}^{L}\right)
^{2}+q_{L_{5}}\left( s_{15}^{L}\right) ^{2}\left( s_{14}^{L}\right)
^{2}\left( s_{24}^{L}\right) ^{2}\right] s_{12}^{L}\right\} ,  \notag \\
C_{\mu _{R}e_{R}} &\equiv &g^{\prime }\left\{ c_{14}^{e}s_{14}^{e}s_{24}^{e}
\left[ q_{e_{4}}-q_{e_{5}}\left( s_{15}^{e}\right) ^{2}\right] +\left[
q_{e_{4}}\left( s_{24}^{e}\right) ^{2}\left( c_{14}^{e}\right)
^{2}+q_{e_{5}}\left( s_{15}^{e}\right) ^{2}\left( s_{14}^{e}\right)
^{2}\left( s_{24}^{e}\right) ^{2}\right] s_{12}^{e}\right\}, 
\label{C}
\end{eqnarray}
where the mixing parameters $s_{12}^{L,e}$ appear after expressing the leptonic fields in the interaction basis in terms of the leptonic fields in the mass eigenstates, considering, for the sake of simplicity, only the mixing in the $1-2$ plane. In addition, we have expanded the quark primed fields in terms of mass eigenstates as follows, 
\be
b'_L=(V'^{\dagger}_{dL})_{31} d_L + (V'^{\dagger}_{dL})_{32} s_L + (V'^{\dagger}_{dL})_{33} b_L 
\ee
and assumed from 
the hierarchy of the CKM matrix that, 
\be
|(V'^{\dagger}_{dL})_{31}|^2 \ll
|(V'^{\dagger}_{dL})_{32} |^2 \ll (V'^{\dagger}_{dL})_{33}^2 \approx 1.
\label{hierarchy1}
\ee
Then $Z'$ exchange generates the effective operators, as in Eq.~\ref{G}.
where the operator corresponds to $C^{NP}_{9\mu}=-C^{NP}_{10\mu}$. For the sake of simplicity, we ignore the contribution of the right-handed muon
operator and we neglect the contribution arising from the mixing between the first and fourth generation of charged leptons, i.e, we set $\theta^{(L,R)}_{14}=0$. Let us note that we are considering a scenario where the fifth family of vector like fermions only couples with the third generation of SM quarks as well as with the first generation of charged leptons, whereas the fourth family will only couple with the second generation of SM charged leptons, thus we are assuming that only $\theta^{Q_L}_{35}$, $\theta^{L_L}_{24}$, $\theta^{e_R}_{24}$, $\theta^{L_L}_{15}$, $\theta^{e_R}_{15}$ are non-zero
% %(see Eqs.\ref{34Q}, \ref{24L}, \ref{24e}), 
 with all other mixing angles being zero (see Section \ref{SU5A4} for a justification of those assumptions in terms of symmetries).
% Thus, we will assume   
% %We then show how such a model is capable of accounting for $R_K$ and $R_{K^*}$.
%  that only $\theta^{Q_L}_{35}$, $\theta^{L_L}_{24}$, $\theta^{e_R}_{24}$, $\theta^{L_L}_{14}$, $\theta^{e_R}_{14}$ are non-zero
% %(see Eqs.\ref{34Q}, \ref{24L}, \ref{24e}), 
% with all other mixing angles being zero (see Section \ref{SU5A4} for a justification of those assumptions in terms of symmetries), the only non-zero mixing matrices in Eq.\ref{Dp} are,

%  We will consider both fourth and fifth vector like families because we will assume that the fifth family only couples with the third generation of SM quarks, whereas the fourThat assumption will be crucial 
% %For the sake of simplicity and 
%\Antonio{In order to completely suppress the large contribution of fourth generation of charged leptons to the $\mu\to e\gamma$ decay, we assume $\theta^R_{24}=0$, i.e, $s^e_{24}=0$, which implies that the right-handed muon operator will not contribute to the $R_K$ and $R_{K^*}$ anomalies.} 
%We ignore the contribution of the right-handed muon operator for simplicity.
%
To explain the $R_K$ and $R_{K^*}$ anomalies we require the coefficient to have the correct sign
and magnitude, as discussed in Eq.~\ref{33}, leading to 
\be
|C_{b_Ls_L} C_{\mu_L\mu_L}| \approx 10^{-3}\left(\frac{M_Z'}{1 \ {\rm TeV}}\right)^2.
\label{G1}
\ee
There are important flavour violating processes such as $B_s-\bar{B}_s$ mixing which can rule out models,
due to the $Z'$ coupling to $bs$. As discussed for example in \cite{Antusch:2017tud}, this leads to the constraint,
\be
|C_{b_Ls_L}|^2 \lesssim 2\times10^{-5}\left(\frac{M_Z'}{1 \ {\rm TeV}}\right)^2.
\label{C1}
\ee

From Eqs.\ref{G1}, \ref{C1} we find the constraint,
\be
\frac{|C_{b_Ls_L}|}{|C_{\mu_L\mu_L}|}\lesssim \frac{1}{50}
\ee
From Eq.\ref{C}, this implies,
\be
\frac{|q_{Q5}(s^{Q}_{35})^2(V'^{\dagger}_{dL})_{32}|}{|q_{L4}(s^{L}_{24})^2|}\lesssim \frac{1}{50}
\ee
This is easily satisfied, since for example if $(V'^{\dagger}_{dL})_{32}\sim V_{ts}\sim \lambda^2\sim (1/5)^2\sim 1/25$ 
then this by itself is almost sufficient to satisfy the constraint.

For example, if we saturate the bound in Eq.\ref{C1}, then Eq.\ref{G1} implies,
\be
|C_{\mu_L\mu_L}| = g'q_{L4}(s^{L}_{24})^2 \approx 0.22 \left(\frac{M_Z'}{1 \ {\rm TeV}}\right)
\label{G2}
\ee
This shows that the mixing angle $\theta^{L}_{24}$ cannot be too small. Note that the LHC limits on the $Z'$ mass are very 
weak since it does not couple to light quarks at leading order, and its coupling to strange quarks is suppressed by a factor of 
$(V'^{\dagger}_{dL})_{32}^2$. 

For a more detailed discussion of the phenomenological constraints on this particular 
model arising from both flavour violating processes such as $B_s-\bar{B}_s$ mixing
and LHC limits on the $Z'$ mass see \cite{Falkowski:2018dsl}. Furthermore, note that the model has very small FCNC in the $Z$ couplings as explained in Ref. \cite{King:2017anf}. In addition, the loop effects of fermions charged under both the SM and extra $U(1)^{\prime}$ groups will generate a small $Z-Z^{\prime}$ mixing of the order of $\frac{M^2_{T}(s^{Q}_{35})^2}{16\pi^2}$, being $M_T$ the mass of the fifth family of quarks. Considering $M_T\approx M_{Z^{\prime}}$, the $Z-Z^{\prime}$ mixing angle will be of the order of $6\times 10^{-3}$, thus leading to suppressed FCNC in the $Z$ couplings.

There are other important constraints due to lepton flavour violating (LFV) processes such as $\mu\rightarrow eee$ 
as recently discussed for example in \cite{Antusch:2017tud}. 
\footnote{We do not consider $\mu-e$ conversion since the $Z'$ does not couple to light quarks at leading order.}
However,
as discussed there, violations of lepton universality does not always lead to lepton flavour violation: it depends
on the mixing angles $\theta_{12}^{L,R}$ arising from the left-handed (L) and right-handed (R) rotations which 
diagonalise the charged lepton Yukawa matrix. This leads to a $Z'\mu e$ flavour changing coupling suppressed
by $\theta_{12}^{L,R}$ and a $Z' ee$ flavour conserving coupling to electrons suppressed by $(\theta_{12}^{L,R})^2$. 
We may estimate the branching ratios for $\mu\rightarrow eee$ by taking the ratio of the $Z'$ exchange diagram squared 
to the $W$ exchange diagram squared,
\be
Br( \mu_L\rightarrow e_Le_Le_L)\approx \left(C_{\mu_L\mu_L}\right)^4 (\theta_{12}^L)^6\left(\frac{M_W}{M_{Z'}}\right)^4
\ee
\be
Br( \mu_R\rightarrow e_Re_Re_R)\approx \left(C_{\mu_R\mu_R}\right)^4 (\theta_{12}^R)^6\left(\frac{M_W}{M_{Z'}}\right)^4
\ee
For typical charged lepton mixing angles such as $\theta_{12}^{L,R}\sim \lambda/3\sim 0.07$, the coefficient in Eq.\ref{G2}
will lead to branching ratios such as
\be
Br( \mu_L\rightarrow e_Le_Le_L)\approx (0.22)^4 (0.07)^6 (0.08)^4\approx 10^{-14}
\ee
below the current experimental limit of $Br( \mu\rightarrow eee)\lesssim 10^{-12}$ but within the range of future experiments.

%  Although the above constraints may be satisfied, any model would be potentially more 
% strongly constrained by the LFV decay $\mu\to e\gamma$, and also 
% $\tau\to e\gamma$, $\tau\to\mu\gamma$.  However, as the formula for the branching ratio for the $\mu\to e\gamma$ decay
% shows (see Appendix \ref{mutoegamma}), such branching ratios also vanish when $ \theta _{12}^{L}=\theta _{12}^{R}=\theta _{14}^{L}=\theta _{14}^{R}=0$,
% as expected.
Although the above constraints may be satisfied, our current framework can lead to the Lepton Flavor Violating (LFV) decay $\mu\to e\gamma$, which is only induced by the $\theta _{14}^{L,R}$ mixing angles in the case of a diagonal SM charged lepton mass matrix, as shown in Appendix \ref{mutoegamma}. Thus, to avoid all LFV decays and at the same time to generate the correct value of the electron mass, we need to also suppress the $\theta _{14}^{L,R}$ mixing angles while at the same time correcting the charged lepton masses.
This can be achieved by adding a fifth vector-like family as discussed in the next section.

Finally, we remark that the models discussed in this paper will be supersymmetric (SUSY). It is well known that SUSY must be broken in realistic models, leading to additional sources of flavour violation coming from 
the SUSY breaking sector via SUSY loop contributions. These have been recently studied for a class of SUSY $SU(5)\times A_4$
models~\cite{Bernigaud:2018qky} which includes the type of model described in Section~\ref{SU5A4}.
Interestingly, according to the model independent analysis based on the region of SUSY parameter space consistent
smuon assisted dark matter~\cite{Bernigaud:2018qky}, the most constraining SUSY loop induced flavour observables are also
$ \mu\rightarrow eee$ and $\mu\to e\gamma$, which are the same modes as discussed above.
Such lepton flavour violating decays could therefore be mediated by either SUSY loops or by $Z'$ exchange in this model.

%  thus giving rise to a correction to he . Thus thus generating 
% % in such a way that the mixing angle between the fifth and the fourth family of leptons will be strongly suppressesTo avoid that d
%
% To avoid all LFV decays, in Section \ref{SU5A4} we shall construct an explicit model 
% in which the SM charged lepton mass matrix is diagonal, and hence $ \theta _{12}^{L}=\theta _{12}^{R}=\theta _{14}^{L}=\theta _{14}^{R}=0$.
%
\section{$SU(5)$ with a vector sector}
\label{SU5}
\begin{table}
\centering
\footnotesize
\captionsetup{width=0.9\textwidth}
\begin{tabular}{| c c c |}
\hline
\multirow{2}{*}{\rule{0pt}{5ex}Field}	& \multicolumn{2}{c |}{Representation/charge} \\
\cline{2-3}
\rule{0pt}{3ex}			& $SU(5)$ & $U(1)'$ \\ [0.75ex]
\hline \hline
\rule{0pt}{3ex}%
$F_i$ 		 & $\bar{\bf 5}$ & 0 \\
$T_i$ 		 & ${\bf 10}$ & 0  \\
\hline
\hline
\rule{0pt}{3ex}%
$H_u$ & ${\bf 5}$ & $0$\\
$H_d$ & $\bar{{\bf 5}}$ & $0$  \\
\hline
\hline
$F_a$ 		 & $\bar{\bf 5}$ & $q_{Fa}$ \\
$\bar{F}_a$ 		 & ${\bf 5}$ & $-q_{Fa}$ \\
$T_a$ 		 & ${\bf 10}$ & $q_{Ta}$  \\
$\bar{T}_a$ 		 & $\bar{\bf 10}$ & $-q_{Ta}$  \\
\hline
\hline
$\phi_{Fa}$ & ${\bf 1}$ & $q_{Fa}$ \\
$\phi_{Ta}$ & ${\bf 1}$ & $q_{Ta}$ \\
\hline
\hline
\end{tabular}
\caption{The $SU(5)$ model considered in this paper.
The indices $i=1,2,3$ while $a=4,5,\ldots$.
}
\label{tab:funfields2}
\end{table}

We now suppose that the SM with a vector sector considered in the previous subsection descends from 
a supersymmetric $SU(5)$ GUT. The three chiral families result from three families of $F_i$ transforming as $\bar{\bf 5}$,
and $T_i$  transforming as ${\bf 10}$, which all carry zero $U(1)'$ charges.
The Higgs $H_u$ and $H_d$ arise from ${\bf 5}$ and $\bar{\bf 5}$ representations, after doublet-triplet splitting (which we do not address).
This results in the $SU(5)$ Yukawa relation, $Y_e=Y_d^T$ in the usual way.

Now we consider adding the previous vector sector to the $SU(5)$ GUT.
In order to violate the $SU(5)$ relation $Y_e=Y_d^T$ 
we will suppose that the fourth vector-like family at low energies results from multiple $\bar{\bf 5}+{\bf 5}$ and ${\bf 10}+\bar{\bf 10}$
at the GUT scale, where each pair has equal and opposite $U(1)'$ charges, but which differ each from another pair. Similar arguments apply for the origin of the fifth family. 
At low energies below the GUT scale, only the matter content of two vector-like families survives
with various $U(1)'$ charges, similarly as in Table \ref{tab:funfields1},
with the remaining components of the multiple $\bar{\bf 5}+{\bf 5}$ and ${\bf 10}+\bar{\bf 10}$ states having GUT scale masses.
Below the GUT scale, the model in Table  \ref{tab:funfields2} leads to the SM plus vector sector in Table  \ref{tab:funfields1}.
Thus the $SU(5)$ plus vector sector can explain the muon anomalies exactly like we discussed in the previous section
(see in particular subsection \ref{pheno}).

We now focus on the $SU(5)$ Yukawa relation, $Y_d=Y_e^T$ and show that it is violated by the $SU(5)$ plus mixing with the vector sector.
At the GUT scale, we identify $Y_e=y^e_{ij}$ and $Y_d=y^d_{ij}$ in Eq.\ref{yuk}. 

The Yukawa terms in $SU(5)$ may be written as,
\begin{equation}
y^u_{ij} H_{{\bf 5}}T_iT_{j}+ y^{\nu}_{ij}H_{{\bf 5}}F_i\nu^c_j+
y^d_{ij} H_{\overline{\bf 5}}T_{i}F_j,
\end{equation}
These give SM Yukawa terms,
\begin{equation}
y^u_{ij} H_uQ_iu^c_j+  y^{\nu}_{ij}H_uL_i\nu^c_j+
y^d_{ij} (H_dQ_id^c_j+H_de^c_iL_j).
\end{equation}
From this equation we identify the charged lepton Yukawa matrix as
\begin{equation}
Y_e=Y_d^T,
\end{equation}
at the GUT scale.
This means that after RG effects are considered we have at low energy,
\begin{equation}
Y_e\approx \frac{1}{3}Y_d^T,
\end{equation}
where QCD corrections lead to an overall scaling factor of about 3 for the quark Yukawa couplings
as compared to those of the leptons.
This implies that 
\begin{equation}
y_{\tau}=\frac{1}{3}y_b, \quad   y_{\mu}=\frac{1}{3}y_s, \quad y_e=\frac{1}{3}y_d .
\end{equation}
Though successful for the third family, this fails for the first and
second families. 

Georgi and Jarlskog (GJ)~\cite{Georgi:1979df} proposed that the (2,2) matrix entry 
of the Yukawa matrices may be given by,
\begin{equation}
y^{d}_{22} H_{\overline{\bf 45}}T_2F_2 ,
\end{equation}
involving a Higgs field $H_{\overline{\bf 45}}$, where 
$H_d$ is the light linear combination of the
electroweak doublets contained in $H_{\bf{\overline 5}}$~and~$H_{\bf{\overline {45}}}$.
This term reduces to 
\begin{equation}
y^{d}_{22} (H_dQ_2d_2^c-3H_de_2^cL_2),
\end{equation}
where the factor of $-3$ is a Clebsch-Gordan
coefficient.
Assuming a zero Yukawa element (texture) in
the (1,1) position, and symmetric and hierarchical Yukawa matrices, this leads to the relations at low energy, 
\begin{equation}
y_{\tau}=\frac{1}{3}y_b, \quad  y_{\mu}=y_s, \quad y_e=\frac{1}{9}y_d,
\label{low}
\end{equation}
which are approximately consistent with the low energy masses.

In our approach we do not wish to consider such large Higgs representations to modify the Yukawa matrices at the GUT scale.
Instead we note that these are not the physical Yukawa matrices due to mixing with the fourth family. By following our discussion given in Section \ref{Yprimebasis}
%generalizing Eq. (\ref{yp2}), considering the mixing with the fourth family, 
we find that the mixing with the fourth family 
%which are in fact given by Eq.\ref{yp} with Eqs.\ref{24L}, \ref{24e}.
%As shown in Eq.\ref{e22} 
%such mixing 
may enhance ${y}'^e_{22}$
compared to its original value ${y}^e_{22}$,
\be
{y}'^e_{22}={y}^e_{22}\cos\theta^{L_L}_{24}\cos\theta^{e_R}_{24}+{y}^e_{24}\cos\theta^{L_L}_{24}\sin\theta^{e_R}_{24}+{y}^e_{42}\sin\theta^{L_L}_{24}\cos\theta^{e_R}_{24}+{y}^e_{44}\sin\theta^{L_L}_{24}\sin\theta^{e_R}_{24}\equiv f  {y}^e_{22},
%{y}'^e_{22}\approx {y}^e_{22}+ \theta^{L_L}_{24}\theta^{e_R}_{24}{y}_{e4}\equiv f  {y}^e_{22},
\label{e221}
\ee
which may be a rather large correction if ${y}^e_{41}\gg {y}^e_{22}$, even for small angle rotations.
We can easily achieve an enhancement by a factor of 3, or indeed any other factor $f$.
Such an enhancement is not present in ${y}'^d_{22}$
due to our choice of zero mixing angles $ \theta^{Q_L}_{24}=\theta^{d_R}_{24}=0$.

Assuming as before a zero Yukawa element (texture) in
the (1,1) position, and symmetric and hierarchical Yukawa matrices, Eq.\ref{e221} leads to the relations at low energy, 
\begin{equation}
y_{\tau}=\frac{1}{3}y_b, \quad  y_{\mu}= \frac{f}{3}y_s, \quad y_e=\frac{1}{3f}y_d.
\end{equation}
These relations are approximately consistent with the low energy masses for $f\approx 2-3$.

It is worth noting that the requirement for enhancing ${y}'^e_{22}$ but not ${y}'^d_{22}$ relies on
the assumption that $\theta^{L_L}_{24}\neq 0$ or $\theta^{e_R}_{24}\neq 0$ but $ \theta^{Q_L}_{24}\theta^{d_R}_{24}=0$.
If we had assumed that the vector-like family originated from a single $\bar{\bf 5}+{\bf 5}$ and ${\bf 10}+\bar{\bf 10}$ representation,
denoted as $F_4+\bar{F}_4$ and $T_4+\bar{T}_4$ then this would constrain the choice of charges for the 
vector-like fourth family to be $\pm q_{F4}$ for the states $L_{L4}$ and $d_{R4}$, together with 
$\pm q_{T4}$ for the states $Q_{L4}$, $u_{R4}$ and $e_{R4}$, and their vector partners.
In particular the vector-like family in Table \ref{tab:funfields1} would have constrained charges $q_{L_4}=-q_{d4}$
and also $q_{Q_4}=-q_{u4}=-q_{e4}$.
This would eventually have led to the constraint on the fourth family mixing that $V_{L_L}=V_{d_R}^{\dagger}$. 
Similarly it would have implied that $V_{Q_L}=V_{u_R}^{\dagger}=V_{e_R}^{\dagger}$.
These relations would imply from Eq.\ref{ytp} that the $SU(5)$ relation at low energy 
would be preserved, $Y'_e\approx \frac{1}{3}Y'^T_d$. Furthermore, for enhancing ${y}'^e_{11}$, we require $\theta^{L_L}_{15}\neq 0$ or $\theta^{e_R}_{15}\neq 0$ but $\theta^{Q_L}_{15}\theta^{d_R}_{15}=0$.

In summary, we need $\theta^{L_L}_{24}\neq 0$ or $\theta^{e_R}_{24}\neq 0$ and $\theta^{L_L}_{15}\neq 0$ or $\theta^{e_R}_{15}\neq 0$ but $\theta^{Q_L}_{24}\theta^{d_R}_{24}=0$ and $\theta^{Q_L}_{15}\theta^{d_R}_{15}=0$. This can be done if the fourth and fifth vector-like families at low energies result from multiple $\bar{\bf 5}+{\bf 5}$ and ${\bf 10}+\bar{\bf 10}$ at the GUT scale, where each pair has equal and opposite $U(1)'$ charges, but which differ each from another pair,
as assumed in Table \ref{tab:funfields2}.
Assuming this, then we have shown that 
the $SU(5)$ theory can account for the muon anomalies $R_{K^{(*)}}$ and obtain $Y_e\neq \frac{1}{3}Y_d^T$
without the need for higher Higgs representations.

 The above discussion assumes that there is a zero Yukawa element (texture) in
the (1,1) position, with symmetric and hierarchical charged lepton Yukawa matrix.
If on the other hand we would assume that the charged lepton Yukawa matrix is diagonal, then we would need to assume
corrections as in both Eq.~\ref{e22} and \ref{e11} in order to account for the correct low energy mass relations
in Eq.~\ref{low}. We will see an example of such a model in the next section.

\section{$SU(5)\times A_4$ with a vector sector}
\label{SU5A4}
In this section we will extend the particle content of our supersymmetric model by adding fourth and fifth generations of fermions in the $\bar{\bf 5}$ and ${\bf 10}$ irreps of $SU(5)$, two right handed Majorana neutrinos, i.e., $\nu_{1R}$, $\nu_{2R}$ and several $SU(5)$ singlet scalar fields. In addition, we will implement the $A_4$ family symmetry, which will be supplemented by the $Z_3\times Z_7$ discrete group. These modifications in our simplified version of our model are done in order to get viable and predictive textures for the fermion sector, that will allow us to successfully describe the current pattern of SM fermion masses and mixing angles, as we will show later in this section. 

The particle content of the model and the field assignments under the $SU(5)\times U(1)'\times A_4\times Z_3\times Z_7$ group are shown in Table \ref{tab:funfields}. Let us note, that we use the $A_4$ family symmetry, since $A_4$ is the smallest discrete group having a three-dimensional irreducible representation and 3 different one-dimensional irreducible representations, which allows to naturally accommodate the three fermion families. Specifically, we grouped the three generations of SM fermionic $\bar{\bf 5}_i\approx F_i$ $(i=1,2,3)$ irreps of $SU(5)$ in an $A_4$ triplet, whereas the three generations of SM fermionic ${\bf 10}_i\sim T_i$ $(i=1,2,3)$ irreps of $SU(5)$ are assigned into $A_4$ trivial singlets. The exotic fermionic fields are also assigned into $A_4$ trivial singlets. As a consequence of the aforementioned fermion assignments under the $A_4\times Z_3\times Z_7$ discrete group, three $A_4$ triplets, $SU(5)$ scalar singlets are needed to provide the masses for the SM down type quarks and charged leptons. 
% Let us note, as we will show later in this section, that the up type quark sector will also contribute to the Cabbibo mixing, but that contribution is very subleading and not enough to generate the value of the Cabbibo angle. Consequently, as a result of the $A_4\times Z_4\times Z_5$ field assignments of our model, the Cabbibo mixing will mainly arise from the down type quark sector, whereas the up quark sector will contribute to the remaining mixing angles. 
%
In addition we need two extra $A_4$ scalar triplets to generate a viable and predictive light active neutrino mass matrix as well as as well as three $A_4$ triplets, $SU(5)$ scalar quintuplets, with different $Z_3$ charges, are required to generate the SM up type quark masses and quark mixing parameters. %, and to guarantee
%  one $A_4$ trivial scalar singlet, charged under the $Z_3\times Z_5$ discrete group, to produce the hierarchical structure of the charged fermion mass matrices that yields the observed charged fermion mass and quark mixing pattern. 
% %that give rises to light active neutrino masses and leptonic mixing angles consistent with the current neutrino oscillation experimental data. 
 %In what follows we comment about the model setup. 
Thus, in view of the above, the $SU(5)$ singlet scalar fields neutral under $U(1)'$, are accommodated into five $A_4$ triplets, i.e., $\xi_e$, $\xi_{\mu}$, $\xi_{\tau}$, $\eta_1$, $\eta_2$ and one $A_4$ trivial singlet, i.e, $\sigma$. Out of the $A_4$ scalar triplets, only $\eta_1$ and $\eta_2$ will participate in the neutrino Yukawa interactions, whereas the remaining $A_4$ triplets will appear in the charged lepton and down type quark Yukawa terms. That separation of the $A_4$ scalar triplets, resulting from the $Z_3\times Z_7$ discrete symmetry, allows us to treat the neutrino and the charged fermion sectors independently.\newline

In addition, the $Z_3$ symmetry allows to have a SM charged lepton mass matrix diagonal, which is crucial to completelly suppress the lepton flavor violating decays. The $Z_7$ symmetry give rises to the hierarchical structure of the charged fermion mass matrices that yields the observed pattern of charged fermion masses and quark mixing angles. Furthermore, we introduce two right handed Majorana neutrinos, i.e., $\nu_{1R}$, $\nu_{2R}$, in order to implement a realistic type I seesaw mechanism at tree level for the generation of the light active neutrino masses. Having only one right handed Majorana neutrino would lead to two massless active neutrinos, which is obviously in contradiction with the experimental data on neutrino oscillations. 
On the other hand, in order to get predictive SM fermion mass matrices consistent with low energy fermion flavor data, we assume the following VEV pattern for the $A_4$ triplet $SU(5)$ singlet scalars: 
%in order implement the  in our model
\begin{eqnarray}
\left\langle \xi _{e}\right\rangle &=&v_{\xi }^{\left( e\right) }\left(
1,0,0\right),\hspace{1cm}\left\langle \xi _{\mu }\right\rangle=v_{\xi}^{\left( \mu \right)
}\left( 0,1,0\right) ,\hspace{1cm}\left\langle \xi _{\tau }\right\rangle
=v_{\xi }^{\left( \tau \right) }\left( 0,0,1\right),\notag \\
\left\langle \eta _{1}\right\rangle &=&v_{\eta _{1}}\left( 0,1,1\right) ,%
\hspace{1cm}\left\langle \eta _{2}\right\rangle =v_{\eta _{2}}e^{i\frac{\phi
_{\nu }}{2}}\left(
1,3,1\right),\hspace{1cm}\left\langle \phi_{F_{4}}\right\rangle=v_{\phi_{F_{4}}}\left( 0,1,0\right),\notag\\
\left\langle \phi_{F_{5}}\right\rangle&=&v_{\phi_{F_{5}}}\left(1,0,0\right).
\end{eqnarray}
where the complex phases $\phi_{\nu }$ is introduced in the VEV pattern of the $A_4$ triplet scalar $\eta _{2}$ in order to successfully reproduce the experimental values of the leptonic mixing angles.
%We will assume, for the sake of simplicity that the Yukawa couplings of our modelLet us note that CP violation in the quark sector will arise from the down
\begin{table}
	\centering
	\footnotesize
	\captionsetup{width=0.9\textwidth}
	\begin{tabular}{| c c c c c c |}
		\hline
		\multirow{2}{*}{\rule{0pt}{5ex}Field}	& \multicolumn{5}{c |}{Representation/charge} \\
		\cline{2-6}
		\rule{0pt}{3ex}			& $SU(5)$ & $U(1)'$ & $A_4$ & $Z_3$ & $Z_{7}$\\ [0.75ex]
		\hline \hline
		\rule{0pt}{3ex}%
		$F$ 		 & $\bar{\bf 5}$ & 0 & ${\bf 3}$ & $0$ & $0$ \\
		$T_1$ 		 & ${\bf 10}$ & 0 & ${\bf 1}$ & $2$ & $3$ \\
		$T_2$ 		 & ${\bf 10}$ & 0 & ${\bf 1}$ & $1$ & $2$ \\
		$T_3$ 		 & ${\bf 10}$ & 0 & ${\bf 1}$ & $0$ & $0$ \\
		$F_4$ 		 & $\bar{\bf 5}$ & $q_{F_4}$ & ${\bf 1}$ & $-1$ & $-2$ \\
		$\bar{F}_4$ 		 & $\bf 5$ & $-q_{F_4}$ & ${\bf 1}$ & $1$ & $2$ \\
		$F_5$ 		 & $\bar{\bf 5}$ & $q_{F_5}$ & ${\bf 1}$ & $-2$ & $-3$ \\
		$\bar{F}_5$ 		 & $\bf 5$ & $-q_{F_5}$ & ${\bf 1}$ & $2$ & $3$ \\
		$T_4$ 		 & ${\bf 10}$ & $q_{T_4}$ & ${\bf 1}$ & $1$ & $2$ \\
		$T_5$ 		 & ${\bf 10}$ & $q_{T_4}$ & ${\bf 1}$ & $0$ & $0$ \\
		$\bar{T}_4$ 		 & $\bar{{\bf 10}}$ & $-q_{T_4}$ & ${\bf 1}$ & $-1$ & $-2$ \\
		$\bar{T}_5$ 		 & $\bar{{\bf 10}}$ & $-q_{T_5}$ & ${\bf 1}$ & $0$ & $0$ \\
		$\nu_{1R}$ 		 & ${\bf 1}$ & 0 & ${\bf 1}$ & $0$ & $-3$ \\
		$\nu_{2R}$ 		 & ${\bf 1}$ & 0 & ${\bf 1}$ & $0$ & $0$ \\
		\hline
		\hline
		\rule{0pt}{3ex}%
		$H^{(1)}_u$ & ${\bf 5}$ & $0$ & ${\bf 1}$ & $2$ & $0$\\
		$H^{(2)}_u$ & ${\bf 5}$ & $0$ & ${\bf 1}$ & $1$ & $0$\\
		$H^{(3)}_u$ & ${\bf 5}$ & $0$ & ${\bf 1}$ & $0$ & $0$\\
		$H^{(1)}_d$ & $\bar{{\bf 5}}$ & $0$ & ${\bf 1}$ &$0$ & $0$ \\
		$H^{(2)}_d$ & $\bar{{\bf 5}}$ & $-q_{F_4}$ & ${\bf 1}$ &$0$ & $5$ \\
		$H^{(3)}_d$ & $\bar{{\bf 5}}$ & $-q_{F_4}$ & ${\bf 1}$ &$0$ & $5$ \\
		$\phi_{F_{4}}$ & ${\bf 1}$ & $q_{F_4}$ & ${\bf 3}$ &$-1$ & $-2$ \\
		$\phi_{F_{5}}$ & ${\bf 1}$ & $q_{F_5}$ & ${\bf 3}$ &$-2$ & $-3$ \\
		$\phi_{T}$ & ${\bf 1}$ & $q_{T_5}$ & ${\bf 1}$ &$0$ & $0$ \\
		$\sigma$ & ${\bf 1}$ & $0$ & ${\bf 1}$ & $0$ & $-1$ \\	
		$\xi_{e}$ & ${\bf 1}$ & $0$ & ${\bf 3}$ &$-2$ & $-3$ \\
		$\xi_{\mu}$ & ${\bf 1}$ & $0$ & ${\bf 3}$ &$-1$ & $-2$ \\
		$\xi_{\tau}$ & ${\bf 1}$ & $0$ & ${\bf 3}$ &$0$ & $0$ \\
		$\eta_{1}$ & ${\bf 1}$ & $0$ & ${\bf 3}$ &$0$ & $3$ \\
		$\eta_{2}$ & ${\bf 1}$ & $0$ & ${\bf 3}$ &$0$ & $0$ \\
		\hline
		\hline
	\end{tabular}
	\caption{The $SU(5)\times A_4$ model considered in this paper. Notice that we included the field $H^{(3)}_d$ with the same quantum numbers of $H^{(2)}_d$ in order to fulfill the anomaly cancellation condition without introducing extra mixing terms between the light and heavy vector like fermions.}
	\label{tab:funfields}
\end{table}
Since the breaking of the $A_4\times Z_3\times Z_7$ discrete group generates the hierarchy among charged fermion masses and quark mixing angles and in order to relate the quark masses with the quark mixing parameters, we set the vacuum expectation values (VEVs) of the $SU(5)$ singlet scalars $\sigma$, $\xi_e$, $\xi_ {\mu}$, $\xi_ {\tau}$, $\eta_s$ ($s=1,2$), $\phi_{F_{4}}$ and $\phi_{F_{5}}$ with respect to the Wolfenstein parameter $\lambda=0.225$ and the model cutoff $\Lambda$, as follows:
\begin{equation}
v_{\phi_{F_{4}}}\sim v_{\phi_{F_{5}}}<<v_{\xi }^{\left( e\right) }\sim \lambda
^{7}\Lambda <<v_{\xi}^{\left( \mu \right) }\sim \lambda ^{5}\Lambda
<<v_{\xi }^{\left( \tau \right) }\sim \lambda ^{3}\Lambda <v_{\sigma}\sim v_{\eta_s}\sim
\lambda\Lambda,%v_{\sigma_1}\sim v_{\sigma_2}\sim v_{\sigma_3}\sim v_{\eta_1}\sim v_{\eta_2}\sim \lambda ^{2}\Lambda.
\label{VEVhierarchy}
\end{equation}
where $s=1,2$. The aforementioned VEV patterns are consistent with the scalar potential minimization equations for a large region parameter space. In particular, the VEV pattern of the $A_4$ scalar triplets $\eta_1$ and $\eta_2$ that participate in the neutrino Yukawa interactions have been derived for the first time in Ref. \cite{King:2013iva} in the framework of an $A_4$ flavor model. Assuming that the scale of breaking of the discrete symmetries is of the order of the GUT scale $\Lambda_{GUT}\approx 10^{16}$ GeV, from Eq. \ref{VEVhierarchy} we find for the model cutoff the estimate $\Lambda\approx 4.4\times 10^{16}$ GeV.

With the above particle content, the following Yukawa terms invariant under the group $SU(5)\times U(1)'\times A_4\times Z_3\times Z_7$ arise:
\begin{eqnarray}
-\mathcal{L}_{\text{Y}} &=&y_{11}^{\left( u\right) }T_{1}T_{1}H^{(1)}_{u}\frac{%
\sigma^{6}}{\Lambda^{6}}+y_{12}^{\left( u\right) }T_{1}T_{2}H^{(3)}_{u}\frac{\sigma^{5}}{\Lambda^{5}}+y_{22}^{\left( u\right) }T_{2}T_{2}H^{(2)}_{u}\frac{%
\sigma^{4}}{\Lambda^{4}}+y_{13}^{\left( u\right) }T_{1}T_{3}H^{(2)}_{u}\frac{%
\sigma^{3}}{\Lambda^{3}}\nonumber \\
&&+y_{23}^{\left( u\right) }T_{2}T_{3}H^{(1)}_{u}\frac{\sigma^2}{\Lambda^2}+y_{33}^{\left( u\right) }T_{3}T_{3}H^{(3)}_{u}+y_{11}^{\left( d\right) }T_{1}FH^{(1)}_{d}\frac{\xi _{e}}{\Lambda }+y_{22}^{\left( d\right) }T_{2}FH^{(1)}_{d}\frac{\xi_{\mu }}{\Lambda }+y_{33}^{\left( d\right) }T_{3}FH^{(1)}_{d}\frac{\xi _{\tau }}{\Lambda }\nonumber\\
&&+y_{24}^{\left( F\right) }T_2F_4H^{(2)}_d\frac{\sigma^5}{\Lambda^5}+z_{24}^{\left( F\right) }T_2F_4H^{(3)}_d\frac{\sigma^5}{\Lambda^5}+x_{24}^{\left( F\right) }\bar{F}_{4}F\phi _{F_{4}}+x_{15}^{\left( F\right) }\bar{F}_{5}F\phi_{F_{5}}+x_{35}^{\left( T\right) }\bar{T}_{5}T_{3}\phi^{*}
_{T}\nonumber \\
&&+\sum_{a=4}^{5}M_{F_{a}}\bar{F}_{a}F_{a}+\sum_{a=4}^{5}M_{T_{a}}\bar{T}_{a}T_{a}+x_{45}^{\left(F\right)}\bar{F}_{4}F_5\frac{\sigma^2\phi_{F_{4}}\phi^*_{F_{5}}}{\Lambda^3}+x_{54}^{\left(F\right)}\bar{F}_{5}F_4\frac{\sigma^2\phi_{F_{5}}\phi^*_{F_{4}}}{\Lambda^3} \nonumber \\
&&+\sum_{s=1}^{2}y_{s}^{\left( \nu \right) }FH^{(3)}_{u}\nu _{sR}\frac{\eta _{s}}{%
\Lambda }+x_{1}^{\left( \nu \right) }\nu_{1R}\overline{\nu_{1R}^{C}}\sigma+M^{\left( \nu \right) }\nu_{2R}\overline{\nu_{2R}^{C}},
\label{Ly}
\end{eqnarray}
where the Yukawa couplings are $\mathcal{O}(1)$ dimensionless parameters, assumed to be real for the sake of simplicity, whereas $M_{F_{a}}$, $M_{T_{a}}$ ($a=4,5$) and $M^{\left( \nu \right) }$ are dimensionful parameters.
%Let us note that from the up quark Yukawa terms, it follows that if one assume a complex phase in the VEV of the $\sigma$ field, that phase will be absorbed by redefinition of the SM fermionic fields ${\bf 10}_i\sim T_i$ $(i=1,2,3)$. %That $A_4$ triplet scalar $\xi _{e}$ will appear in the down type quark and charged lepton Yukawa interactions. 
%Thus, the down type quark Yukawa term $y_{12}^{\left( d\right) }T_{1}FH_{d}\frac{\xi _{e}}{\Lambda }$ is the only source of CP violation in the quark sector.\newline

On the other hand, it is worth mentioning that the lightest of the physical neutral scalar states of $H^{(1)}_{u}$, $H^{(2)}_{u}$, $H^{(3)}_{u}$, $H^{(1)}_{d}$, $H^{(2)}_{d}$ and $H^{(3)}_{d}$ is the SM-like $125$ GeV Higgs discovered at the LHC. As clearly seen from Eq. \ref{Ly}, the top quark mass mainly arises from $H^{(3)}_{u}$. Consequently, the dominant contribution to the SM-like $125$ GeV Higgs mainly arises from the CP even neutral state of the $SU(2)$ doublet part of $H^{(3)}_{u}$. In addition, let us note that the scalar potential of our model has many free parameters, which allows freedom to assume that the remaining scalars are heavy and outside the LHC reach. In addition, the loop effects of the heavy scalars contributing to precision observables can be suppressed by making an appropriate choice of the free parameters in the scalar potential. These adjustments do not affect the physical observables in the quark and lepton sectors, which are determined mainly by the Yukawa couplings. \newline
%$y_{ij}^{\left(u\right) }$ ($i,j=1,2,3$), $y_{12}^{\left(d\right)}$, $y_{21}^{\left(d\right)}$, $y_{22}^{\left(d\right)}$, $y_{23}^{\left(d\right)}$, $x_{ij}^{\left(u,d\right) }$

From the Yukawa interactions given above, it follows that the SM mass matrices for quarks and charged leptons are given by:
\begin{eqnarray}
M_{U} &=&\left( 
\begin{array}{ccc}
a_{11}^{\left( u\right) }\lambda ^{6} & a_{12}^{\left( u\right) }\lambda ^{5}
& a_{13}^{\left( u\right) }\lambda ^{3} \\ 
a_{12}^{\left( u\right) }\lambda ^{5} & a_{22}^{\left( u\right) }\lambda ^{4}
& a_{23}^{\left( u\right) }\lambda ^{2} \\ 
a_{13}^{\left( u\right) }\lambda ^{3} & a_{23}^{\left( u\right) }\lambda ^{2}
& a_{33}^{\left( u\right) }%
\end{array}%
\right) \frac{v}{\sqrt{2}},\nonumber \\%\hspace{0.3cm}
M_{D}&=&\left( 
\begin{array}{ccc}
a_{11}^{\left( d\right) }\lambda^{7} & 0 & 0 \\ 
0 & a_{22}^{\left( d\right) }\lambda ^{5}
& 0\\ 
0 & 0 & a_{33}^{\left( d\right) }\lambda ^{3}%
\end{array}%
\right) \frac{v}{\sqrt{2}},\nonumber\\
%\end{eqnarray}
%\begin{eqnarray}
M_{l} &=&\left( 
\begin{array}{ccc}
a_{11}^{\left( l\right) }\lambda^{7} & 0 & 0 \\ 
0 & a_{22}^{\left( l\right) }\lambda ^{5} & 0 \\ 
0 & 0 & a_{33}^{\left( l\right) }\lambda^{3}
\end{array}%
\right) \frac{v}{\sqrt{2}},\nonumber\\
a_{ij}^{\left( l\right) }&\approx&\frac{\kappa}{3}\left[ 1 + \delta _{i2}\delta _{j2} (f_2-1)+\delta _{i1}\delta _{j1} (f_1-1)  \right]a_{ji}^{\left(d\right)},
\label{Mq}
\end{eqnarray}
where $v=246$ GeV is the electroweak symmetry breaking scale, the factor of $3$ includes the QCD corrections, 
the $\kappa$ parameter is introduced to account for the threshold corrections to the down type quarks and charged lepton mass matrices \cite{Ross:2007az}, the factors $f_1$ and $f_2$ consider the effects of the mixings with the fourth and fifth families, respectively of charged leptons as in Eqs.~\ref{e22} and \ref{e11}. Let us note that we have assumed  as follows from an extension of our discussion given in Section \ref{Yprimebasis}, with appropiate modifications of the Eqs. \ref{e11} and \ref{e22}, that the factors $f_1$ and $f_2$ are given by:
\begin{eqnarray}
f_1&\approx&\cos\theta_{15}^{L},\hspace{1cm}\tan\theta_{15}^{L}\approx-\frac{x^{\left(F\right)}_{15}v_{\phi_{F_5}}}{M_{F_5}},\\
f_2&\approx&\cos\theta_{24}^{L}+y^{\left(F\right)}_{24}\frac{\sqrt{2}v_{H^{(2)}_d}}{v}\sin\theta_{24}^{L},\hspace{1cm}\tan\theta_{24}^{L}\approx-\frac{x^{\left(F\right)}_{24}v_{\phi_{F_4}}}{M_{F_4}}
\notag\\
%\frac{\sqrt{2}M_{F_4}}{v}\sin\theta_{24}^{L}\sin\theta_{24}^{R}
% \tan 2\theta_{24}^{R}&\approx&\lambda^5\frac{y^{\left(d\right)}_{22}x^{\left(F\right)}_{24}v_{H^{(2)}_d}v_{\phi_{F_4}}+y^{\left(F\right)}_{24}v_{H^{(2)}_d}M_{F_4}}{M^2_{F_4}+\left(x^{\left(F\right)}_{24}\right)^2v^2_{\phi_{F_4}}},\notag  
\end{eqnarray}
% Let us note, that the factors $f_1$ and $f_2$ are equal to $x^{F}_{15}\sin\theta_{15}^{L}\sin\theta_{15}^{R}$ and $x^{F}_{24}\sin\theta_{24}^{L}\sin\theta_{24}^{R}$ (which follows by extending our discussion given in Section \ref{Yprimebasis} and modifying Eqs. \ref{e11} and \ref{e22}, accordingly), respectively, and are crucial to generate the right values of the electron and muon masses without spoiling our predictions for the SM down type quark mass spectrum. Furthermore
 Then, considering $M_{F_4}\sim M_{F_5}\sim v_{\phi_{F_4}}\sim v_{\phi_{F_5}}\sim\mathcal{O}(1)$ TeV and $x^{\left(F\right)}_{15}\sim x^{\left(F\right)}_{24}\sim\mathcal{O}(1)$, we find that the factors $f_1$ and $f_2$ will be of order unity, which is crucial to generate the right values of the electron and muon masses without spoiling our predictions for the SM down type quark mass spectrum.

The mechanism described above works because 
the fifth generation of vector like leptons only mixes with the first family of charged leptons. Thus, as a result of this mixing, the 11 entry of the charged lepton mass matrix will receive a correction proportional to $\sin\theta^{L_L}_{15}\sin\theta^{e_R}_{15}$ instead of the quantity $\theta^{L_L}_{14}\theta^{e_R}_{14}$ shown in Eq. (\ref{e11}), thus yielding the right value of the electron mass (without spoiling the predictions of the down quark mass) and at the same time preventing the $\mu\to e\gamma$ decay. Thus the present flavor model has the features $\theta^{L,R}_{14}=\theta^{L,R}_{25}=0$, $\theta^{R}_{15}\approx 0$, $\theta^{R}_{24}\approx 0$ and $\theta^{L}_{15}\neq 0$ and $\theta^{L}_{24}\neq 0$. In this model, due to the discrete symmetry assignments, the mass matrices for SM down type quarks and charged leptons are diagonal and the right values of the electron and muon masses arise from the $\theta^{L}_{15}$ and $\theta^{L}_{24}$ mixing angles, respectively and the mixing between the fourth and fifth generation of vector like leptons is very tiny, thus allowing to have a realistic SM fermion mass spectrum and strongly suppressing the $\mu\to e\gamma$ rate. %Note that the $F_4$, $\bar{F}_4$
 
% $\theta _{14}^{L}$

Additionally, as seen from the Yukawa terms given in Eq. \ref{Ly}, considering $v_{\phi_{F_{4}}}\approx v_{\phi_{F_{5}}}\approx \mathcal{O}(1)$ TeV and assuming that the scale of breaking of the discrete symmetries is of the order of the GUT scale $\Lambda_{GUT}\approx 10^{16}$ GeV, we find that for dimensionless coupling of order unity, the mass mixing term between the 4th and the 5th generations of charged fermions is of the order of $10^{-10}$ GeV. Considering 4th and the 5th generations of charged leptons contained in the $5$, $\bar{5}$ $SU(5)$ representations have masses around $\mathcal{O}(1)$ TeV, we find a mixing angle between these fermions to be $\theta_{45}\approx 10^{-13}$, which implies that branching fractions for the charged lepton flavor violating decays induced by this mixing will be very tiny and well below their corresponding experimentally upper bound. Furthermore, as seen from Eq. \ref{Mq} and Yukawa terms $x_{24}^{\left( F\right) }\bar{F}_{4}F\phi _{F_{4}}$, $x_{15}^{\left( F\right) }\bar{F}_{5}F\phi_{F_{5}}$, $y_{24}^{\left( F\right) }T_2F_4H^{(2)}_d\frac{\sigma^5}{\Lambda^5}$ and \Antonio{$z_{24}^{\left( F\right) }T_2F_4H^{(3)}_d\frac{\sigma^5}{\Lambda^5}$} shown in Eq. \ref{Ly}, %In addition, 
%as indicated by 
% from the aforementioned Yukawa terms, we find that 
the SM charged lepton mass matrix is diagonal and $\theta _{24}^{L}\neq 0$, $\theta _{15}^{L}\neq 0$, respectively, whereas $\theta _{14}^{L,R}=\theta _{25}^{L,R}=0$, $\theta^{R}_{15}\approx 0$, $\theta^{R}_{24}\approx 0$, thus preventing contributions to the $\mu\to e\gamma$ decay rate arising from these mixing angles, as follows from Appendix \ref{mutoegamma}. 
 %which yields a vanishing contribution to the $\mu\to e\gamma$ decay rate arising from these mixing angles, as follows from Appendix \ref{mutoegamma}. 
Besides that, it is worth mentioning that we are considering incomplete $SU(5)$ multiplets for the fourth and fifth generations of fermions, which can be justified by assuming that the exotic down type quark fields contained in the $5$ and $\bar{5}$ irreps of $SU(5)$,  $F_4$, $F_5$, $\bar{F}_4$, $\bar{F}_5$ as well as the charged exotic leptons and down type quarks included in the $10$, $\bar{10}$ irreps of $SU(5)$ $T_4$, $T_5$, $\bar{T}_4$, $\bar{T}_5$, have masses much larger than the TeV scale, whereas the remaining fermions inside these representations do acquire TeV scale masses. That assumption will guarantee that $\theta^{Q}_{24}=\theta^{d}_{24}=\theta^{Q}_{15}=\theta^{d}_{15}=\theta^{d}_{35}=\theta^{e}_{35}=0$, $\theta^{e}_{15}\approx 0$, $\theta^{e}_{24}\approx 0$ despite the fact $\theta^{L}_{24}\neq 0$, $\theta^{L}_{15}\neq 0$ and $\theta^{u}_{35}\neq 0$.

Since we assume that the dimensionless Yukawa couplings appearing in Eq. (\ref{Ly}) are roughly of the same order of magnitude and we consider the
VEVs $v_{H^{(1)}_u}$, $v_{H^{(2)}_u}$, $v_{H^{(3)}_u}$, $v_{H^{(1)}_d}$ and $v_{H^{(2)}_d}$ of the order of the electroweak scale $v\simeq
246$ GeV, the hierarchy of charged fermion masses and quark mixing matrix
elements arises from the breaking of the $A_4\times Z_3\times Z_{7}$ symmetry. Let us note that despite the fact that the running of Yukawa couplings from the GUT scale up to the electroweak scale is not explicitly included in our calculations, our effective Yukawa couplings can accommodate for the renormalization groups effects, since these effective Yukawa couplings depend not only on the Yukawa couplings but also on the VEVs of the scalar fields participating in the Yukawa interactions and those VEVs can be adjusted to account for these effects. This freedom in adjusting the VEVs of the scalars fields participating in the Yukawa interactions is due to the large number of parameters in the scalar potential. Furthermore, we recall that we adjust the corresponding effective Yukawa couplings instead of the Yukawa couplings to fit the physical observables in the quark and lepton sector to their experimental values at the $M_Z$ scale.

The charged lepton and quark masses \cite{Bora:2012tx,Xing:2007fb}, the
quark mixing angles and Jarskog invariant \cite{Olive:2016xmw} can be well
reproduced in terms of natural parameters of order one, as shown in Table %
\ref{Tab}, starting from the following benchmark point: 
\begin{equation}
\begin{array}{c}
a_{11}^{\left( u\right) }\simeq 1.884 + 0.387i\,,\hspace{1cm}a_{12}^{\left( u\right)
}\simeq -1.933-0.211i,,\hspace{1cm}a_{22}^{\left( u\right) }\simeq 1.974- 0.023i\,, \\ 
a_{33}^{\left( u\right) }\simeq 0.989\,,\hspace{1cm}a_{13}^{\left( u\right)
}\simeq 0.691+0.277i\,,\hspace{1cm}a_{23}^{\left( u\right) }\simeq -0.788+0.014i\,, \\ 
a_{11}^{\left( l\right) }\simeq 0.095\,,\hspace{1cm}a_{22}^{\left( l\right)
}\simeq 1.016,\hspace{1cm}a_{33}^{\left( l\right) }\simeq 0.879,\\
\kappa\simeq 1.862,\hspace{1cm}f_{1}\simeq-0.729,\hspace{1cm}f_2\simeq 1.871.
\end{array}
\label{eq:bm-values}
\end{equation}%
\begin{table}[tbh]
\begin{center}
\begin{tabular}{c|l|l}
\hline\hline
Observable & Model value & Experimental value \\ \hline
$m_{e}(MeV)$ & \quad $0.487$ & \quad $0.487$ \\ \hline
$m_{\mu }(MeV)$ & \quad $102.8$ & \quad $102.8\pm 0.0003$ \\ \hline
$m_{\tau }(GeV)$ & \quad $1.75$ & \quad $1.75\pm 0.0003$ \\ \hline
$m_{u}(\mathrm{MeV})$ & \quad $1.45$ & \quad $1.45_{-0.45}^{+0.56}$ \\ \hline
$m_{c}(\mathrm{MeV})$ & \quad $635$ & \quad $635\pm 86$ \\ \hline
$m_{t}(\mathrm{GeV})$ & \quad $172.1$ & \quad $172.1\pm 0.6\pm 0.9$ \\ \hline
$m_{d}(\mathrm{MeV})$ & \quad $2.9$ & \quad $2.9_{-0.4}^{+0.5}$ \\ \hline
$m_{s}(\mathrm{MeV})$ & \quad $57.7$ & \quad $57.7_{-15.7}^{+16.8}$ \\ \hline
$m_{b}(\mathrm{GeV})$ & \quad $2.82$ & \quad $2.82_{-0.04}^{+0.09}$ \\ \hline
$\sin \theta^{(q)}_{12}$ & \quad $0.225$ & \quad $0.225$ \\ \hline
$\sin \theta^{(q)}_{23}$ & \quad $0.0414$ & \quad $0.0414$ \\ \hline
$\sin \theta^{(q)}_{13}$ & \quad $0.00355$ & \quad $0.00357$ \\ \hline
$J$ & \quad $2.99\times 10^{-5}$ & \quad $2.96_{-0.16}^{+0.20}\times 10^{-5}$ \\ \hline\hline
\end{tabular}%
\end{center}
\caption{Model and experimental values of the charged fermion masses and CKM
parameters.}
\label{Tab}
\end{table}
In Table \ref{Tab} we show the model and experimental values for the
physical observables of the quark sector. We use the $M_{Z}$-scale
experimental values of the quark masses given by Ref. \cite{Bora:2012tx}
(which are similar to those in \cite{Xing:2007fb}). The experimental values
of the CKM parameters are taken from Ref. \cite{Olive:2016xmw}. As indicated
by Table \ref{Tab}, the obtained quark masses, quark mixing angles, and CP
violating phase are consistent with the low energy quark flavor data. 
%The correlation between the strange and down quark masses is shown in Figure \ref{mq}. To generate this plot we performed small variations around the best-fit values requiring that the obtained values for the charged leptons masses agree with their experimental data. 
As shown from Table \ref{Tab}, the obtained values for the SM down type quark masses are inside the $1\sigma$ experimentally allowed range. In addition, our obtained values for the SM up type quark masses are inside the $1\sigma$ experimentally allowed range, as indicated in Table \ref{Tab}. \newline
%\begin{center}
% \begin{figure}[tbp]
% \hspace{4cm}\includegraphics[width=0.49\textwidth]{msvsmd.pdf}
% \caption{Correlation between the strange and down quark masses.}
% \label{mq}
% \end{figure}
%\end{center}

On the other hand, from the neutrino Yukawa interactions, we find that the Dirac and Majorna neutrino mass matrices are given by:
\begin{equation}
m_{\nu D}=\left( 
\begin{array}{cc}
0 & b \\ 
a & 3b \\ 
a & b%
\end{array}%
\right) ,\hspace{1cm}\hspace{1cm}M_{R}=\left( 
\begin{array}{cc}
M_{atm} & 0 \\ 
0 & M_{sol}%
\end{array}%
\right) ,\hspace{1cm}\hspace{1cm}b=\left\vert b\right\vert e^{i\frac{\phi
_{\nu }}{2}}.
\end{equation}
Since the right handed Majorana neutrinos $\nu_{1R}$ and $\nu_{2R}$ acquire very large masses, the light active neutrino masses are generated via tree level type I seesaw mechanism and thus the light neutrino mass matrix takes the following form:
\begin{equation}
m_{\nu }=m_{\nu D}M_{R}^{-1}m_{\nu D}^{T}=m_{\nu a}\left( 
\begin{array}{ccc}
0 & 0 & 0 \\ 
0 & 1 & 1 \\ 
0 & 1 & 1%
\end{array}%
\right) +m_{\nu b}e^{i\phi _{\nu }}\left( 
\begin{array}{ccc}
1 & 3 & 1 \\ 
3 & 9 & 3 \\ 
1 & 3 & 1%
\end{array}%
\right) ,
\end{equation}
where $m_{\nu a}$ and $m_{\nu b}$ are given by:
\begin{equation}
m_{\nu a}=\frac{a^{2}}{M_{atm}},\hspace{1cm}\hspace{1cm}m_{\nu b}=\frac{b^{2}}{%
M_{sol}}.
\end{equation}
The neutrino mass squared splittings, light active neutrino masses, leptonic mixing angles and CP violating phase for the scenario of normal neutrino mass hierarchy can be very well reproduced, as shown in Table \ref{Neutrinos}, for the following benchmark point:
\begin{equation}
m_{\nu a}\simeq 26.57\mbox{meV},\hspace{1cm}m_{\nu b}\simeq 2.684\,\mbox{meV},%
\hspace{1cm}\phi _{\nu }=120^{\circ }.
\end{equation}
In addition, we find that the light active neutrino masses are:
\begin{equation}
m_1=0,\hspace{1cm}m_2=8.59\mbox{meV}\hspace{1cm}m_3=49.81\mbox{meV}.
\end{equation}

\begin{table}[tbh]
\begin{center}
\resizebox{15cm}{!}{
%\footnote
\renewcommand{\arraystretch}{1.2}
\begin{tabular}{c|l|l|l|l|l}
\hline\hline
Observable & Model & bpf $\pm 1\sigma$ \cite{deSalas:2017kay} & bpf $\pm 1\sigma$ \cite{Esteban:2016qun}  & $3\sigma$ range \cite{deSalas:2017kay} & $3\sigma$ range \cite{Esteban:2016qun}\\ \hline
% $m_{1}(meV)$ & \quad $0$ & \quad $0$  & & \\ \hline
% $m_{2}(meV)$ & \quad $8.59$ & \quad $8.66\pm 0.10$  & & \\ \hline
% $m_{3}(meV)$ & \quad $49.81$ & \quad $49.57\pm 0.47$  & & \\ \hline
$\Delta m_{21}^{2}$ [$10^{-5}$eV$^{2}$] & \quad $7.38$ & \quad $7.55_{-0.16}^{+0.20}$ & \quad $7.40_{-0.20}^{+0.21}$  & \quad $7.05-8.14$ & \quad $6.80-8.02$\\ \hline
$\Delta m_{31}^{2}$ [$10^{-3}$eV$^{2}$] & \quad $2.48$ & \quad $2.50\pm
0.03$ & \quad $2.494_{-0.031}^{+0.033}$  & \quad $2.41-2.60$ & \quad $2.399-2.593$\\ \hline
$\theta^{(l)}_{12} (^{\circ })$  & \quad $34.32$ & \quad $34.5_{-1.0}^{+1.2}$ & \quad $36.62_{-0.76}^{+0.78}$ & \quad $31.5-38.0$ & \quad $31.42-36.05$\\ \hline
$\theta^{(l)}_{13} (^{\circ })$ & \quad $8.67$ & \quad $8.45_{-0.14}^{+0.16}$ & \quad $8.54\pm 0.15$  & \quad $8.0-8.9$ & \quad $8.09-8.98$\\ \hline
$\theta^{(l)}_{23} (^{\circ })$  & \quad $45.77$ & \quad $47.9_{-1.7}^{+1.0}$ & \quad $47.2_{-3.9}^{+1.9}$   & \quad $41.8-50.7$ & \quad $40.3-51.5$\\ \hline
$\delta^{(l)}_{CP} (^{\circ })$ &  $-86.67$ & \quad $-142_{-27}^{+38}$ & \quad $-108_{-31}^{+43}$  & \quad $157-349$ & \quad $144-374$\\ %\hline
\hline%\hline
$(\alpha_3-\alpha_2) (^{\circ })$ & $-71.90$ &  \qquad   - &  \qquad -&  \qquad -& \qquad  -\\ \hline\hline
\end{tabular}}%
\end{center}
\caption{Model and experimental values of the light active neutrino masses, leptonic mixing angles and CP violating phase for the scenario of normal (NH) neutrino mass hierarchy. The difference $\alpha_3-\alpha_2$ between the Majorana phases predicted by the model is also shown. The experimental values are taken from Refs. \cite{deSalas:2017kay,Esteban:2016qun}}
\label{Neutrinos}
\end{table}

From Table \ref{Neutrinos}, it follows that the neutrino mass squared splittings, i.e, $\Delta m_{21}^{2}$ and $\Delta m_{31}^{2}$, the leptonic mixing angles $\theta^{(l)}_{12}$, $\theta^{(l)}_{23}$, $\theta^{(l)}_{13}$ and the Dirac leptonic CP violating phase are consistent with neutrino oscillation experimental data for the scenario of normal neutrino mass hierarchy. Let us note that, for the inverted neutrino mass hierarchy, the obtained leptonic mixing parameters are very much outside the $3\sigma$ experimentally allowed range. Consequently, our model is only viable for the scenario of normal neutrino mass hierarchy.

Another important observable, worth to be determined in this model, is the effective Majorana neutrino mass parameter of neutrinoless double beta decay, which give us information on the Majorana nature of neutrinos. The amplitude for this process is directly proportional to the effective Majorana mass parameter, which is defined as:
\begin{equation}
  m_{ee}=\left\vert \sum_{j}U_{ek}^{2}m_{\nu _{k}}\right\vert=\left\vert m_{\nu _{1}}c^{2}_{12}c^{2}_{13}+m_{\nu _{2}}s^{2}_{12}c^{2}_{13}e^{i\alpha_{21}}+m_{\nu _{3}}s^{2}_{13}e^{i\left(\alpha_{31}-2\delta^{(l)}_{CP}\right)}\right\vert ,  \label{mee}
\end{equation}
where $U_{ej}$ and $m_{\nu _{k}}$ are the the PMNS leptonic mixing matrix elements and the neutrino Majorana masses, respectively. Furthermore, $s_{ij}=\sin\theta^{(l)}_{ij}$, $c_{ij}=\cos\theta^{(l)}_{ij}$, $\alpha_{ij}=\alpha_i-\alpha_j$, being $\alpha_i$ the Majorana phases, with $i\neq j$ and $i,j=1,2,3$. 
Note that since $m_{\nu _{1}}=0$ in our model, then $m_{ee}$ only depends on the relative phase 
$\alpha_{32}-2\delta^{(l)}_{CP}$ where $\alpha_{32}=\alpha_3-\alpha_2$.

\begin{figure}[tbp]
\includegraphics[width=0.49\textwidth]{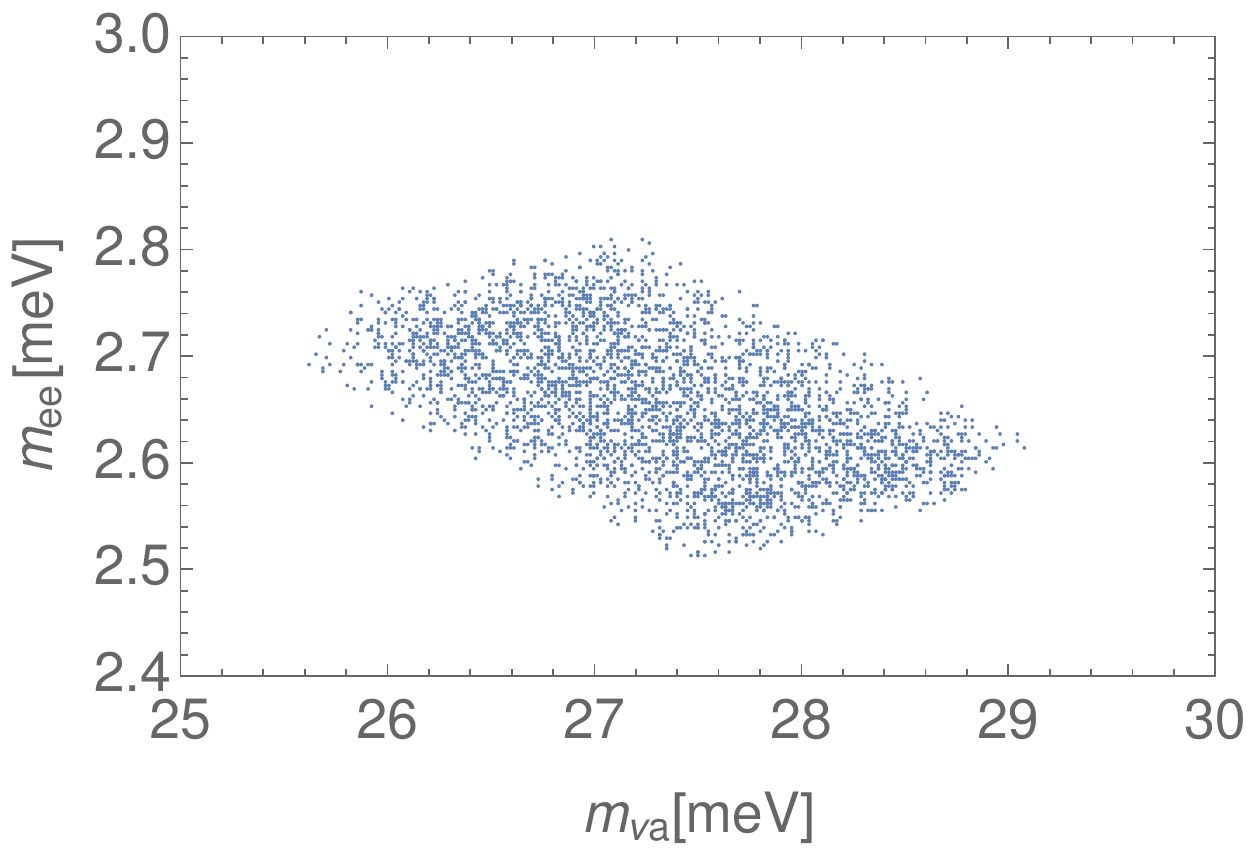}\includegraphics[width=0.49\textwidth]{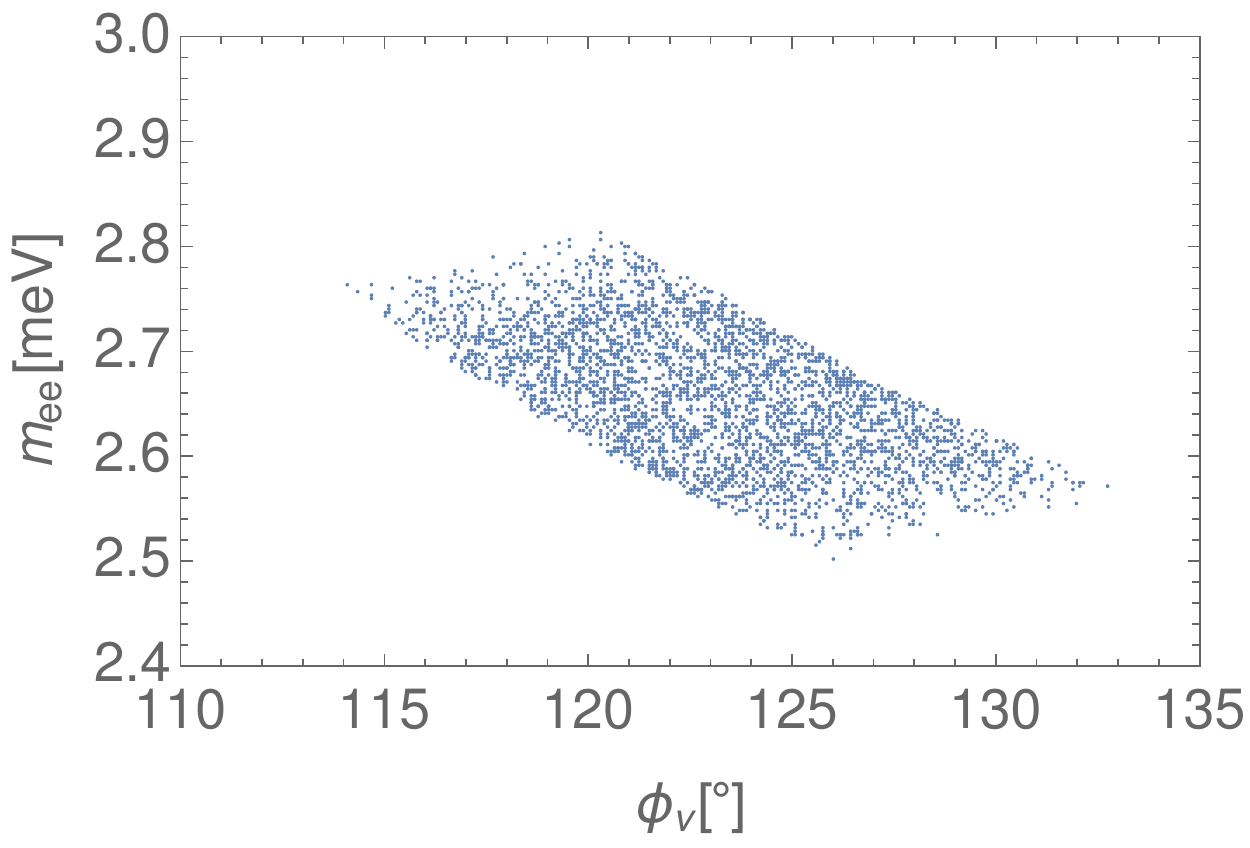}\newline
\includegraphics[width=0.49\textwidth]{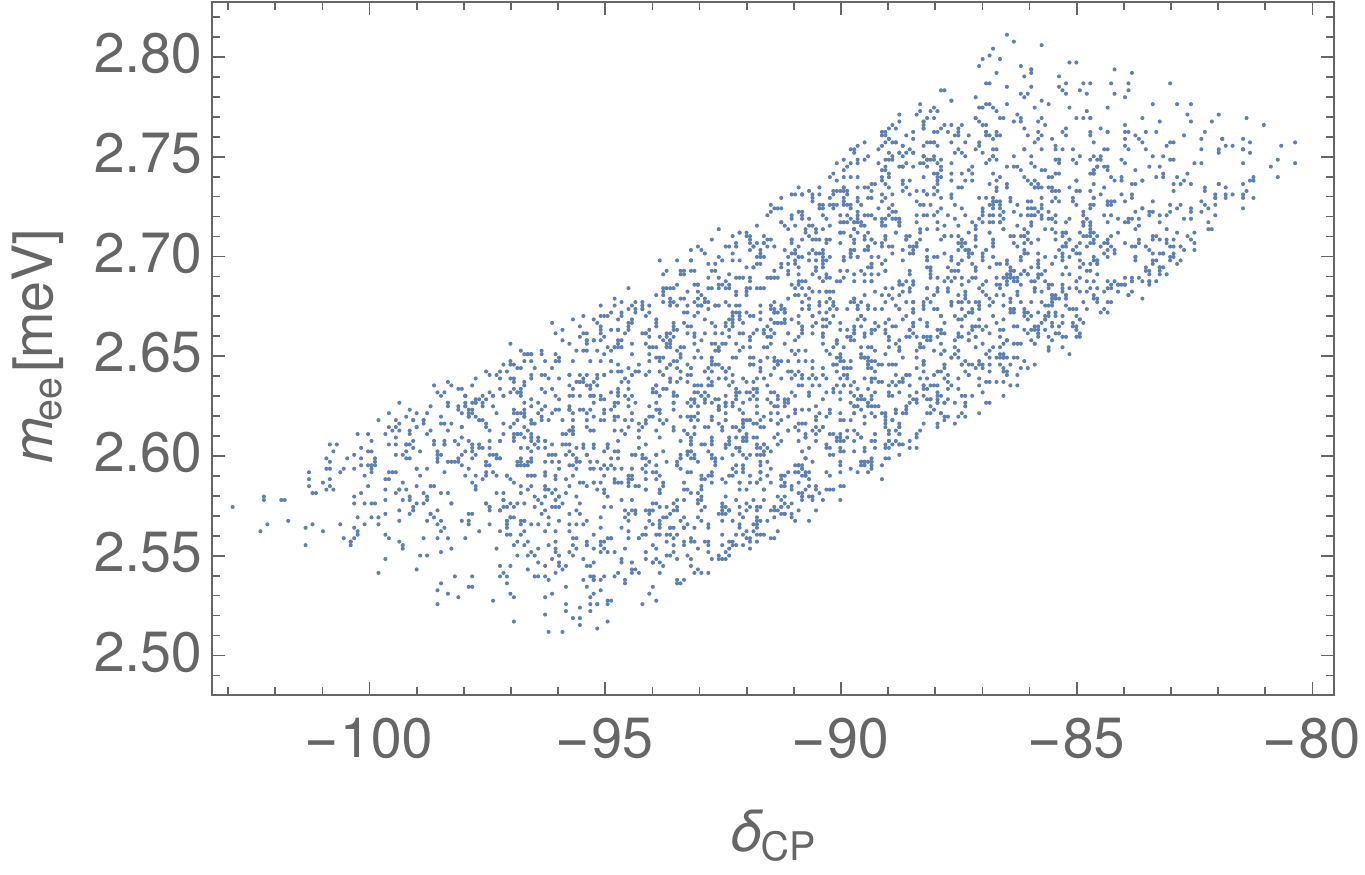}
\caption{Effective Majorana neutrino mass parameter as functions of the $m_{\nu a}$, $\phi_{\nu}$ parameters and leptonic Dirac CP violating phase $\delta_{CP}$.} 
\label{meefig}
\end{figure}
Figure \ref{meefig} shows the effective Majorana neutrino mass parameter as functions of the $m_{\nu a}$, $\phi_{\nu}$ and $\delta_{CP}$ parameters (here $\delta_{CP}$ is the leptonic Dirac CP violating phase). To obtain the plots of Figure \ref{meefig}, the parameters $m_{\nu a}$, $\phi_ {\nu}$ and $\delta_{CP}$ were randomly generated in a range of values where the neutrino mass squared splittings and leptonic mixing parameters are inside the $3\sigma$ experimentally allowed range. As indicated by Figure \ref{meefig},  our model predicts effective Majorana neutrino mass parameter in the range $2.5$ meV $\lesssim m_{ee}\lesssim$ $2.8$ meV, for the scenario of normal neutrino mass hierarchy.

Our obtained range of values for the effective Majorana neutrino mass parameter is beyond the reach of the present and forthcoming $0\nu \beta \beta $-decay experiments. The current most stringent experimental upper limit on the effective Majorana neutrino mass parameter $m_{ee}\leq 160$ meV is set by 
$T_{1/2}^{0\nu
\beta \beta }(^{136}\mathrm{Xe})\geq 1.1\times 10^{26}$ yr at 90\% C.L.
from the KamLAND-Zen experiment \cite{KamLAND-Zen:2016pfg}.

\section{Conclusion}
\label{conclusion}

In this paper we have shown that $SU(5)$ GUTs with multiple vector-like families at the GUT scale which transform under
a gauged $U(1)'$ (under which the three chiral families are neutral) can result from two vector-like families at low energies which 
can induce non-universal and flavourful $Z'$ couplings,
which can account for the B physics anomalies in $R_{K^{(*)}}$. In such theories, we have shown that the same physics 
which explains $R_{K^{(*)}}$
also correct the Yukawa relation $Y_e=Y_d^T$ in the muon sector without the need for higher Higgs representations.

To illustrate the mechanism, we have constructed a concrete a model based on $SU(5)\times A_4 \times Z_3\times Z_7$ with two vector-like families at the GUT scale,
and two right-handed neutrinos,
leading to successful fit to quark and lepton (including neutrino) masses, mixing angles and CP phases,
 where the constraints from lepton flavour violation require $Y_e$ to be diagonal. 
This particular model predicts normal neutrino mass ordering with the inverted ordering disfavoured by our fit,
and an effective Majorana neutrino mass parameter in the range $2.5$ meV $\lesssim m_{ee}\lesssim$ $2.8$ meV, for the scenario of normal neutrino mass hierarchy.

In conclusion, we have shown that the idea of a flavourful $Z'$ arising from mixing with a vector-like families, can be extended to $SU(5)$ GUTs.
In such theories, we have shown that 
the physics responsible for explaining the B physics anomalies in $R_{K^{(*)}}$ as a result of modified couplings in the muon sector
can also lead to violation of the $SU(5)$ Yukawa relations $Y_e=Y_d^T$ in the muon sector without the need for higher Higgs representations.

\subsection*{Acknowledgements}

SFK acknowledges the STFC Consolidated Grant ST/L000296/1 and the European Union's Horizon 2020 Research and Innovation programme under Marie Sk\l{}odowska-Curie grant agreements Elusives ITN No.\ 674896 and InvisiblesPlus RISE No.\ 690575 and would like to thank AECH and the 
Universidad T\'{e}cnica Federico Santa Mar\'{\i}a for their hospitality.
AECH has been supported by Chilean grants Fondecyt No. 1170803 and CONICYT PIA/Basal FB0821. SFK thanks Universidad Técnica Federico Santa
Mar\'{\i}a for hospitality, where this work was started. The visit of SFK to Universidad T\'{e}cnica Federico Santa Mar\'{\i}a was supported by Chilean grant Fondecyt No. 1170803.

\appendix

\section{The product rules for $A_4$}

\label{A4}The $A_{4}$ group, which is the group of even permutations of four elements, is the smallest discrete group having one three-dimensional representation, i.e., $\mathbf{3}$ as well as three inequivalent one-dimensional representations, i.e., $\mathbf{1}$, $\mathbf{1}^{\prime }$ and $\mathbf{1}^{\prime \prime }$, satisfying the
following product rules:
\begin{eqnarray}
&&\hspace{18mm}\mathbf{3}\otimes \mathbf{3}=\mathbf{3}_{s}\oplus \mathbf{3}%
_{a}\oplus \mathbf{1}\oplus \mathbf{1}^{\prime }\oplus \mathbf{1}^{\prime
\prime },  \label{A4-singlet-multiplication} \\[0.12in]
&&\mathbf{1}\otimes \mathbf{1}=\mathbf{1},\hspace{5mm}\mathbf{1}^{\prime
}\otimes \mathbf{1}^{\prime \prime }=\mathbf{1},\hspace{5mm}\mathbf{1}%
^{\prime }\otimes \mathbf{1}^{\prime }=\mathbf{1}^{\prime \prime },\hspace{%
5mm}\mathbf{1}^{\prime \prime }\otimes \mathbf{1}^{\prime \prime }=\mathbf{1}%
^{\prime },  \notag
\end{eqnarray}%
Considering $\left( x_{1},y_{1},z_{1}\right) $ and $\left(
x_{2},y_{2},z_{2}\right) $ as the basis vectors for two $A_{4}$-triplets $%
\mathbf{3}$, the following relations are fullfilled:
\begin{eqnarray}
&&\left( \mathbf{3}\otimes \mathbf{3}\right) _{\mathbf{1}%
}=x_{1}y_{1}+x_{2}y_{2}+x_{3}y_{3}, \notag \\
&&\left( \mathbf{3}\otimes \mathbf{3}\right) _{\mathbf{1}^{\prime
}}=x_{1}y_{1}+\omega x_{2}y_{2}+\omega ^{2}x_{3}y_{3},\notag \\
&&\left( \mathbf{3}\otimes \mathbf{3}\right) _{\mathbf{1}^{\prime \prime
}}=x_{1}y_{1}+\omega ^{2}x_{2}y_{2}+\omega x_{3}y_{3}\notag \\
&&\left( \mathbf{3}\otimes \mathbf{3}\right) _{\mathbf{3}_{s}}=\left(
x_{2}y_{3}+x_{3}y_{2},x_{3}y_{1}+x_{1}y_{3},x_{1}y_{2}+x_{2}y_{1}\right),\notag\\
&&\left( \mathbf{3}\otimes \mathbf{3}\right) _{\mathbf{3}_{a}}=\left(
x_{2}y_{3}-x_{3}y_{2},x_{3}y_{1}-x_{1}y_{3},x_{1}y_{2}-x_{2}y_{1}\right),\label{triplet-vectors}
\end{eqnarray}%
where $\omega =e^{i\frac{2\pi }{3}}$. The representation $\mathbf{1}$ is
trivial, while the non-trivial $\mathbf{1}^{\prime }$ and $\mathbf{1}%
^{\prime \prime }$ are complex conjugate to each other. Some reviews of
discrete symmetries in particle physics are found in Refs. \cite%
{Ishimori:2010au,Altarelli:2010gt,King:2013eh,King:2014nza,King:2017guk}.

%\newpage

\section{Branching ratio of $\mu\to e\gamma$}
\label{mutoegamma}
The branching ratio of the $\mu\to e\gamma$ decay in our model, for the scenario where the charged lepton masses are much smaller than the $Z^{\prime}$ mass is given by \cite{Chiang:2011cv,Raby:2017igl}:
%, we get the following estimate:
\begin{eqnarray}
Br(\mu\to e\gamma)&=&\frac{m_{\mu }^{3}}{2304\pi ^{4}\Gamma_{\mu }M_{Z^{\prime }}^{4}}\left[
\left\vert 3C_{e_{R}E_{R}}C_{\mu _{L}E_{L}}m_{E}+C_{e_{R}\mu _{R}}\left(
3C_{\mu _{L}\mu _{L}}-C_{\mu _{R}\mu _{R}}\right) m_{\mu }\right\vert
\right. ^{2}\notag \\
&&+\left. \left\vert 3C_{e_{L}E_{L}}C_{\mu _{R}E_{R}}m_{E}+C_{e_{L}\mu
_{L}}\left( 3C_{\mu _{R}\mu _{R}}-C_{\mu _{L}\mu _{L}}\right) m_{\mu
}\right\vert ^{2}\right] 
%\left(\theta^L_{12}+\theta^R_{12}\right)^2\approx 4\times 10^{-13}
\end{eqnarray}
% Br(\mu\to e\gamma)\approx\frac{\alpha_{em}G^2_F}{4\pi^4\Gamma_{\mu}}\frac{M^4_Wm^5_{\mu}}{M^4_{Z'}}\left(\frac{\left(C_{\mu_L\mu_L}+C_{\mu_R\mu_R}\right)\sin\theta_W}{2g'}\right)^4\left(\theta^L_{12}+\theta^R_{12}\right)^2\approx 4\times 10^{-13}
where:
\begin{eqnarray}
C_{\mu _{L}E_{L}}&=&g^{\prime }q_{L4}\sin \theta _{24}^{L},%
\hspace{1.5cm}C_{\mu _{R}E_{R}}=g^{\prime }q_{e4}\sin \theta
_{24}^{R}\notag\\
C_{e_{L}E_{L}}&=&\sin \theta _{12}^{L}C_{\mu
_{L}E_{L}}=g^{\prime }q_{L4}\sin \theta _{12}^{L}\sin \theta _{24}^{L},%
\hspace{1.5cm}\notag \\
C_{e_{R}E_{R}}&=&\sin \theta _{12}^{R}C_{\mu
_{R}E_{R}}=g^{\prime }q_{e4}\sin \theta _{12}^{R}\sin \theta _{24}^{R},\notag \\
C_{e_{L}\mu_{L}}&=&g'q_{L4}\left(\sin \theta _{12}^{L}\sin^2\theta _{24}^{L}\cos^2\theta _{24}^{L}+\sin\theta _{14}^{L}\sin\theta _{24}^{L}\cos\theta _{14}^{L}\right),\notag \\
C_{e_{R}\mu_{R}}&=&g'q_{e4}\left(\sin \theta _{12}^{R}\sin^2\theta _{24}^{R}\cos^2\theta _{24}^{R}+\sin\theta _{14}^{R}\sin\theta _{24}^{R}\cos\theta _{14}^{R}\right),
\end{eqnarray}
being $\Gamma_{\mu}=\frac{G^2_Fm^5_{\mu}}{192\pi^3}=3\times 10^{-19}$ GeV the total muon decay width. The generalization to the fifth generation of fermions is straightforward and is made by replacing $\theta^{L,R}_{n4}$ by $\theta^{L,R}_{n5}$ ($n=1,2$). Note that the branching ratio becomes zero for a diagonal SM charged lepton mass matrix provided that $\theta _{14}^{L}=\theta _{14}^{R}=\theta _{25}^{L}=\theta _{25}^{R}=0$, which is the case of our flavor model described in section \ref{SU5A4}.%\vspace{-1.0cm}

% when the charged lepton mixing between the electron and muon vanishes, $ \theta _{12}^{L}=\theta _{12}^{R}=0$,
% as expected.
% C_{e_{L}\mu_{L}}&=&g'q_{L4}\left[\sin \theta _{12}^{L}\sin^2\theta _{24}^{L}\cos^2\theta _{24}^{L}+s^{L}_{14}s^{L}_{24}c^{L}_{14}\right],\notag \\
% C_{e_{R}\mu_{R}}&=&g'q_{e4}\left[\sin \theta _{12}^{R}(s^{e}_{24})^2(c^{e}_{24})^2+s^{e}_{14}s^{e}_{24}c^{e}_{14}\right],
%,\hspace{1.5cm}C_{E_{L}E_{L}}=g^{\prime}q_{L4}


\begin{thebibliography}{99}
\setlength{\itemsep}{0em}
%\cite{Langacker:2008yv}
\bibitem{Langacker:2008yv} 
  P.~Langacker,
  %``The Physics of Heavy $Z^\prime$ Gauge Bosons,''
  Rev.\ Mod.\ Phys.\  {\bf 81}, 1199 (2009)
  doi:10.1103/RevModPhys.81.1199
  [arXiv:0801.1345 [hep-ph]].
  %%CITATION = doi:10.1103/RevModPhys.81.1199;%%
  %894 citations counted in INSPIRE as of 06 Mar 2018



%\cite{Descotes-Genon:2013wba}
\bibitem{Descotes-Genon:2013wba} 
  S.~Descotes-Genon, J.~Matias and J.~Virto,
  %``Understanding the $B\to K^*\mu^+\mu^-$ Anomaly,''
  Phys.\ Rev.\ D {\bf 88}, 074002 (2013)
  doi:10.1103/PhysRevD.88.074002
  [arXiv:1307.5683 [hep-ph]].
  %%CITATION = doi:10.1103/PhysRevD.88.074002;%%
  %310 citations counted in INSPIRE as of 06 Mar 2018



%\cite{Altmannshofer:2013foa}
\bibitem{Altmannshofer:2013foa} 
  W.~Altmannshofer and D.~M.~Straub,
  %``New Physics in $B \to K^*\mu\mu$?,''
  Eur.\ Phys.\ J.\ C {\bf 73}, 2646 (2013)
  doi:10.1140/epjc/s10052-013-2646-9
  [arXiv:1308.1501 [hep-ph]].
  %%CITATION = doi:10.1140/epjc/s10052-013-2646-9;%%
  %229 citations counted in INSPIRE as of 06 Mar 2018



%\cite{Ghosh:2014awa}
\bibitem{Ghosh:2014awa} 
  D.~Ghosh, M.~Nardecchia and S.~A.~Renner,
  %``Hint of Lepton Flavour Non-Universality in $B$ Meson Decays,''
  JHEP {\bf 1412}, 131 (2014)
  doi:10.1007/JHEP12(2014)131
  [arXiv:1408.4097 [hep-ph]].
  %%CITATION = doi:10.1007/JHEP12(2014)131;%%
  %120 citations counted in INSPIRE as of 06 Mar 2018



%\cite{Aaij:2014ora}
\bibitem{Aaij:2014ora} 
  R.~Aaij {\it et al.} [LHCb Collaboration],
  %``Test of lepton universality using $B^{+}\rightarrow K^{+}\ell^{+}\ell^{-}$ decays,''
  Phys.\ Rev.\ Lett.\  {\bf 113}, 151601 (2014)
  doi:10.1103/PhysRevLett.113.151601
  [arXiv:1406.6482 [hep-ex]].
  %%CITATION = doi:10.1103/PhysRevLett.113.151601;%%
  %519 citations counted in INSPIRE as of 06 Mar 2018


 \bibitem{Bifani}
S.~Bifani for the LHCb Collaboration, {\it Search for new physics with $b \to s \ell^+ \ell^-$ decays at LHCb}, CERN Seminar, 18 April 2017,
{\tt https://cds.cern.ch/record/2260258}.


%\cite{Hiller:2017bzc}
\bibitem{Hiller:2017bzc} 
  G.~Hiller and I.~Nisandzic,
  %``$R_K$ and $R_{K^{\ast}}$ beyond the standard model,''
  Phys.\ Rev.\ D {\bf 96}, no. 3, 035003 (2017)
  doi:10.1103/PhysRevD.96.035003
  [arXiv:1704.05444 [hep-ph]].
  %%CITATION = doi:10.1103/PhysRevD.96.035003;%%
  %66 citations counted in INSPIRE as of 06 Mar 2018


%\cite{Ciuchini:2017mik}
\bibitem{Ciuchini:2017mik}
  M.~Ciuchini, A.~M.~Coutinho, M.~Fedele, E.~Franco, A.~Paul, L.~Silvestrini and M.~Valli,
  %``On Flavourful Easter eggs for New Physics hunger and Lepton Flavour Universality violation,''
  Eur.\ Phys.\ J.\ C {\bf 77} (2017) no.10,  688
  doi:10.1140/epjc/s10052-017-5270-2
  [arXiv:1704.05447 [hep-ph]].
  %%CITATION = doi:10.1140/epjc/s10052-017-5270-2;%%
  %67 citations counted in INSPIRE as of 22 Mar 2018



%\cite{Geng:2017svp}
\bibitem{Geng:2017svp} 
  L.~S.~Geng, B.~Grinstein, S.~Jäger, J.~Martin Camalich, X.~L.~Ren and R.~X.~Shi,
  %``Towards the discovery of new physics with lepton-universality ratios of $b\to s\ell\ell$ decays,''
  Phys.\ Rev.\ D {\bf 96}, no. 9, 093006 (2017)
  doi:10.1103/PhysRevD.96.093006
  [arXiv:1704.05446 [hep-ph]].
  %%CITATION = doi:10.1103/PhysRevD.96.093006;%%
  %75 citations counted in INSPIRE as of 06 Mar 2018



%\cite{Capdevila:2017bsm}
\bibitem{Capdevila:2017bsm} 
  B.~Capdevila, A.~Crivellin, S.~Descotes-Genon, J.~Matias and J.~Virto,
  %``Patterns of New Physics in $b\to s\ell^+\ell^-$ transitions in the light of recent data,''
  JHEP {\bf 1801}, 093 (2018)
  doi:10.1007/JHEP01(2018)093
  [arXiv:1704.05340 [hep-ph]].
  %%CITATION = doi:10.1007/JHEP01(2018)093;%%
  %100 citations counted in INSPIRE as of 06 Mar 2018



%\cite{Ghosh:2017ber}
\bibitem{Ghosh:2017ber} 
  D.~Ghosh,
  %``Explaining the $R_K$ and $R_{K^*}$ anomalies,''
  Eur.\ Phys.\ J.\ C {\bf 77}, no. 10, 694 (2017)
  doi:10.1140/epjc/s10052-017-5282-y
  [arXiv:1704.06240 [hep-ph]].
  %%CITATION = doi:10.1140/epjc/s10052-017-5282-y;%%
  %35 citations counted in INSPIRE as of 06 Mar 2018



%\cite{Bardhan:2017xcc}
\bibitem{Bardhan:2017xcc} 
  D.~Bardhan, P.~Byakti and D.~Ghosh,
  %``Role of Tensor operators in $R_K$ and $R_{K^*}$,''
  Phys.\ Lett.\ B {\bf 773}, 505 (2017)
  doi:10.1016/j.physletb.2017.08.062
  [arXiv:1705.09305 [hep-ph]].
  %%CITATION = doi:10.1016/j.physletb.2017.08.062;%%
  %20 citations counted in INSPIRE as of 06 Mar 2018



%\cite{Glashow:2014iga}
\bibitem{Glashow:2014iga} 
  S.~L.~Glashow, D.~Guadagnoli and K.~Lane,
  %``Lepton Flavor Violation in $B$ Decays?,''
  Phys.\ Rev.\ Lett.\  {\bf 114}, 091801 (2015)
  doi:10.1103/PhysRevLett.114.091801
  [arXiv:1411.0565 [hep-ph]].
  %%CITATION = doi:10.1103/PhysRevLett.114.091801;%%
  %135 citations counted in INSPIRE as of 06 Mar 2018



%\cite{DAmico:2017mtc}
\bibitem{DAmico:2017mtc} 
  G.~D'Amico, M.~Nardecchia, P.~Panci, F.~Sannino, A.~Strumia, R.~Torre and A.~Urbano,
  %``Flavour anomalies after the $R_{K^*}$ measurement,''
  JHEP {\bf 1709}, 010 (2017)
  doi:10.1007/JHEP09(2017)010
  [arXiv:1704.05438 [hep-ph]].
  %%CITATION = doi:10.1007/JHEP09(2017)010;%%
  %82 citations counted in INSPIRE as of 06 Mar 2018



%\cite{Descotes-Genon:2015uva}
\bibitem{Descotes-Genon:2015uva}
  S.~Descotes-Genon, L.~Hofer, J.~Matias and J.~Virto,
  %``Global analysis of $b\to s\ell\ell$ anomalies,''
  JHEP {\bf 1606} (2016) 092
  doi:10.1007/JHEP06(2016)092
  [arXiv:1510.04239 [hep-ph]].
  %%CITATION = doi:10.1007/JHEP06(2016)092;%%
  %237 citations counted in INSPIRE as of 22 Mar 2018


%\cite{Calibbi:2015kma}
\bibitem{Calibbi:2015kma}
  L.~Calibbi, A.~Crivellin and T.~Ota,
  %``Effective Field Theory Approach to b→sℓℓ(′), B→K(*)ν$\overline{ν}$ and B→D(*)τν with Third Generation Couplings,''
  Phys.\ Rev.\ Lett.\  {\bf 115} (2015) 181801
  doi:10.1103/PhysRevLett.115.181801
  [arXiv:1506.02661 [hep-ph]].
  %%CITATION = doi:10.1103/PhysRevLett.115.181801;%%
  %140 citations counted in INSPIRE as of 22 Mar 2018



%\cite{Crivellin:2015era}
\bibitem{Crivellin:2015era}
  A.~Crivellin, L.~Hofer, J.~Matias, U.~Nierste, S.~Pokorski and J.~Rosiek,
  %``Lepton-flavour violating $B$ decays in generic $Z'$ models,''
  Phys.\ Rev.\ D {\bf 92} (2015) no.5,  054013
  doi:10.1103/PhysRevD.92.054013
  [arXiv:1504.07928 [hep-ph]].
  %%CITATION = doi:10.1103/PhysRevD.92.054013;%%
  %96 citations counted in INSPIRE as of 22 Mar 2018


%\cite{Crivellin:2015lwa}
\bibitem{Crivellin:2015lwa}
  A.~Crivellin, G.~D'Ambrosio and J.~Heeck,
  %``Addressing the LHC flavor anomalies with horizontal gauge symmetries,''
  Phys.\ Rev.\ D {\bf 91} (2015) no.7,  075006
  doi:10.1103/PhysRevD.91.075006
  [arXiv:1503.03477 [hep-ph]].
  %%CITATION = doi:10.1103/PhysRevD.91.075006;%%
  %183 citations counted in INSPIRE as of 22 Mar 2018



%\cite{Bonilla:2017lsq}
\bibitem{Bonilla:2017lsq} 
  C.~Bonilla, T.~Modak, R.~Srivastava and J.~W.~F.~Valle,
  %``$U(1)_{B_3-3L_\mu}$ gauge symmetry as the simplest description of $b\to s$ anomalies,''
  arXiv:1705.00915 [hep-ph].
  %%CITATION = ARXIV:1705.00915;%%
  %30 citations counted in INSPIRE as of 22 Mar 2018



%\cite{CarcamoHernandez:2019cbd}
\bibitem{CarcamoHernandez:2019cbd} 
  A.~E.~Cárcamo Hernández, S.~Kovalenko, R.~Pasechnik and I.~Schmidt,
  %``Sequentially loop-generated quark and lepton mass hierarchies in an extended Inert Higgs Doublet model,''
  arXiv:1901.02764 [hep-ph].
  %%CITATION = ARXIV:1901.02764;%%
  %3 citations counted in INSPIRE as of 25 Feb 2019



%\cite{CarcamoHernandez:2019xkb}
\bibitem{CarcamoHernandez:2019xkb} 
  A.~E.~Cárcamo Hernández, S.~Kovalenko, R.~Pasechnik and I.~Schmidt,
  %``Phenomenology of an extended IDM with loop-generated fermion mass hierarchies,''
  arXiv:1901.09552 [hep-ph].
  %%CITATION = ARXIV:1901.09552;%%



%\cite{DelleRose:2019ukt}
\bibitem{DelleRose:2019ukt} 
  L.~Delle Rose, S.~Khalil, S.~J.~D.~King and S.~Moretti,
  %``$R_K$ and $R_{K^*}$ in an Aligned 2HDM with Right-Handed Neutrinos,''
  arXiv:1903.11146 [hep-ph].
  %%CITATION = ARXIV:1903.11146;%%




%\cite{Descotes-Genon:2017ptp}
\bibitem{Descotes-Genon:2017ptp} 
  S.~Descotes-Genon, M.~Moscati and G.~Ricciardi,
  %``Nonminimal 331 model for lepton flavor universality violation in $b{\rightarrow}s{\ell}{\ell}$ decays,''
  Phys.\ Rev.\ D {\bf 98}, no. 11, 115030 (2018)
  doi:10.1103/PhysRevD.98.115030
  [arXiv:1711.03101 [hep-ph]].
  %%CITATION = doi:10.1103/PhysRevD.98.115030;%%
  %7 citations counted in INSPIRE as of 07 Jan 2019



%\cite{Assad:2017iib}
\bibitem{Assad:2017iib} 
  N.~Assad, B.~Fornal and B.~Grinstein,
  %``Baryon Number and Lepton Universality Violation in Leptoquark and Diquark Models,''
  Phys.\ Lett.\ B {\bf 777}, 324 (2018)
  doi:10.1016/j.physletb.2017.12.042
  [arXiv:1708.06350 [hep-ph]].
  %%CITATION = doi:10.1016/j.physletb.2017.12.042;%%
  %63 citations counted in INSPIRE as of 25 Feb 2019



%\cite{King:2017anf}
\bibitem{King:2017anf} 
  S.~F.~King,
  %``Flavourful Z$^{′}$ models for $ {R}_{K^{\left(\ast \right)}} $,''
  JHEP {\bf 1708}, 019 (2017)
  doi:10.1007/JHEP08(2017)019
  [arXiv:1706.06100 [hep-ph]].
  %%CITATION = doi:10.1007/JHEP08(2017)019;%%
  %13 citations counted in INSPIRE as of 06 Mar 2018



%\cite{Romao:2017qnu}
\bibitem{Romao:2017qnu} 
  M.~C.~Romao, S.~F.~King and G.~K.~Leontaris,
  %``Non-universal $Z'$ from Fluxed GUTs,''
  arXiv:1710.02349 [hep-ph].
  %%CITATION = ARXIV:1710.02349;%%
  %1 citations counted in INSPIRE as of 06 Mar 2018



%\cite{Antusch:2017tud}
\bibitem{Antusch:2017tud} 
  S.~Antusch, C.~Hohl, S.~F.~King and V.~Susic,
  %``Non-universal Z' from SO(10) GUTs with vector-like family and the origin of neutrino masses,''
  arXiv:1712.05366 [hep-ph].
  %%CITATION = ARXIV:1712.05366;%%

  
  %\cite{Falkowski:2018dsl}
\bibitem{Falkowski:2018dsl}
  A.~Falkowski, S.~F.~King, E.~Perdomo and M.~Pierre,
  %``Flavourful $Z'$ portal for vector-like neutrino Dark Matter and $R_{K^{(*)}}$,''
  arXiv:1803.04430 [hep-ph].
  %%CITATION = ARXIV:1803.04430;%%
  


%\cite{King:1999mb}
\bibitem{King:1999mb} 
  S.~F.~King,
  %``Large mixing angle MSW and atmospheric neutrinos from single right-handed neutrino dominance and U(1) family symmetry,''
  Nucl.\ Phys.\ B {\bf 576}, 85 (2000)
  doi:10.1016/S0550-3213(00)00109-7
  [hep-ph/9912492].
  %%CITATION = doi:10.1016/S0550-3213(00)00109-7;%%
  %226 citations counted in INSPIRE as of 12 Mar 2019



%\cite{King:2002nf}
\bibitem{King:2002nf} 
  S.~F.~King,
  %``Constructing the large mixing angle MNS matrix in seesaw models with right-handed neutrino dominance,''
  JHEP {\bf 0209}, 011 (2002)
  doi:10.1088/1126-6708/2002/09/011
  [hep-ph/0204360].
  %%CITATION = doi:10.1088/1126-6708/2002/09/011;%%
  %198 citations counted in INSPIRE as of 12 Mar 2019


%\cite{Ishimori:2010au}
\bibitem{Ishimori:2010au} 
  H.~Ishimori, T.~Kobayashi, H.~Ohki, Y.~Shimizu, H.~Okada and M.~Tanimoto,
  %``Non-Abelian Discrete Symmetries in Particle Physics,''
  Prog.\ Theor.\ Phys.\ Suppl.\  {\bf 183}, 1 (2010)
  doi:10.1143/PTPS.183.1
  [arXiv:1003.3552 [hep-th]].
  %%CITATION = doi:10.1143/PTPS.183.1;%%
  %508 citations counted in INSPIRE as of 06 Mar 2018



%\cite{Altarelli:2010gt}
\bibitem{Altarelli:2010gt} 
  G.~Altarelli and F.~Feruglio,
  %``Discrete Flavor Symmetries and Models of Neutrino Mixing,''
  Rev.\ Mod.\ Phys.\  {\bf 82}, 2701 (2010)
  doi:10.1103/RevModPhys.82.2701
  [arXiv:1002.0211 [hep-ph]].
  %%CITATION = doi:10.1103/RevModPhys.82.2701;%%
  %577 citations counted in INSPIRE as of 06 Mar 2018



%\cite{King:2013eh}
\bibitem{King:2013eh} 
  S.~F.~King and C.~Luhn,
  %``Neutrino Mass and Mixing with Discrete Symmetry,''
  Rept.\ Prog.\ Phys.\  {\bf 76}, 056201 (2013)
  doi:10.1088/0034-4885/76/5/056201
  [arXiv:1301.1340 [hep-ph]].
  %%CITATION = doi:10.1088/0034-4885/76/5/056201;%%
  %399 citations counted in INSPIRE as of 06 Mar 2018



%\cite{King:2014nza}
\bibitem{King:2014nza} 
  S.~F.~King, A.~Merle, S.~Morisi, Y.~Shimizu and M.~Tanimoto,
  %``Neutrino Mass and Mixing: from Theory to Experiment,''
  New J.\ Phys.\  {\bf 16}, 045018 (2014)
  doi:10.1088/1367-2630/16/4/045018
  [arXiv:1402.4271 [hep-ph]].
  %%CITATION = doi:10.1088/1367-2630/16/4/045018;%%
  %195 citations counted in INSPIRE as of 06 Mar 2018



%\cite{King:2017guk}
\bibitem{King:2017guk} 
  S.~F.~King,
  %``Unified Models of Neutrinos, Flavour and CP Violation,''
  Prog.\ Part.\ Nucl.\ Phys.\  {\bf 94}, 217 (2017)
  doi:10.1016/j.ppnp.2017.01.003
  [arXiv:1701.04413 [hep-ph]].
  %%CITATION = doi:10.1016/j.ppnp.2017.01.003;%%
  %29 citations counted in INSPIRE as of 06 Mar 2018



%\cite{Gerard:1982mm}
\bibitem{Gerard:1982mm} 
  J.~M.~Gerard,
  %``FERMION MASS SPECTRUM IN SU(2)-L x U(1),''
  Z.\ Phys.\ C {\bf 18}, 145 (1983).
  doi:10.1007/BF01572477
  %%CITATION = doi:10.1007/BF01572477;%%
  %6 citations counted in INSPIRE as of 04 Oct 2018




%\cite{Kubo:2003iw}
\bibitem{Kubo:2003iw} 
  J.~Kubo, A.~Mondragon, M.~Mondragon and E.~Rodriguez-Jauregui,
  %``The Flavor symmetry,''
  Prog.\ Theor.\ Phys.\  {\bf 109}, 795 (2003)
  Erratum: [Prog.\ Theor.\ Phys.\  {\bf 114}, 287 (2005)]
  doi:10.1143/PTP.109.795
  [hep-ph/0302196].
  %%CITATION = doi:10.1143/PTP.109.795;%%
  %194 citations counted in INSPIRE as of 30 Mar 2019




%\cite{Kubo:2003pd}
\bibitem{Kubo:2003pd} 
  J.~Kubo,
  %``Majorana phase in minimal S(3) invariant extension of the standard model,''
  Phys.\ Lett.\ B {\bf 578}, 156 (2004)
  Erratum: [Phys.\ Lett.\ B {\bf 619}, 387 (2005)]
  doi:10.1016/j.physletb.2005.06.013, 10.1016/j.physletb.2003.10.048
  [hep-ph/0309167].
  %%CITATION = doi:10.1016/j.physletb.2005.06.013, 10.1016/j.physletb.2003.10.048;%%
  %67 citations counted in INSPIRE as of 06 Mar 2018



%\cite{Kobayashi:2003fh}
\bibitem{Kobayashi:2003fh} 
  T.~Kobayashi, J.~Kubo and H.~Terao,
  %``Exact S(3) symmetry solving the supersymmetric flavor problem,''
  Phys.\ Lett.\ B {\bf 568}, 83 (2003)
  doi:10.1016/j.physletb.2003.03.002
  [hep-ph/0303084].
  %%CITATION = doi:10.1016/j.physletb.2003.03.002;%%
  %59 citations counted in INSPIRE as of 06 Mar 2018



%\cite{Chen:2004rr}
\bibitem{Chen:2004rr} 
  S.~L.~Chen, M.~Frigerio and E.~Ma,
  %``Large neutrino mixing and normal mass hierarchy: A Discrete understanding,''
  Phys.\ Rev.\ D {\bf 70}, 073008 (2004)
  Erratum: [Phys.\ Rev.\ D {\bf 70}, 079905 (2004)]
  doi:10.1103/PhysRevD.70.079905, 10.1103/PhysRevD.70.073008
  [hep-ph/0404084].
  %%CITATION = doi:10.1103/PhysRevD.70.079905, 10.1103/PhysRevD.70.073008;%%
  %139 citations counted in INSPIRE as of 06 Mar 2018



%\cite{Mondragon:2007af}
\bibitem{Mondragon:2007af} 
  A.~Mondragon, M.~Mondragon and E.~Peinado,
  %``Lepton masses, mixings and FCNC in a minimal S(3)-invariant extension of the Standard Model,''
  Phys.\ Rev.\ D {\bf 76}, 076003 (2007)
  doi:10.1103/PhysRevD.76.076003
  [arXiv:0706.0354 [hep-ph]].
  %%CITATION = doi:10.1103/PhysRevD.76.076003;%%
  %81 citations counted in INSPIRE as of 06 Mar 2018



%\cite{Mondragon:2008gm}
\bibitem{Mondragon:2008gm} 
  A.~Mondragon, M.~Mondragon and E.~Peinado,
  %``Lepton flavour violating processes in an S(3)-symmetric model,''
  Rev.\ Mex.\ Fis.\  {\bf 54}, no. 3, 81 (2008)
  [Rev.\ Mex.\ Fis.\ Suppl.\  {\bf 54}, 0181 (2008)]
  [arXiv:0805.3507 [hep-ph]].
  %%CITATION = ARXIV:0805.3507;%%
  %11 citations counted in INSPIRE as of 06 Mar 2018



%\cite{Bhattacharyya:2010hp}
\bibitem{Bhattacharyya:2010hp} 
  G.~Bhattacharyya, P.~Leser and H.~Pas,
  %``Exotic Higgs boson decay modes as a harbinger of $S_3$ flavor symmetry,''
  Phys.\ Rev.\ D {\bf 83}, 011701 (2011)
  doi:10.1103/PhysRevD.83.011701
  [arXiv:1006.5597 [hep-ph]].
  %%CITATION = doi:10.1103/PhysRevD.83.011701;%%
  %57 citations counted in INSPIRE as of 06 Mar 2018



%\cite{Dong:2011vb}
\bibitem{Dong:2011vb} 
  P.~V.~Dong, H.~N.~Long, C.~H.~Nam and V.~V.~Vien,
  %``The $S_3$ flavor symmetry in 3-3-1 models,''
  Phys.\ Rev.\ D {\bf 85}, 053001 (2012)
  doi:10.1103/PhysRevD.85.053001
  [arXiv:1111.6360 [hep-ph]].
  %%CITATION = doi:10.1103/PhysRevD.85.053001;%%
  %52 citations counted in INSPIRE as of 06 Mar 2018



%\cite{Dias:2012bh}
\bibitem{Dias:2012bh} 
  A.~G.~Dias, A.~C.~B.~Machado and C.~C.~Nishi,
  %``An $S_3$ Model for Lepton Mass Matrices with Nearly Minimal Texture,''
  Phys.\ Rev.\ D {\bf 86}, 093005 (2012)
  doi:10.1103/PhysRevD.86.093005
  [arXiv:1206.6362 [hep-ph]].
  %%CITATION = doi:10.1103/PhysRevD.86.093005;%%
  %18 citations counted in INSPIRE as of 06 Mar 2018



%\cite{Meloni:2012ci}
\bibitem{Meloni:2012ci} 
  D.~Meloni,
  %``$S_3$ as a flavour symmetry for quarks and leptons after the Daya Bay result on $\theta_{13}$,''
  JHEP {\bf 1205}, 124 (2012)
  doi:10.1007/JHEP05(2012)124
  [arXiv:1203.3126 [hep-ph]].
  %%CITATION = doi:10.1007/JHEP05(2012)124;%%
  %27 citations counted in INSPIRE as of 06 Mar 2018



%\cite{Canales:2012dr}
\bibitem{Canales:2012dr} 
  F.~Gonzalez Canales, A.~Mondragon and M.~Mondragon,
  %``The $S_3$ Flavour Symmetry: Neutrino Masses and Mixings,''
  Fortsch.\ Phys.\  {\bf 61}, 546 (2013)
  doi:10.1002/prop.201200121
  [arXiv:1205.4755 [hep-ph]].
  %%CITATION = doi:10.1002/prop.201200121;%%
  %22 citations counted in INSPIRE as of 06 Mar 2018



%\cite{Canales:2013cga}
\bibitem{Canales:2013cga} 
  F.~González Canales, A.~Mondragón, M.~Mondragón, U.~J.~Saldaña Salazar and L.~Velasco-Sevilla,
  %``Quark sector of S3 models: classification and comparison with experimental data,''
  Phys.\ Rev.\ D {\bf 88}, 096004 (2013)
  doi:10.1103/PhysRevD.88.096004
  [arXiv:1304.6644 [hep-ph]].
  %%CITATION = doi:10.1103/PhysRevD.88.096004;%%
  %43 citations counted in INSPIRE as of 06 Mar 2018



%\cite{Ma:2013zca}
\bibitem{Ma:2013zca} 
  E.~Ma and B.~Melic,
  %``Updated $S_{3}$ model of quarks,''
  Phys.\ Lett.\ B {\bf 725}, 402 (2013)
  doi:10.1016/j.physletb.2013.07.015
  [arXiv:1303.6928 [hep-ph]].
  %%CITATION = doi:10.1016/j.physletb.2013.07.015;%%
  %30 citations counted in INSPIRE as of 06 Mar 2018



%\cite{Kajiyama:2013sza}
\bibitem{Kajiyama:2013sza} 
  Y.~Kajiyama, H.~Okada and K.~Yagyu,
  %``Electron/Muon Specific Two Higgs Doublet Model,''
  Nucl.\ Phys.\ B {\bf 887}, 358 (2014)
  doi:10.1016/j.nuclphysb.2014.08.009
  [arXiv:1309.6234 [hep-ph]].
  %%CITATION = doi:10.1016/j.nuclphysb.2014.08.009;%%
  %45 citations counted in INSPIRE as of 06 Mar 2018



%\cite{Hernandez:2013hea}
\bibitem{Hernandez:2013hea} 
  A.~E.~Cárcamo Hernández, R.~Martinez and F.~Ochoa,
  %``Fermion masses and mixings in the 3-3-1 model with right-handed neutrinos based on the $S_3$ flavor symmetry,''
  Eur.\ Phys.\ J.\ C {\bf 76}, no. 11, 634 (2016)
  doi:10.1140/epjc/s10052-016-4480-3
  [arXiv:1309.6567 [hep-ph]].
  %%CITATION = doi:10.1140/epjc/s10052-016-4480-3;%%
  %42 citations counted in INSPIRE as of 06 Mar 2018



%\cite{Ma:2014qra}
\bibitem{Ma:2014qra} 
  E.~Ma and R.~Srivastava,
  %``Dirac or inverse seesaw neutrino masses with $B-L$ gauge symmetry and $S_3$ flavor symmetry,''
  Phys.\ Lett.\ B {\bf 741}, 217 (2015)
  doi:10.1016/j.physletb.2014.12.049
  [arXiv:1411.5042 [hep-ph]].
  %%CITATION = doi:10.1016/j.physletb.2014.12.049;%%
  %45 citations counted in INSPIRE as of 06 Mar 2018



%\cite{Hernandez:2014vta}
\bibitem{Hernandez:2014vta} 
  A.~E.~Cárcamo Hernández, R.~Martinez and J.~Nisperuza,
  %``$S_3$ discrete group as a source of the quark mass and mixing pattern in $331$ models,''
  Eur.\ Phys.\ J.\ C {\bf 75}, no. 2, 72 (2015)
  doi:10.1140/epjc/s10052-015-3278-z
  [arXiv:1401.0937 [hep-ph]].
  %%CITATION = doi:10.1140/epjc/s10052-015-3278-z;%%
  %34 citations counted in INSPIRE as of 06 Mar 2018



%\cite{Hernandez:2014lpa}
\bibitem{Hernandez:2014lpa} 
  A.~E.~Cárcamo Hernández, E.~Cataño Mur and R.~Martinez,
  %``Lepton masses and mixing in $SU(3)_{C}\otimes SU(3)_{L}\otimes U(1)_{X}$ models with a $S_3$ flavor symmetry,''
  Phys.\ Rev.\ D {\bf 90}, no. 7, 073001 (2014)
  doi:10.1103/PhysRevD.90.073001
  [arXiv:1407.5217 [hep-ph]].
  %%CITATION = doi:10.1103/PhysRevD.90.073001;%%
  %23 citations counted in INSPIRE as of 06 Mar 2018



%\cite{Gupta:2014nba}
\bibitem{Gupta:2014nba} 
  S.~Gupta, C.~S.~Kim and P.~Sharma,
  %``Radiative and seesaw threshold corrections to the $S_3$ symmetric neutrino mass matrix,''
  Phys.\ Lett.\ B {\bf 740}, 353 (2015)
  doi:10.1016/j.physletb.2014.12.005
  [arXiv:1408.0172 [hep-ph]].
  %%CITATION = doi:10.1016/j.physletb.2014.12.005;%%
  %5 citations counted in INSPIRE as of 06 Mar 2018



%\cite{Hernandez:2015dga}
\bibitem{Hernandez:2015dga} 
  A.~E.~Cárcamo Hernández, I.~de Medeiros Varzielas and E.~Schumacher,
  %``Fermion and scalar phenomenology of a two-Higgs-doublet model with $S_3$,''
  Phys.\ Rev.\ D {\bf 93}, no. 1, 016003 (2016)
  doi:10.1103/PhysRevD.93.016003
  [arXiv:1509.02083 [hep-ph]].
  %%CITATION = doi:10.1103/PhysRevD.93.016003;%%
  %22 citations counted in INSPIRE as of 06 Mar 2018



%\cite{Hernandez:2015zeh}
\bibitem{Hernandez:2015zeh} 
  A.~E.~Cárcamo Hernández, I.~de Medeiros Varzielas and N.~A.~Neill,
  %``Novel Randall-Sundrum model with $S_{3}$ flavor symmetry,''
  Phys.\ Rev.\ D {\bf 94}, no. 3, 033011 (2016)
  doi:10.1103/PhysRevD.94.033011
  [arXiv:1511.07420 [hep-ph]].
  %%CITATION = doi:10.1103/PhysRevD.94.033011;%%
  %12 citations counted in INSPIRE as of 06 Mar 2018



%\cite{Hernandez:2016rbi}
\bibitem{Hernandez:2016rbi} 
  A.~E.~Cárcamo Hernández, I.~de Medeiros Varzielas and E.~Schumacher,
  %``The $750\,\text{GeV}$ diphoton resonance in the light of a 2HDM with $S_3$ flavour symmetry,''
  arXiv:1601.00661 [hep-ph].
  %%CITATION = ARXIV:1601.00661;%%
  %59 citations counted in INSPIRE as of 06 Mar 2018

%\cite{Hernandez:2015hrt}
\bibitem{Hernandez:2015hrt} 
  A.~E.~Cárcamo Hernández,
  %``A novel and economical explanation for SM fermion masses and mixings,''
  Eur.\ Phys.\ J.\ C {\bf 76}, no. 9, 503 (2016)
  doi:10.1140/epjc/s10052-016-4351-y
  [arXiv:1512.09092 [hep-ph]].
  %%CITATION = doi:10.1140/epjc/s10052-016-4351-y;%%
  %85 citations counted in INSPIRE as of 20 Mar 2018


%\cite{CarcamoHernandez:2016pdu}
\bibitem{CarcamoHernandez:2016pdu} 
  A.~E.~Cárcamo Hernández, S.~Kovalenko and I.~Schmidt,
  %``Radiatively generated hierarchy of lepton and quark masses,''
  JHEP {\bf 1702}, 125 (2017)
  doi:10.1007/JHEP02(2017)125
  [arXiv:1611.09797 [hep-ph]].
  %%CITATION = doi:10.1007/JHEP02(2017)125;%%
  %7 citations counted in INSPIRE as of 06 Mar 2018



%\cite{Arbelaez:2016mhg}
\bibitem{Arbelaez:2016mhg} 
  C.~Arbeláez, A.~E.~Cárcamo Hernández, S.~Kovalenko and I.~Schmidt,
  %``Radiative Seesaw-type Mechanism of Fermion Masses and Non-trivial Quark Mixing,''
  Eur.\ Phys.\ J.\ C {\bf 77}, no. 6, 422 (2017)
  doi:10.1140/epjc/s10052-017-4948-9
  [arXiv:1602.03607 [hep-ph]].
  %%CITATION = doi:10.1140/epjc/s10052-017-4948-9;%%
  %42 citations counted in INSPIRE as of 06 Mar 2018



%\cite{Gomez-Izquierdo:2017rxi}
\bibitem{Gomez-Izquierdo:2017rxi} 
  J.~C.~Gómez-Izquierdo,
  %``Non-minimal flavored ${S}_{3}\otimes {Z}_{2}$ left–right symmetric model,''
  Eur.\ Phys.\ J.\ C {\bf 77}, no. 8, 551 (2017)
  doi:10.1140/epjc/s10052-017-5094-0
  [arXiv:1701.01747 [hep-ph]].
  %%CITATION = doi:10.1140/epjc/s10052-017-5094-0;%%
  %4 citations counted in INSPIRE as of 06 Mar 2018



%\cite{Cruz:2017add}
\bibitem{Cruz:2017add} 
  A.~A.~Cruz and M.~Mondragón,
  %``Neutrino masses, mixing, and leptogenesis in an S3 model,''
  arXiv:1701.07929 [hep-ph].
  %%CITATION = ARXIV:1701.07929;%%
  %3 citations counted in INSPIRE as of 06 Mar 2018


%\cite{CarcamoHernandez:2018vdj}
\bibitem{CarcamoHernandez:2018vdj} 
  A.~E.~Cárcamo Hernández, J.~Vignatti and A.~Zerwekh,
  %``A model of strongly coupled heavy vector resonances for fermion masses and mixings,''
  arXiv:1807.05321 [hep-ph].
  %%CITATION = ARXIV:1807.05321;%%



%\cite{Ma:2001dn}
\bibitem{Ma:2001dn} 
  E.~Ma and G.~Rajasekaran,
  %``Softly broken A(4) symmetry for nearly degenerate neutrino masses,''
  Phys.\ Rev.\ D {\bf 64}, 113012 (2001)
  doi:10.1103/PhysRevD.64.113012
  [hep-ph/0106291].
  %%CITATION = doi:10.1103/PhysRevD.64.113012;%%
  %637 citations counted in INSPIRE as of 06 Mar 2018



%\cite{deMedeirosVarzielas:2005qg}
\bibitem{deMedeirosVarzielas:2005qg} 
  I.~de Medeiros Varzielas, S.~F.~King and G.~G.~Ross,
  %``Tri-bimaximal neutrino mixing from discrete subgroups of SU(3) and SO(3) family symmetry,''
  Phys.\ Lett.\ B {\bf 644}, 153 (2007)
  doi:10.1016/j.physletb.2006.11.015
  [hep-ph/0512313].
  %%CITATION = doi:10.1016/j.physletb.2006.11.015;%%
  %196 citations counted in INSPIRE as of 06 Mar 2018



%\cite{He:2006dk}
\bibitem{He:2006dk} 
  X.~G.~He, Y.~Y.~Keum and R.~R.~Volkas,
  %``A(4) flavor symmetry breaking scheme for understanding quark and neutrino mixing angles,''
  JHEP {\bf 0604}, 039 (2006)
  doi:10.1088/1126-6708/2006/04/039
  [hep-ph/0601001].
  %%CITATION = doi:10.1088/1126-6708/2006/04/039;%%
  %256 citations counted in INSPIRE as of 06 Mar 2018



%\cite{Chen:2009um}
\bibitem{Chen:2009um} 
  M.~C.~Chen and S.~F.~King,
  %``A4 See-Saw Models and Form Dominance,''
  JHEP {\bf 0906}, 072 (2009)
  doi:10.1088/1126-6708/2009/06/072
  [arXiv:0903.0125 [hep-ph]].
  %%CITATION = doi:10.1088/1126-6708/2009/06/072;%%
  %124 citations counted in INSPIRE as of 06 Mar 2018



%\cite{Burrows:2010wz}
\bibitem{Burrows:2010wz} 
  T.~J.~Burrows and S.~F.~King,
  %``$A_4$ x SU(5) SUSY GUT of Flavour in 8d,''
  Nucl.\ Phys.\ B {\bf 842}, 107 (2011)
  doi:10.1016/j.nuclphysb.2010.08.018
  [arXiv:1007.2310 [hep-ph]].
  %%CITATION = doi:10.1016/j.nuclphysb.2010.08.018;%%
  %38 citations counted in INSPIRE as of 06 Mar 2018



%\cite{King:2011ab}
\bibitem{King:2011ab} 
  S.~F.~King and C.~Luhn,
  %``A4 models of tri-bimaximal-reactor mixing,''
  JHEP {\bf 1203}, 036 (2012)
  doi:10.1007/JHEP03(2012)036
  [arXiv:1112.1959 [hep-ph]].
  %%CITATION = doi:10.1007/JHEP03(2012)036;%%
  %61 citations counted in INSPIRE as of 06 Mar 2018



%\cite{Antusch:2011ic}
\bibitem{Antusch:2011ic} 
  S.~Antusch, S.~F.~King, C.~Luhn and M.~Spinrath,
  %``Trimaximal mixing with predicted $\theta_{13}$ from a new type of constrained sequential dominance,''
  Nucl.\ Phys.\ B {\bf 856}, 328 (2012)
  doi:10.1016/j.nuclphysb.2011.11.009
  [arXiv:1108.4278 [hep-ph]].
  %%CITATION = doi:10.1016/j.nuclphysb.2011.11.009;%%
  %115 citations counted in INSPIRE as of 06 Mar 2018



%\cite{Ahn:2012tv}
\bibitem{Ahn:2012tv} 
  Y.~H.~Ahn and S.~K.~Kang,
  %``Non-zero $\theta_{13}$ and CP violation in a model with $A_4$ flavor symmetry,''
  Phys.\ Rev.\ D {\bf 86}, 093003 (2012)
  doi:10.1103/PhysRevD.86.093003
  [arXiv:1203.4185 [hep-ph]].
  %%CITATION = doi:10.1103/PhysRevD.86.093003;%%
  %46 citations counted in INSPIRE as of 06 Mar 2018



%\cite{Cooper:2012wf}
\bibitem{Cooper:2012wf} 
  I.~K.~Cooper, S.~F.~King and C.~Luhn,
  %``A4xSU(5) SUSY GUT of Flavour with Trimaximal Neutrino Mixing,''
  JHEP {\bf 1206}, 130 (2012)
  doi:10.1007/JHEP06(2012)130
  [arXiv:1203.1324 [hep-ph]].
  %%CITATION = doi:10.1007/JHEP06(2012)130;%%
  %43 citations counted in INSPIRE as of 06 Mar 2018



%\cite{Memenga:2013vc}
\bibitem{Memenga:2013vc} 
  N.~Memenga, W.~Rodejohann and H.~Zhang,
  %``$A_4$ flavor symmetry model for Dirac neutrinos and sizable $U_{e3}$,''
  Phys.\ Rev.\ D {\bf 87}, no. 5, 053021 (2013)
  doi:10.1103/PhysRevD.87.053021
  [arXiv:1301.2963 [hep-ph]].
  %%CITATION = doi:10.1103/PhysRevD.87.053021;%%
  %28 citations counted in INSPIRE as of 06 Mar 2018



%\cite{King:2013hoa}
\bibitem{King:2013hoa} 
  S.~F.~King,
  %``A model of quark and lepton mixing,''
  JHEP {\bf 1401}, 119 (2014)
  doi:10.1007/JHEP01(2014)119
  [arXiv:1311.3295 [hep-ph]].
  %%CITATION = doi:10.1007/JHEP01(2014)119;%%
  %31 citations counted in INSPIRE as of 06 Mar 2018



%\cite{King:2013iva}
\bibitem{King:2013iva} 
  S.~F.~King,
  %``Minimal predictive see-saw model with normal neutrino mass hierarchy,''
  JHEP {\bf 1307}, 137 (2013)
  doi:10.1007/JHEP07(2013)137
  [arXiv:1304.6264 [hep-ph]].
  %%CITATION = doi:10.1007/JHEP07(2013)137;%%
  %45 citations counted in INSPIRE as of 06 Mar 2018



%\cite{Ding:2013bpa}
\bibitem{Ding:2013bpa} 
  G.~J.~Ding, S.~F.~King and A.~J.~Stuart,
  %``Generalised CP and $A_4$ Family Symmetry,''
  JHEP {\bf 1312}, 006 (2013)
  doi:10.1007/JHEP12(2013)006
  [arXiv:1307.4212 [hep-ph]].
  %%CITATION = doi:10.1007/JHEP12(2013)006;%%
  %73 citations counted in INSPIRE as of 06 Mar 2018



%\cite{Felipe:2013vwa}
\bibitem{Felipe:2013vwa} 
  R.~Gonzalez Felipe, H.~Serodio and J.~P.~Silva,
  %``Neutrino masses and mixing in A4 models with three Higgs doublets,''
  Phys.\ Rev.\ D {\bf 88}, no. 1, 015015 (2013)
  doi:10.1103/PhysRevD.88.015015
  [arXiv:1304.3468 [hep-ph]].
  %%CITATION = doi:10.1103/PhysRevD.88.015015;%%
  %23 citations counted in INSPIRE as of 06 Mar 2018



%\cite{Varzielas:2012ai}
\bibitem{Varzielas:2012ai} 
  I.~de Medeiros Varzielas and D.~Pidt,
  %``UV completions of flavour models and large theta_{13},''
  JHEP {\bf 1303}, 065 (2013)
  doi:10.1007/JHEP03(2013)065
  [arXiv:1211.5370 [hep-ph]].
  %%CITATION = doi:10.1007/JHEP03(2013)065;%%
  %25 citations counted in INSPIRE as of 06 Mar 2018



%\cite{Ishimori:2012fg}
\bibitem{Ishimori:2012fg} 
  H.~Ishimori and E.~Ma,
  %``New Simple $A_4$ Neutrino Model for Nonzero $\theta_{13}$ and Large $\delta_{CP}$,''
  Phys.\ Rev.\ D {\bf 86}, 045030 (2012)
  doi:10.1103/PhysRevD.86.045030
  [arXiv:1205.0075 [hep-ph]].
  %%CITATION = doi:10.1103/PhysRevD.86.045030;%%
  %51 citations counted in INSPIRE as of 06 Mar 2018



%\cite{King:2013hj}
\bibitem{King:2013hj} 
  S.~F.~King, S.~Morisi, E.~Peinado and J.~W.~F.~Valle,
  %``Quark-Lepton Mass Relation in a Realistic $A_4$ Extension of the Standard Model,''
  Phys.\ Lett.\ B {\bf 724}, 68 (2013)
  doi:10.1016/j.physletb.2013.05.067
  [arXiv:1301.7065 [hep-ph]].
  %%CITATION = doi:10.1016/j.physletb.2013.05.067;%%
  %37 citations counted in INSPIRE as of 06 Mar 2018



%\cite{Antusch:2013wn}
\bibitem{Antusch:2013wn} 
  S.~Antusch, S.~F.~King and M.~Spinrath,
  %``Spontaneous CP violation in $A_4 \times SU(5)$ with Constrained Sequential Dominance 2,''
  Phys.\ Rev.\ D {\bf 87}, no. 9, 096018 (2013)
  doi:10.1103/PhysRevD.87.096018
  [arXiv:1301.6764 [hep-ph]].
  %%CITATION = doi:10.1103/PhysRevD.87.096018;%%
  %31 citations counted in INSPIRE as of 06 Mar 2018



%\cite{Hernandez:2013dta}
\bibitem{Hernandez:2013dta} 
  A.~E.~Carcamo Hernandez, I.~de Medeiros Varzielas, S.~G.~Kovalenko, H.~Päs and I.~Schmidt,
  %``Lepton masses and mixings in an $A_4$ multi-Higgs model with a radiative seesaw mechanism,''
  Phys.\ Rev.\ D {\bf 88}, no. 7, 076014 (2013)
  doi:10.1103/PhysRevD.88.076014
  [arXiv:1307.6499 [hep-ph]].
  %%CITATION = doi:10.1103/PhysRevD.88.076014;%%
  %60 citations counted in INSPIRE as of 06 Mar 2018



%\cite{Babu:2002dz}
\bibitem{Babu:2002dz} 
  K.~S.~Babu, E.~Ma and J.~W.~F.~Valle,
  %``Underlying A(4) symmetry for the neutrino mass matrix and the quark mixing matrix,''
  Phys.\ Lett.\ B {\bf 552}, 207 (2003)
  doi:10.1016/S0370-2693(02)03153-2
  [hep-ph/0206292].
  %%CITATION = doi:10.1016/S0370-2693(02)03153-2;%%
  %623 citations counted in INSPIRE as of 06 Mar 2018



%\cite{Altarelli:2005yx}
\bibitem{Altarelli:2005yx} 
  G.~Altarelli and F.~Feruglio,
  %``Tri-bimaximal neutrino mixing, A(4) and the modular symmetry,''
  Nucl.\ Phys.\ B {\bf 741}, 215 (2006)
  doi:10.1016/j.nuclphysb.2006.02.015
  [hep-ph/0512103].
  %%CITATION = doi:10.1016/j.nuclphysb.2006.02.015;%%
  %533 citations counted in INSPIRE as of 06 Mar 2018



%\cite{Varzielas:2008jm}
\bibitem{Varzielas:2008jm} 
  I.~de Medeiros Varzielas, G.~G.~Ross and M.~Serna,
  %``Quasi-degenerate neutrinos and tri-bi-maximal mixing,''
  Phys.\ Rev.\ D {\bf 80}, 073002 (2009)
  doi:10.1103/PhysRevD.80.073002
  [arXiv:0811.2226 [hep-ph]].
  %%CITATION = doi:10.1103/PhysRevD.80.073002;%%
  %21 citations counted in INSPIRE as of 06 Mar 2018



%\cite{Gupta:2011ct}
\bibitem{Gupta:2011ct} 
  S.~Gupta, A.~S.~Joshipura and K.~M.~Patel,
  %``Minimal extension of tri-bimaximal mixing and generalized Z_2 X Z_2 symmetries,''
  Phys.\ Rev.\ D {\bf 85}, 031903 (2012)
  doi:10.1103/PhysRevD.85.031903
  [arXiv:1112.6113 [hep-ph]].
  %%CITATION = doi:10.1103/PhysRevD.85.031903;%%
  %73 citations counted in INSPIRE as of 06 Mar 2018



%\cite{Morisi:2013eca}
\bibitem{Morisi:2013eca} 
  S.~Morisi, M.~Nebot, K.~M.~Patel, E.~Peinado and J.~W.~F.~Valle,
  %``Quark-Lepton Mass Relation and CKM mixing in an A4 Extension of the Minimal Supersymmetric Standard Model,''
  Phys.\ Rev.\ D {\bf 88}, 036001 (2013)
  doi:10.1103/PhysRevD.88.036001
  [arXiv:1303.4394 [hep-ph]].
  %%CITATION = doi:10.1103/PhysRevD.88.036001;%%
  %28 citations counted in INSPIRE as of 06 Mar 2018



%\cite{Altarelli:2005yp}
\bibitem{Altarelli:2005yp} 
  G.~Altarelli and F.~Feruglio,
  %``Tri-bimaximal neutrino mixing from discrete symmetry in extra dimensions,''
  Nucl.\ Phys.\ B {\bf 720}, 64 (2005)
  doi:10.1016/j.nuclphysb.2005.05.005
  [hep-ph/0504165].
  %%CITATION = doi:10.1016/j.nuclphysb.2005.05.005;%%
  %529 citations counted in INSPIRE as of 06 Mar 2018



%\cite{Kadosh:2010rm}
\bibitem{Kadosh:2010rm} 
  A.~Kadosh and E.~Pallante,
  %``An A(4) flavor model for quarks and leptons in warped geometry,''
  JHEP {\bf 1008}, 115 (2010)
  doi:10.1007/JHEP08(2010)115
  [arXiv:1004.0321 [hep-ph]].
  %%CITATION = doi:10.1007/JHEP08(2010)115;%%
  %50 citations counted in INSPIRE as of 06 Mar 2018



%\cite{Kadosh:2013nra}
\bibitem{Kadosh:2013nra} 
  A.~Kadosh,
  %``$\Theta_13$ and charged Lepton Flavor Violation in "warped" $A_4$ models,''
  JHEP {\bf 1306}, 114 (2013)
  doi:10.1007/JHEP06(2013)114
  [arXiv:1303.2645 [hep-ph]].
  %%CITATION = doi:10.1007/JHEP06(2013)114;%%
  %10 citations counted in INSPIRE as of 06 Mar 2018



%\cite{delAguila:2010vg}
\bibitem{delAguila:2010vg} 
  F.~del Aguila, A.~Carmona and J.~Santiago,
  %``Neutrino Masses from an A4 Symmetry in Holographic Composite Higgs Models,''
  JHEP {\bf 1008}, 127 (2010)
  doi:10.1007/JHEP08(2010)127
  [arXiv:1001.5151 [hep-ph]].
  %%CITATION = doi:10.1007/JHEP08(2010)127;%%
  %76 citations counted in INSPIRE as of 06 Mar 2018



%\cite{Campos:2014lla}
\bibitem{Campos:2014lla} 
  M.~D.~Campos, A.~E.~Cárcamo Hernández, S.~Kovalenko, I.~Schmidt and E.~Schumacher,
  %``Fermion masses and mixings in an $SU(5)$ grand unified model with an extra flavor symmetry,''
  Phys.\ Rev.\ D {\bf 90}, no. 1, 016006 (2014)
  doi:10.1103/PhysRevD.90.016006
  [arXiv:1403.2525 [hep-ph]].
  %%CITATION = doi:10.1103/PhysRevD.90.016006;%%
  %28 citations counted in INSPIRE as of 06 Mar 2018



%\cite{Vien:2014pta}
\bibitem{Vien:2014pta} 
  V.~V.~Vien and H.~N.~Long,
  %``Neutrino mixing with nonzero $\theta_{13}$ and CP violation in the 3-3-1 model based on $A_4$ flavor symmetry,''
  Int.\ J.\ Mod.\ Phys.\ A {\bf 30}, no. 21, 1550117 (2015)
  doi:10.1142/S0217751X15501171
  [arXiv:1405.4665 [hep-ph]].
  %%CITATION = doi:10.1142/S0217751X15501171;%%
  %13 citations counted in INSPIRE as of 06 Mar 2018



%\cite{King:2014iia}
\bibitem{King:2014iia} 
  S.~F.~King,
  %``A to Z of Flavour with Pati-Salam,''
  JHEP {\bf 1408}, 130 (2014)
  doi:10.1007/JHEP08(2014)130
  [arXiv:1406.7005 [hep-ph]].
  %%CITATION = doi:10.1007/JHEP08(2014)130;%%
  %35 citations counted in INSPIRE as of 06 Mar 2018



%\cite{Joshipura:2015dsa}
\bibitem{Joshipura:2015dsa} 
  A.~S.~Joshipura and K.~M.~Patel,
  %``Generalized $\mu$-$\tau$ symmetry and discrete subgroups of O(3),''
  Phys.\ Lett.\ B {\bf 749}, 159 (2015)
  doi:10.1016/j.physletb.2015.07.062
  [arXiv:1507.01235 [hep-ph]].
  %%CITATION = doi:10.1016/j.physletb.2015.07.062;%%
  %23 citations counted in INSPIRE as of 06 Mar 2018



%\cite{Hernandez:2015tna}
\bibitem{Hernandez:2015tna} 
  A.~E.~Cárcamo Hernández and R.~Martinez,
  %``A predictive 3-3-1 model with $A_4$ flavor symmetry,''
  Nucl.\ Phys.\ B {\bf 905}, 337 (2016)
  doi:10.1016/j.nuclphysb.2016.02.025
  [arXiv:1501.05937 [hep-ph]].
  %%CITATION = doi:10.1016/j.nuclphysb.2016.02.025;%%
  %26 citations counted in INSPIRE as of 06 Mar 2018



%\cite{Bjorkeroth:2015ora}
\bibitem{Bjorkeroth:2015ora} 
  F.~Björkeroth, F.~J.~de Anda, I.~de Medeiros Varzielas and S.~F.~King,
  %``Towards a complete A$_{4} \times$ SU(5) SUSY GUT,''
  JHEP {\bf 1506}, 141 (2015)
  doi:10.1007/JHEP06(2015)141
  [arXiv:1503.03306 [hep-ph]].
  %%CITATION = doi:10.1007/JHEP06(2015)141;%%
  %29 citations counted in INSPIRE as of 06 Mar 2018



%\cite{Chattopadhyay:2017zvs}
\bibitem{Chattopadhyay:2017zvs} 
  P.~Chattopadhyay and K.~M.~Patel,
  %``Discrete symmetries for electroweak natural type-I seesaw mechanism,''
  Nucl.\ Phys.\ B {\bf 921}, 487 (2017)
  doi:10.1016/j.nuclphysb.2017.06.008
  [arXiv:1703.09541 [hep-ph]].
  %%CITATION = doi:10.1016/j.nuclphysb.2017.06.008;%%
  %3 citations counted in INSPIRE as of 06 Mar 2018



%\cite{CarcamoHernandez:2017kra}
\bibitem{CarcamoHernandez:2017kra} 
  A.~E.~Cárcamo Hernández and H.~N.~Long,
  %``A highly predictive $A_{4}$ flavour 3-3-1 model with radiative inverse seesaw mechanism,''
  arXiv:1705.05246 [hep-ph], {\it J.
Phys. G: Nucl. Part. Phys}. {\bf 45}  (2018) 045001
DOI: 10.1088/1361-6471/aaace7.
  %%CITATION = ARXIV:1705.05246;%%
    %6 citations counted in INSPIRE as of 06 Mar 2018



%\cite{CarcamoHernandez:2017cwi}
\bibitem{CarcamoHernandez:2017cwi} 
  A.~E.~Cárcamo Hernández, S.~Kovalenko, H.~N.~Long and I.~Schmidt,
  %``A variant of 3-3-1 model for the generation of the SM fermion mass and mixing pattern,''
  arXiv:1705.09169 [hep-ph].
  %%CITATION = ARXIV:1705.09169;%%
  %4 citations counted in INSPIRE as of 06 Mar 2018



%\cite{CentellesChulia:2017koy}
\bibitem{CentellesChulia:2017koy} 
  S.~Centelles Chuliá, R.~Srivastava and J.~W.~F.~Valle,
  %``Generalized Bottom-Tau unification, neutrino oscillations and dark matter: predictions from a lepton quarticity flavor approach,''
  Phys.\ Lett.\ B {\bf 773}, 26 (2017)
  doi:10.1016/j.physletb.2017.07.065
  [arXiv:1706.00210 [hep-ph]].
  %%CITATION = doi:10.1016/j.physletb.2017.07.065;%%
  %9 citations counted in INSPIRE as of 06 Mar 2018



%\cite{Bjorkeroth:2017tsz}
\bibitem{Bjorkeroth:2017tsz} 
  F.~Björkeroth, E.~J.~Chun and S.~F.~King,
  %``Accidental Peccei–Quinn symmetry from discrete flavour symmetry and Pati–Salam,''
  Phys.\ Lett.\ B {\bf 777}, 428 (2018)
  doi:10.1016/j.physletb.2017.12.058
  [arXiv:1711.05741 [hep-ph]].
  %%CITATION = doi:10.1016/j.physletb.2017.12.058;%%
  %1 citations counted in INSPIRE as of 06 Mar 2018



%\cite{Belyaev:2018vkl}
\bibitem{Belyaev:2018vkl} 
  A.~S.~Belyaev, S.~F.~King and P.~B.~Schaefers,
  %``Muon g-2 and Dark Matter suggest Non-Universal Gaugino Masses: $\mathbf{SU(5)\times A_4}$ case study at the LHC,''
  arXiv:1801.00514 [hep-ph].
  %%CITATION = ARXIV:1801.00514;%%


%\cite{King:2018fke}
\bibitem{King:2018fke} 
  S.~F.~King and Y.~L.~Zhou,
  %``Spontaneous breaking of $SO(3)$ to finite family symmetries with supersymmetry - an $A_4$ model,''
  arXiv:1809.10292 [hep-ph].
  %%CITATION = ARXIV:1809.10292;%%
  %4 citations counted in INSPIRE as of 16 Nov 2018


%\cite{Bernigaud:2018qky}
\bibitem{Bernigaud:2018qky} 
  J.~Bernigaud, B.~Herrmann, S.~F.~King and S.~J.~Rowley,
  %``Non-minimal flavour violation in $A_4\times SU(5)$ SUSY GUTs with smuon assisted dark matter,''
  arXiv:1812.07463 [hep-ph].
  %%CITATION = ARXIV:1812.07463;%%


%\cite{deAnda:2018ecu}
\bibitem{deAnda:2018ecu} 
  F.~J.~de Anda, S.~F.~King and E.~Perdomo,
  %``$SU(5)$ Grand Unified Theory with $A_4$ Modular Symmetry,''
  arXiv:1812.05620 [hep-ph].
  %%CITATION = ARXIV:1812.05620;%%
  %2 citations counted in INSPIRE as of 07 Jan 2019



%\cite{CarcamoHernandez:2019pmy}
\bibitem{CarcamoHernandez:2019pmy} 
  A.~E.~Cárcamo Hernández, J.~M.~González and U.~J.~Saldaña-Salazar,
  %``Viable low-scale model with Universal and Inverse Seesaw Mechanisms,''
  arXiv:1904.09993 [hep-ph].
  %%CITATION = ARXIV:1904.09993;%%



%\cite{Patel:2010hr}
\bibitem{Patel:2010hr} 
  K.~M.~Patel,
  %``An SO(10)XS4 Model of Quark-Lepton Complementarity,''
  Phys.\ Lett.\ B {\bf 695}, 225 (2011)
  doi:10.1016/j.physletb.2010.11.024
  [arXiv:1008.5061 [hep-ph]].
  %%CITATION = doi:10.1016/j.physletb.2010.11.024;%%
  %50 citations counted in INSPIRE as of 06 Mar 2018



%\cite{Morisi:2011pm}
\bibitem{Morisi:2011pm} 
  S.~Morisi, K.~M.~Patel and E.~Peinado,
  %``Model for T2K indication with maximal atmospheric angle and tri-maximal solar angle,''
  Phys.\ Rev.\ D {\bf 84}, 053002 (2011)
  doi:10.1103/PhysRevD.84.053002
  [arXiv:1107.0696 [hep-ph]].
  %%CITATION = doi:10.1103/PhysRevD.84.053002;%%
  %87 citations counted in INSPIRE as of 06 Mar 2018



%\cite{Mohapatra:2012tb}
\bibitem{Mohapatra:2012tb} 
  R.~N.~Mohapatra and C.~C.~Nishi,
  %``$S_4$ Flavored CP Symmetry for Neutrinos,''
  Phys.\ Rev.\ D {\bf 86}, 073007 (2012)
  doi:10.1103/PhysRevD.86.073007
  [arXiv:1208.2875 [hep-ph]].
  %%CITATION = doi:10.1103/PhysRevD.86.073007;%%
  %75 citations counted in INSPIRE as of 06 Mar 2018



%\cite{BhupalDev:2012nm}
\bibitem{BhupalDev:2012nm} 
  P.~S.~Bhupal Dev, B.~Dutta, R.~N.~Mohapatra and M.~Severson,
  %``$\theta_{13}$ and Proton Decay in a Minimal $SO(10) \times S_4$ model of Flavor,''
  Phys.\ Rev.\ D {\bf 86}, 035002 (2012)
  doi:10.1103/PhysRevD.86.035002
  [arXiv:1202.4012 [hep-ph]].
  %%CITATION = doi:10.1103/PhysRevD.86.035002;%%
  %58 citations counted in INSPIRE as of 06 Mar 2018



%\cite{Hagedorn:2012ut}
\bibitem{Hagedorn:2012ut} 
  C.~Hagedorn, S.~F.~King and C.~Luhn,
  %``SUSY S$_{4} \times$ SU(5) revisited,''
  Phys.\ Lett.\ B {\bf 717}, 207 (2012)
  doi:10.1016/j.physletb.2012.09.026
  [arXiv:1205.3114 [hep-ph]].
  %%CITATION = doi:10.1016/j.physletb.2012.09.026;%%
  %33 citations counted in INSPIRE as of 06 Mar 2018



%\cite{Varzielas:2012pa}
\bibitem{Varzielas:2012pa} 
  I.~de Medeiros Varzielas and L.~Lavoura,
  %``Flavour models for $TM_{1}$ lepton mixing,''
  J.\ Phys.\ G {\bf 40}, 085002 (2013)
  doi:10.1088/0954-3899/40/8/085002
  [arXiv:1212.3247 [hep-ph]].
  %%CITATION = doi:10.1088/0954-3899/40/8/085002;%%
  %36 citations counted in INSPIRE as of 06 Mar 2018



%\cite{Ding:2013hpa}
\bibitem{Ding:2013hpa} 
  G.~J.~Ding, S.~F.~King, C.~Luhn and A.~J.~Stuart,
  %``Spontaneous CP violation from vacuum alignment in $S_4$ models of leptons,''
  JHEP {\bf 1305}, 084 (2013)
  doi:10.1007/JHEP05(2013)084
  [arXiv:1303.6180 [hep-ph]].
  %%CITATION = doi:10.1007/JHEP05(2013)084;%%
  %84 citations counted in INSPIRE as of 06 Mar 2018



%\cite{Ishimori:2010fs}
\bibitem{Ishimori:2010fs} 
  H.~Ishimori, Y.~Shimizu, M.~Tanimoto and A.~Watanabe,
  %``Neutrino masses and mixing from $S_{4}$ flavor twisting,''
  Phys.\ Rev.\ D {\bf 83}, 033004 (2011)
  doi:10.1103/PhysRevD.83.033004
  [arXiv:1010.3805 [hep-ph]].
  %%CITATION = doi:10.1103/PhysRevD.83.033004;%%
  %50 citations counted in INSPIRE as of 06 Mar 2018



%\cite{Ding:2013eca}
\bibitem{Ding:2013eca} 
  G.~J.~Ding and Y.~L.~Zhou,
  %``Dirac Neutrinos with $S_4$ Flavor Symmetry in Warped Extra Dimensions,''
  Nucl.\ Phys.\ B {\bf 876}, 418 (2013)
  doi:10.1016/j.nuclphysb.2013.08.011
  [arXiv:1304.2645 [hep-ph]].
  %%CITATION = doi:10.1016/j.nuclphysb.2013.08.011;%%
  %23 citations counted in INSPIRE as of 06 Mar 2018



%\cite{Hagedorn:2011un}
\bibitem{Hagedorn:2011un} 
  C.~Hagedorn and M.~Serone,
  %``Leptons in Holographic Composite Higgs Models with Non-Abelian Discrete Symmetries,''
  JHEP {\bf 1110}, 083 (2011)
  doi:10.1007/JHEP10(2011)083
  [arXiv:1106.4021 [hep-ph]].
  %%CITATION = doi:10.1007/JHEP10(2011)083;%%
  %26 citations counted in INSPIRE as of 06 Mar 2018



%\cite{Campos:2014zaa}
\bibitem{Campos:2014zaa} 
  M.~D.~Campos, A.~E.~Cárcamo Hernández, H.~Päs and E.~Schumacher,
  %``Higgs $\rightarrow$ $\mu\tau$ as an indication for $S_4$ flavor symmetry,''
  Phys.\ Rev.\ D {\bf 91}, no. 11, 116011 (2015)
  doi:10.1103/PhysRevD.91.116011
  [arXiv:1408.1652 [hep-ph]].
  %%CITATION = doi:10.1103/PhysRevD.91.116011;%%
  %65 citations counted in INSPIRE as of 06 Mar 2018



%\cite{Dong:2010zu}
\bibitem{Dong:2010zu} 
  P.~V.~Dong, H.~N.~Long, D.~V.~Soa and V.~V.~Vien,
  %``The 3-3-1 model with $S_4$ flavor symmetry,''
  Eur.\ Phys.\ J.\ C {\bf 71}, 1544 (2011)
  doi:10.1140/epjc/s10052-011-1544-2
  [arXiv:1009.2328 [hep-ph]].
  %%CITATION = doi:10.1140/epjc/s10052-011-1544-2;%%
  %51 citations counted in INSPIRE as of 06 Mar 2018



%\cite{VanVien:2015xha}
\bibitem{VanVien:2015xha} 
  V.~V.~Vien, H.~N.~Long and D.~P.~Khoi,
  %``Neutrino Mixing with Non-Zero $\theta_{13}$ and CP Violation in the 3-3-1 Model Based on $S_4$ Flavor Symmetry,''
  Int.\ J.\ Mod.\ Phys.\ A {\bf 30}, no. 17, 1550102 (2015)
  doi:10.1142/S0217751X1550102X
  [arXiv:1506.06063 [hep-ph]].
  %%CITATION = doi:10.1142/S0217751X1550102X;%%
  %10 citations counted in INSPIRE as of 06 Mar 2018



%\cite{Dimou:2015cmw}
\bibitem{Dimou:2015cmw} 
  M.~Dimou, S.~F.~King and C.~Luhn,
  %``Phenomenological implications of an SU(5)×S$_4$×U(1) SUSY GUT of flavor,''
  Phys.\ Rev.\ D {\bf 93}, no. 7, 075026 (2016)
  doi:10.1103/PhysRevD.93.075026
  [arXiv:1512.09063 [hep-ph]].
  %%CITATION = doi:10.1103/PhysRevD.93.075026;%%
  %5 citations counted in INSPIRE as of 06 Mar 2018



%\cite{King:2016yvg}
\bibitem{King:2016yvg} 
  S.~F.~King and C.~Luhn,
  %``Littlest Seesaw model from S$_{4} \times$ U(1),''
  JHEP {\bf 1609}, 023 (2016)
  doi:10.1007/JHEP09(2016)023
  [arXiv:1607.05276 [hep-ph]].
  %%CITATION = doi:10.1007/JHEP09(2016)023;%%
  %11 citations counted in INSPIRE as of 06 Mar 2018



%\cite{Bjorkeroth:2017ybg}
\bibitem{Bjorkeroth:2017ybg} 
  F.~Björkeroth, F.~J.~de Anda, S.~F.~King and E.~Perdomo,
  %``A natural S$_{4}$ × SO(10) model of flavour,''
  JHEP {\bf 1710}, 148 (2017)
  doi:10.1007/JHEP10(2017)148
  [arXiv:1705.01555 [hep-ph]].
  %%CITATION = doi:10.1007/JHEP10(2017)148;%%
  %8 citations counted in INSPIRE as of 06 Mar 2018



%\cite{deAnda:2017yeb}
\bibitem{deAnda:2017yeb} 
  F.~J.~de Anda, S.~F.~King and E.~Perdomo,
  %``$\mathbf{SO(10)}\times \mathbf{S_4}$ grand unified theory of flavour and leptogenesis,''
  JHEP {\bf 1712}, 075 (2017)
  doi:10.1007/JHEP12(2017)075
  [arXiv:1710.03229 [hep-ph]].
  %%CITATION = doi:10.1007/JHEP12(2017)075;%%
  %1 citations counted in INSPIRE as of 06 Mar 2018


%\cite{deAnda:2018oik}
\bibitem{deAnda:2018oik} 
  F.~J.~de Anda and S.~F.~King,
  %``An $S_4 \times SU(5)$ SUSY GUT of avour in 6d,''
  arXiv:1803.04978 [hep-ph].
  %%CITATION = ARXIV:1803.04978;%%


%\cite{CarcamoHernandez:2019eme}
\bibitem{CarcamoHernandez:2019eme} 
  A.~E.~Cárcamo Hernández and S.~F.~King,
  %``Littlest Inverse Seesaw Model,''
  arXiv:1903.02565 [hep-ph].
  %%CITATION = ARXIV:1903.02565;%%


%\cite{Frampton:1994rk}
\bibitem{Frampton:1994rk} 
  P.~H.~Frampton and T.~W.~Kephart,
  %``Simple nonAbelian finite flavor groups and fermion masses,''
  Int.\ J.\ Mod.\ Phys.\ A {\bf 10}, 4689 (1995)
  doi:10.1142/S0217751X95002187
  [hep-ph/9409330].
  %%CITATION = doi:10.1142/S0217751X95002187;%%
  %157 citations counted in INSPIRE as of 06 Mar 2018



%\cite{Grimus:2003kq}
\bibitem{Grimus:2003kq} 
  W.~Grimus and L.~Lavoura,
  %``A Discrete symmetry group for maximal atmospheric neutrino mixing,''
  Phys.\ Lett.\ B {\bf 572}, 189 (2003)
  doi:10.1016/j.physletb.2003.08.032
  [hep-ph/0305046].
  %%CITATION = doi:10.1016/j.physletb.2003.08.032;%%
  %249 citations counted in INSPIRE as of 06 Mar 2018



%\cite{Grimus:2004rj}
\bibitem{Grimus:2004rj} 
  W.~Grimus, A.~S.~Joshipura, S.~Kaneko, L.~Lavoura and M.~Tanimoto,
  %``Lepton mixing angle $\theta_{13} = 0$ with a horizontal symmetry $D_4$,''
  JHEP {\bf 0407}, 078 (2004)
  doi:10.1088/1126-6708/2004/07/078
  [hep-ph/0407112].
  %%CITATION = doi:10.1088/1126-6708/2004/07/078;%%
  %119 citations counted in INSPIRE as of 06 Mar 2018



%\cite{Frigerio:2004jg}
\bibitem{Frigerio:2004jg} 
  M.~Frigerio, S.~Kaneko, E.~Ma and M.~Tanimoto,
  %``Quaternion family symmetry of quarks and leptons,''
  Phys.\ Rev.\ D {\bf 71}, 011901 (2005)
  doi:10.1103/PhysRevD.71.011901
  [hep-ph/0409187].
  %%CITATION = doi:10.1103/PhysRevD.71.011901;%%
  %81 citations counted in INSPIRE as of 06 Mar 2018



%\cite{Adulpravitchai:2008yp}
\bibitem{Adulpravitchai:2008yp} 
  A.~Adulpravitchai, A.~Blum and C.~Hagedorn,
  %``A Supersymmetric D4 Model for mu-tau Symmetry,''
  JHEP {\bf 0903}, 046 (2009)
  doi:10.1088/1126-6708/2009/03/046
  [arXiv:0812.3799 [hep-ph]].
  %%CITATION = doi:10.1088/1126-6708/2009/03/046;%%
  %48 citations counted in INSPIRE as of 06 Mar 2018



%\cite{Ishimori:2008gp}
\bibitem{Ishimori:2008gp} 
  H.~Ishimori, T.~Kobayashi, H.~Ohki, Y.~Omura, R.~Takahashi and M.~Tanimoto,
  %``D(4) Flavor Symmetry for Neutrino Masses and Mixing,''
  Phys.\ Lett.\ B {\bf 662}, 178 (2008)
  doi:10.1016/j.physletb.2008.03.007
  [arXiv:0802.2310 [hep-ph]].
  %%CITATION = doi:10.1016/j.physletb.2008.03.007;%%
  %45 citations counted in INSPIRE as of 06 Mar 2018



%\cite{Hagedorn:2010mq}
\bibitem{Hagedorn:2010mq} 
  C.~Hagedorn and R.~Ziegler,
  %``$\mu-\tau$ Symmetry and Charged Lepton Mass Hierarchy in a Supersymmetric $D_4$ Model,''
  Phys.\ Rev.\ D {\bf 82}, 053011 (2010)
  doi:10.1103/PhysRevD.82.053011
  [arXiv:1007.1888 [hep-ph]].
  %%CITATION = doi:10.1103/PhysRevD.82.053011;%%
  %16 citations counted in INSPIRE as of 06 Mar 2018



%\cite{Meloni:2011cc}
\bibitem{Meloni:2011cc} 
  D.~Meloni, S.~Morisi and E.~Peinado,
  %``Stability of dark matter from the D4xZ2 flavor group,''
  Phys.\ Lett.\ B {\bf 703}, 281 (2011)
  doi:10.1016/j.physletb.2011.07.084
  [arXiv:1104.0178 [hep-ph]].
  %%CITATION = doi:10.1016/j.physletb.2011.07.084;%%
  %25 citations counted in INSPIRE as of 06 Mar 2018



%\cite{Vien:2013zra}
\bibitem{Vien:2013zra} 
  V.~V.~Vien and H.~N.~Long,
  %``The $D_4$ flavor symmery in 3-3-1 model with neutral leptons,''
  Int.\ J.\ Mod.\ Phys.\ A {\bf 28}, 1350159 (2013)
  doi:10.1142/S0217751X13501595
  [arXiv:1312.5034 [hep-ph]].
  %%CITATION = doi:10.1142/S0217751X13501595;%%
  %25 citations counted in INSPIRE as of 06 Mar 2018



%\cite{Babu:2004tn}
\bibitem{Babu:2004tn} 
  K.~S.~Babu and J.~Kubo,
  %``Dihedral families of quarks, leptons and Higgses,''
  Phys.\ Rev.\ D {\bf 71}, 056006 (2005)
  doi:10.1103/PhysRevD.71.056006
  [hep-ph/0411226].
  %%CITATION = doi:10.1103/PhysRevD.71.056006;%%
  %126 citations counted in INSPIRE as of 06 Mar 2018



%\cite{Kajiyama:2005rk}
\bibitem{Kajiyama:2005rk} 
  Y.~Kajiyama, E.~Itou and J.~Kubo,
  %``NonAbelian discrete family symmetry to soften the SUSY flavor problem and to suppress proton decay,''
  Nucl.\ Phys.\ B {\bf 743}, 74 (2006)
  doi:10.1016/j.nuclphysb.2006.02.042
  [hep-ph/0511268].
  %%CITATION = doi:10.1016/j.nuclphysb.2006.02.042;%%
  %55 citations counted in INSPIRE as of 06 Mar 2018



%\cite{Kajiyama:2007pr}
\bibitem{Kajiyama:2007pr} 
  Y.~Kajiyama,
  %``R-parity violation and non-Abelian discrete family symmetry,''
  JHEP {\bf 0704}, 007 (2007)
  doi:10.1088/1126-6708/2007/04/007
  [hep-ph/0702056 [HEP-PH]].
  %%CITATION = doi:10.1088/1126-6708/2007/04/007;%%
  %15 citations counted in INSPIRE as of 06 Mar 2018



%\cite{Kifune:2007fj}
\bibitem{Kifune:2007fj} 
  N.~Kifune, J.~Kubo and A.~Lenz,
  %``Flavor Changing Neutral Higgs Bosons in a Supersymmetric Extension based on a $Q_{6}$ Family Symmetry,''
  Phys.\ Rev.\ D {\bf 77}, 076010 (2008)
  doi:10.1103/PhysRevD.77.076010
  [arXiv:0712.0503 [hep-ph]].
  %%CITATION = doi:10.1103/PhysRevD.77.076010;%%
  %33 citations counted in INSPIRE as of 06 Mar 2018



%\cite{Babu:2009nn}
\bibitem{Babu:2009nn} 
  K.~S.~Babu and Y.~Meng,
  %``Flavor Violation in Supersymmetric Q(6) Model,''
  Phys.\ Rev.\ D {\bf 80}, 075003 (2009)
  doi:10.1103/PhysRevD.80.075003
  [arXiv:0907.4231 [hep-ph]].
  %%CITATION = doi:10.1103/PhysRevD.80.075003;%%
  %28 citations counted in INSPIRE as of 06 Mar 2018



%\cite{Kawashima:2009jv}
\bibitem{Kawashima:2009jv} 
  K.~Kawashima, J.~Kubo and A.~Lenz,
  %``Testing the new CP phase in a Supersymmetric Model with Q(6) Family Symmetry by B(s) Mixing,''
  Phys.\ Lett.\ B {\bf 681}, 60 (2009)
  doi:10.1016/j.physletb.2009.09.064
  [arXiv:0907.2302 [hep-ph]].
  %%CITATION = doi:10.1016/j.physletb.2009.09.064;%%
  %22 citations counted in INSPIRE as of 06 Mar 2018



%\cite{Kaburaki:2010xc}
\bibitem{Kaburaki:2010xc} 
  Y.~Kaburaki, K.~Konya, J.~Kubo and A.~Lenz,
  %``Triangle Relation of Dark Matter, EDM and CP Violation in B0 Mixing in a Supersymmetric Q6 Model,''
  Phys.\ Rev.\ D {\bf 84}, 016007 (2011)
  doi:10.1103/PhysRevD.84.016007
  [arXiv:1012.2435 [hep-ph]].
  %%CITATION = doi:10.1103/PhysRevD.84.016007;%%
  %20 citations counted in INSPIRE as of 06 Mar 2018



%\cite{Babu:2011mv}
\bibitem{Babu:2011mv} 
  K.~S.~Babu, K.~Kawashima and J.~Kubo,
  %``Variations on the Supersymmetric $Q_6$ Model of Flavor,''
  Phys.\ Rev.\ D {\bf 83}, 095008 (2011)
  doi:10.1103/PhysRevD.83.095008
  [arXiv:1103.1664 [hep-ph]].
  %%CITATION = doi:10.1103/PhysRevD.83.095008;%%
  %28 citations counted in INSPIRE as of 06 Mar 2018



%\cite{Araki:2011zg}
\bibitem{Araki:2011zg} 
  T.~Araki and Y.~F.~Li,
  %``Q_6 flavor symmetry model for the extension of the minimal standard model by three right-handed sterile neutrinos,''
  Phys.\ Rev.\ D {\bf 85}, 065016 (2012)
  doi:10.1103/PhysRevD.85.065016
  [arXiv:1112.5819 [hep-ph]].
  %%CITATION = doi:10.1103/PhysRevD.85.065016;%%
  %51 citations counted in INSPIRE as of 06 Mar 2018



%\cite{Gomez-Izquierdo:2013uaa}
\bibitem{Gomez-Izquierdo:2013uaa} 
  J.~C.~Gómez-Izquierdo, F.~González-Canales and M.~Mondragon,
  %``$Q_{6}$ as the flavor symmetry in a non-minimal SUSY $SU(5)$ model,''
  Eur.\ Phys.\ J.\ C {\bf 75}, no. 5, 221 (2015)
  doi:10.1140/epjc/s10052-015-3440-7
  [arXiv:1312.7385 [hep-ph]].
  %%CITATION = doi:10.1140/epjc/s10052-015-3440-7;%%
  %10 citations counted in INSPIRE as of 06 Mar 2018



%\cite{Gomez-Izquierdo:2017med}
\bibitem{Gomez-Izquierdo:2017med} 
  J.~C.~Gómez-Izquierdo, F.~Gonzalez-Canales and M.~Mondragón,
  %``On ${\bf Q}_{6}$ flavor symmetry and the breaking of $\mu \leftrightarrow \tau$ symmetry,''
  Int.\ J.\ Mod.\ Phys.\ A {\bf 32}, no. 28-29, 1750171 (2017)
  doi:10.1142/S0217751X17501718
  [arXiv:1705.06324 [hep-ph]].
  %%CITATION = doi:10.1142/S0217751X17501718;%%
  %3 citations counted in INSPIRE as of 06 Mar 2018



%\cite{Luhn:2007sy}
\bibitem{Luhn:2007sy} 
  C.~Luhn, S.~Nasri and P.~Ramond,
  %``Tri-bimaximal neutrino mixing and the family symmetry semidirect product of Z(7) and Z(3),''
  Phys.\ Lett.\ B {\bf 652}, 27 (2007)
  doi:10.1016/j.physletb.2007.06.059
  [arXiv:0706.2341 [hep-ph]].
  %%CITATION = doi:10.1016/j.physletb.2007.06.059;%%
  %153 citations counted in INSPIRE as of 06 Mar 2018



%\cite{Hagedorn:2008bc}
\bibitem{Hagedorn:2008bc} 
  C.~Hagedorn, M.~A.~Schmidt and A.~Y.~Smirnov,
  %``Lepton Mixing and Cancellation of the Dirac Mass Hierarchy in SO(10) GUTs with Flavor Symmetries T(7) and Sigma(81),''
  Phys.\ Rev.\ D {\bf 79}, 036002 (2009)
  doi:10.1103/PhysRevD.79.036002
  [arXiv:0811.2955 [hep-ph]].
  %%CITATION = doi:10.1103/PhysRevD.79.036002;%%
  %74 citations counted in INSPIRE as of 06 Mar 2018



%\cite{Cao:2010mp}
\bibitem{Cao:2010mp} 
  Q.~H.~Cao, S.~Khalil, E.~Ma and H.~Okada,
  %``Observable $T_7$ Lepton Flavor Symmetry at the Large Hadron Collider,''
  Phys.\ Rev.\ Lett.\  {\bf 106}, 131801 (2011)
  doi:10.1103/PhysRevLett.106.131801
  [arXiv:1009.5415 [hep-ph]].
  %%CITATION = doi:10.1103/PhysRevLett.106.131801;%%
  %38 citations counted in INSPIRE as of 06 Mar 2018



%\cite{Luhn:2012bc}
\bibitem{Luhn:2012bc} 
  C.~Luhn, K.~M.~Parattu and A.~Wingerter,
  %``A Minimal Model of Neutrino Flavor,''
  JHEP {\bf 1212}, 096 (2012)
  doi:10.1007/JHEP12(2012)096
  [arXiv:1210.1197 [hep-ph]].
  %%CITATION = doi:10.1007/JHEP12(2012)096;%%
  %30 citations counted in INSPIRE as of 06 Mar 2018



%\cite{Kajiyama:2013lja}
\bibitem{Kajiyama:2013lja} 
  Y.~Kajiyama, H.~Okada and K.~Yagyu,
  %``$T_7$ Flavor Model in Three Loop Seesaw and Higgs Phenomenology,''
  JHEP {\bf 1310}, 196 (2013)
  doi:10.1007/JHEP10(2013)196
  [arXiv:1307.0480 [hep-ph]].
  %%CITATION = doi:10.1007/JHEP10(2013)196;%%
  %48 citations counted in INSPIRE as of 06 Mar 2018



%\cite{Bonilla:2014xla}
\bibitem{Bonilla:2014xla} 
  C.~Bonilla, S.~Morisi, E.~Peinado and J.~W.~F.~Valle,
  %``Relating quarks and leptons with the $T_7$ flavour group,''
  Phys.\ Lett.\ B {\bf 742}, 99 (2015)
  doi:10.1016/j.physletb.2015.01.017
  [arXiv:1411.4883 [hep-ph]].
  %%CITATION = doi:10.1016/j.physletb.2015.01.017;%%
  %22 citations counted in INSPIRE as of 06 Mar 2018



%\cite{Vien:2014gza}
\bibitem{Vien:2014gza} 
  V.~V.~Vien and H.~N.~Long,
  %``The $T_7$ flavor symmetry in 3-3-1 model with neutral leptons,''
  JHEP {\bf 1404}, 133 (2014)
  doi:10.1007/JHEP04(2014)133
  [arXiv:1402.1256 [hep-ph]].
  %%CITATION = doi:10.1007/JHEP04(2014)133;%%
  %28 citations counted in INSPIRE as of 06 Mar 2018



%\cite{Vien:2015koa}
\bibitem{Vien:2015koa} 
  V.~V.~Vien,
  %``$T_7$ flavor symmetry scheme for understanding neutrino mass and mixing in 3-3-1 model with neutral leptons,''
  Mod.\ Phys.\ Lett.\ A {\bf 29}, 28 (2014)
  doi:10.1142/S0217732314501399
  [arXiv:1508.02585 [hep-ph]].
  %%CITATION = doi:10.1142/S0217732314501399;%%
  %6 citations counted in INSPIRE as of 06 Mar 2018



%\cite{Hernandez:2015cra}
\bibitem{Hernandez:2015cra} 
  A.~E.~Cárcamo Hernández and R.~Martinez,
  %``Fermion mass and mixing pattern in a minimal T7 flavor 331 model,''
  J.\ Phys.\ G {\bf 43}, no. 4, 045003 (2016)
  doi:10.1088/0954-3899/43/4/045003
  [arXiv:1501.07261 [hep-ph]].
  %%CITATION = doi:10.1088/0954-3899/43/4/045003;%%
  %22 citations counted in INSPIRE as of 06 Mar 2018



%\cite{Arbelaez:2015toa}
\bibitem{Arbelaez:2015toa} 
  C.~Arbeláez, A.~E.~Cárcamo Hernández, S.~Kovalenko and I.~Schmidt,
  %``Adjoint $SU(5)$ GUT model with $T_{7}$ flavor symmetry,''
  Phys.\ Rev.\ D {\bf 92}, no. 11, 115015 (2015)
  doi:10.1103/PhysRevD.92.115015
  [arXiv:1507.03852 [hep-ph]].
  %%CITATION = doi:10.1103/PhysRevD.92.115015;%%
  %12 citations counted in INSPIRE as of 06 Mar 2018



%\cite{Ding:2011qt}
\bibitem{Ding:2011qt} 
  G.~J.~Ding,
  %``Tri-Bimaximal Neutrino Mixing and the $T_{13}$ Flavor Symmetry,''
  Nucl.\ Phys.\ B {\bf 853}, 635 (2011)
  doi:10.1016/j.nuclphysb.2011.08.012
  [arXiv:1105.5879 [hep-ph]].
  %%CITATION = doi:10.1016/j.nuclphysb.2011.08.012;%%
  %11 citations counted in INSPIRE as of 06 Mar 2018



%\cite{Hartmann:2011dn}
\bibitem{Hartmann:2011dn} 
  C.~Hartmann,
  %``The Frobenius group T13 and the canonical see-saw mechanism applied to neutrino mixing,''
  Phys.\ Rev.\ D {\bf 85}, 013012 (2012)
  doi:10.1103/PhysRevD.85.013012
  [arXiv:1109.5143 [hep-ph]].
  %%CITATION = doi:10.1103/PhysRevD.85.013012;%%
  %6 citations counted in INSPIRE as of 06 Mar 2018



%\cite{Hartmann:2011pq}
\bibitem{Hartmann:2011pq} 
  C.~Hartmann and A.~Zee,
  %``Neutrino Mixing and the Frobenius Group T13,''
  Nucl.\ Phys.\ B {\bf 853}, 105 (2011)
  doi:10.1016/j.nuclphysb.2011.07.023
  [arXiv:1106.0333 [hep-ph]].
  %%CITATION = doi:10.1016/j.nuclphysb.2011.07.023;%%
  %9 citations counted in INSPIRE as of 06 Mar 2018



%\cite{Kajiyama:2010sb}
\bibitem{Kajiyama:2010sb} 
  Y.~Kajiyama and H.~Okada,
  %``T(13) Flavor Symmetry and Decaying Dark Matter,''
  Nucl.\ Phys.\ B {\bf 848}, 303 (2011)
  doi:10.1016/j.nuclphysb.2011.02.020
  [arXiv:1011.5753 [hep-ph]].
  %%CITATION = doi:10.1016/j.nuclphysb.2011.02.020;%%
  %32 citations counted in INSPIRE as of 06 Mar 2018



%\cite{Aranda:2000tm}
\bibitem{Aranda:2000tm} 
  A.~Aranda, C.~D.~Carone and R.~F.~Lebed,
  %``Maximal neutrino mixing from a minimal flavor symmetry,''
  Phys.\ Rev.\ D {\bf 62}, 016009 (2000)
  doi:10.1103/PhysRevD.62.016009
  [hep-ph/0002044].
  %%CITATION = doi:10.1103/PhysRevD.62.016009;%%
  %116 citations counted in INSPIRE as of 06 Mar 2018



%\cite{Sen:2007vx}
\bibitem{Sen:2007vx} 
  S.~Sen,
  %``Quark masses in supersymmetric SU(3)(color) x SU(3)(W) x U(1)(X) model with discrete T-prime flavor symmetry,''
  Phys.\ Rev.\ D {\bf 76}, 115020 (2007)
  doi:10.1103/PhysRevD.76.115020
  [arXiv:0710.2734 [hep-ph]].
  %%CITATION = doi:10.1103/PhysRevD.76.115020;%%
  %16 citations counted in INSPIRE as of 06 Mar 2018



%\cite{Aranda:2007dp}
\bibitem{Aranda:2007dp} 
  A.~Aranda,
  %``Neutrino mixing from the double tetrahedral group T-prime,''
  Phys.\ Rev.\ D {\bf 76}, 111301 (2007)
  doi:10.1103/PhysRevD.76.111301
  [arXiv:0707.3661 [hep-ph]].
  %%CITATION = doi:10.1103/PhysRevD.76.111301;%%
  %75 citations counted in INSPIRE as of 06 Mar 2018



%\cite{Chen:2007afa}
\bibitem{Chen:2007afa} 
  M.~C.~Chen and K.~T.~Mahanthappa,
  %``CKM and Tri-bimaximal MNS Matrices in a $SU(5) \times ^{(d)}T$ Model,''
  Phys.\ Lett.\ B {\bf 652}, 34 (2007)
  doi:10.1016/j.physletb.2007.06.064
  [arXiv:0705.0714 [hep-ph]].
  %%CITATION = doi:10.1016/j.physletb.2007.06.064;%%
  %215 citations counted in INSPIRE as of 06 Mar 2018



%\cite{Frampton:2008bz}
\bibitem{Frampton:2008bz} 
  P.~H.~Frampton, T.~W.~Kephart and S.~Matsuzaki,
  %``Simplified Renormalizable T-prime Model for Tribimaximal Mixing and Cabibbo Angle,''
  Phys.\ Rev.\ D {\bf 78}, 073004 (2008)
  doi:10.1103/PhysRevD.78.073004
  [arXiv:0807.4713 [hep-ph]].
  %%CITATION = doi:10.1103/PhysRevD.78.073004;%%
  %62 citations counted in INSPIRE as of 06 Mar 2018



%\cite{Eby:2011ph}
\bibitem{Eby:2011ph} 
  D.~A.~Eby, P.~H.~Frampton, X.~G.~He and T.~W.~Kephart,
  %``Quartification with T' Flavor,''
  Phys.\ Rev.\ D {\bf 84}, 037302 (2011)
  doi:10.1103/PhysRevD.84.037302
  [arXiv:1103.5737 [hep-ph]].
  %%CITATION = doi:10.1103/PhysRevD.84.037302;%%
  %15 citations counted in INSPIRE as of 06 Mar 2018



%\cite{Frampton:2013lva}
\bibitem{Frampton:2013lva} 
  P.~H.~Frampton, C.~M.~Ho and T.~W.~Kephart,
  %``Heterotic discrete flavor model,''
  Phys.\ Rev.\ D {\bf 89}, no. 2, 027701 (2014)
  doi:10.1103/PhysRevD.89.027701
  [arXiv:1305.4402 [hep-ph]].
  %%CITATION = doi:10.1103/PhysRevD.89.027701;%%
  %11 citations counted in INSPIRE as of 06 Mar 2018



%\cite{Chen:2013wba}
\bibitem{Chen:2013wba} 
  M.~C.~Chen, J.~Huang, K.~T.~Mahanthappa and A.~M.~Wijangco,
  %``Large \theta_13 in a SUSY SU(5)xT' Model,''
  JHEP {\bf 1310}, 112 (2013)
  doi:10.1007/JHEP10(2013)112
  [arXiv:1307.7711 [hep-ph]].
  %%CITATION = doi:10.1007/JHEP10(2013)112;%%
  %29 citations counted in INSPIRE as of 06 Mar 2018


%\cite{Vien:2018otl}
\bibitem{Vien:2018otl} 
  V.~V.~Vien, H.~N.~Long and A.~E.~Cárcamo Hernández,
  %``Lepton masses and mixings in a $T'$ flavoured 3-3-1 model with type I and II seesaw mechanisms,''
  Mod.\ Phys.\ Lett.\ A {\bf 34}, 1950005 (2019)
  doi:10.1142/S0217732319500056
  [arXiv:1812.07263 [hep-ph]].
  %%CITATION = doi:10.1142/S0217732319500056;%%



%\cite{Branco:1983tn}
\bibitem{Branco:1983tn} 
  G.~C.~Branco, J.~M.~Gerard and W.~Grimus,
  %``Geometrical T Violation,''
  Phys.\ Lett.\  {\bf 136B}, 383 (1984).
  doi:10.1016/0370-2693(84)92024-0
  %%CITATION = doi:10.1016/0370-2693(84)92024-0;%%
  %126 citations counted in INSPIRE as of 04 Oct 2018



%\cite{deMedeirosVarzielas:2006fc}
\bibitem{deMedeirosVarzielas:2006fc} 
  I.~de Medeiros Varzielas, S.~F.~King and G.~G.~Ross,
  %``Neutrino tri-bi-maximal mixing from a non-Abelian discrete family symmetry,''
  Phys.\ Lett.\ B {\bf 648}, 201 (2007)
  doi:10.1016/j.physletb.2007.03.009
  [hep-ph/0607045].
  %%CITATION = doi:10.1016/j.physletb.2007.03.009;%%
  %233 citations counted in INSPIRE as of 06 Mar 2018



%\cite{Ma:2007wu}
\bibitem{Ma:2007wu} 
  E.~Ma,
  %``Near tribimaximal neutrino mixing with Delta(27) symmetry,''
  Phys.\ Lett.\ B {\bf 660}, 505 (2008)
  doi:10.1016/j.physletb.2007.12.060
  [arXiv:0709.0507 [hep-ph]].
  %%CITATION = doi:10.1016/j.physletb.2007.12.060;%%
  %96 citations counted in INSPIRE as of 06 Mar 2018



%\cite{Bazzocchi:2009qg}
\bibitem{Bazzocchi:2009qg} 
  F.~Bazzocchi and I.~de Medeiros Varzielas,
  %``Tri-bi-maximal mixing in viable family symmetry unified model with extended seesaw,''
  Phys.\ Rev.\ D {\bf 79}, 093001 (2009)
  doi:10.1103/PhysRevD.79.093001
  [arXiv:0902.3250 [hep-ph]].
  %%CITATION = doi:10.1103/PhysRevD.79.093001;%%
  %32 citations counted in INSPIRE as of 06 Mar 2018



%\cite{Varzielas:2012nn}
\bibitem{Varzielas:2012nn} 
  I.~de Medeiros Varzielas, D.~Emmanuel-Costa and P.~Leser,
  %``Geometrical CP Violation from Non-Renormalisable Scalar Potentials,''
  Phys.\ Lett.\ B {\bf 716}, 193 (2012)
  doi:10.1016/j.physletb.2012.08.008
  [arXiv:1204.3633 [hep-ph]].
  %%CITATION = doi:10.1016/j.physletb.2012.08.008;%%
  %52 citations counted in INSPIRE as of 06 Mar 2018



%\cite{Bhattacharyya:2012pi}
\bibitem{Bhattacharyya:2012pi} 
  G.~Bhattacharyya, I.~de Medeiros Varzielas and P.~Leser,
  %``A common origin of fermion mixing and geometrical CP violation, and its test through Higgs physics at the LHC,''
  Phys.\ Rev.\ Lett.\  {\bf 109}, 241603 (2012)
  doi:10.1103/PhysRevLett.109.241603
  [arXiv:1210.0545 [hep-ph]].
  %%CITATION = doi:10.1103/PhysRevLett.109.241603;%%
  %65 citations counted in INSPIRE as of 06 Mar 2018



%\cite{Ferreira:2012ri}
\bibitem{Ferreira:2012ri} 
  P.~M.~Ferreira, W.~Grimus, L.~Lavoura and P.~O.~Ludl,
  %``Maximal CP Violation in Lepton Mixing from a Model with Delta(27) flavour Symmetry,''
  JHEP {\bf 1209}, 128 (2012)
  doi:10.1007/JHEP09(2012)128
  [arXiv:1206.7072 [hep-ph]].
  %%CITATION = doi:10.1007/JHEP09(2012)128;%%
  %60 citations counted in INSPIRE as of 30 Mar 2019




%\cite{Ma:2013xqa}
\bibitem{Ma:2013xqa} 
  E.~Ma,
  %``Neutrino Mixing and Geometric CP Violation with Delta(27) Symmetry,''
  Phys.\ Lett.\ B {\bf 723}, 161 (2013)
  doi:10.1016/j.physletb.2013.05.011
  [arXiv:1304.1603 [hep-ph]].
  %%CITATION = doi:10.1016/j.physletb.2013.05.011;%%
  %27 citations counted in INSPIRE as of 06 Mar 2018



%\cite{Nishi:2013jqa}
\bibitem{Nishi:2013jqa} 
  C.~C.~Nishi,
  %``Generalized $CP$ symmetries in $\Delta(27)$ flavor models,''
  Phys.\ Rev.\ D {\bf 88}, no. 3, 033010 (2013)
  doi:10.1103/PhysRevD.88.033010
  [arXiv:1306.0877 [hep-ph]].
  %%CITATION = doi:10.1103/PhysRevD.88.033010;%%
  %44 citations counted in INSPIRE as of 06 Mar 2018



%\cite{Varzielas:2013sla}
\bibitem{Varzielas:2013sla} 
  I.~de Medeiros Varzielas and D.~Pidt,
  %``Towards realistic models of quark masses with geometrical CP violation,''
  J.\ Phys.\ G {\bf 41}, 025004 (2014)
  doi:10.1088/0954-3899/41/2/025004
  [arXiv:1307.0711 [hep-ph]].
  %%CITATION = doi:10.1088/0954-3899/41/2/025004;%%
  %34 citations counted in INSPIRE as of 06 Mar 2018



%\cite{Aranda:2013gga}
\bibitem{Aranda:2013gga} 
  A.~Aranda, C.~Bonilla, S.~Morisi, E.~Peinado and J.~W.~F.~Valle,
  %``Dirac neutrinos from flavor symmetry,''
  Phys.\ Rev.\ D {\bf 89}, no. 3, 033001 (2014)
  doi:10.1103/PhysRevD.89.033001
  [arXiv:1307.3553 [hep-ph]].
  %%CITATION = doi:10.1103/PhysRevD.89.033001;%%
  %34 citations counted in INSPIRE as of 06 Mar 2018



%\cite{Ma:2014eka}
\bibitem{Ma:2014eka} 
  E.~Ma and A.~Natale,
  %``Scotogenic $Z_2$ or $U(1)_D$ Model of Neutrino Mass with $\Delta(27)$ Symmetry,''
  Phys.\ Lett.\ B {\bf 734}, 403 (2014)
  doi:10.1016/j.physletb.2014.05.070
  [arXiv:1403.6772 [hep-ph]].
  %%CITATION = doi:10.1016/j.physletb.2014.05.070;%%
  %33 citations counted in INSPIRE as of 06 Mar 2018



%\cite{Abbas:2014ewa}
\bibitem{Abbas:2014ewa} 
  M.~Abbas and S.~Khalil,
  %``Fermion masses and mixing in $\Delta(27)$ flavour model,''
  Phys.\ Rev.\ D {\bf 91}, no. 5, 053003 (2015)
  doi:10.1103/PhysRevD.91.053003
  [arXiv:1406.6716 [hep-ph]].
  %%CITATION = doi:10.1103/PhysRevD.91.053003;%%
  %9 citations counted in INSPIRE as of 06 Mar 2018



%\cite{Abbas:2015zna}
\bibitem{Abbas:2015zna} 
  M.~Abbas, S.~Khalil, A.~Rashed and A.~Sil,
  %``Neutrino masses and deviation from tribimaximal mixing in Δ(27) model with inverse seesaw mechanism,''
  Phys.\ Rev.\ D {\bf 93}, no. 1, 013018 (2016)
  doi:10.1103/PhysRevD.93.013018
  [arXiv:1508.03727 [hep-ph]].
  %%CITATION = doi:10.1103/PhysRevD.93.013018;%%
  %10 citations counted in INSPIRE as of 06 Mar 2018



%\cite{Varzielas:2015aua}
\bibitem{Varzielas:2015aua} 
  I.~de Medeiros Varzielas,
  %``$\Delta(27)$ family symmetry and neutrino mixing,''
  JHEP {\bf 1508}, 157 (2015)
  doi:10.1007/JHEP08(2015)157
  [arXiv:1507.00338 [hep-ph]].
  %%CITATION = doi:10.1007/JHEP08(2015)157;%%
  %18 citations counted in INSPIRE as of 06 Mar 2018



%\cite{Bjorkeroth:2015uou}
\bibitem{Bjorkeroth:2015uou} 
  F.~Björkeroth, F.~J.~de Anda, I.~de Medeiros Varzielas and S.~F.~King,
  %``Towards a complete $\Delta(27) \times SO(10)$ SUSY GUT,''
  Phys.\ Rev.\ D {\bf 94}, no. 1, 016006 (2016)
  doi:10.1103/PhysRevD.94.016006
  [arXiv:1512.00850 [hep-ph]].
  %%CITATION = doi:10.1103/PhysRevD.94.016006;%%
  %24 citations counted in INSPIRE as of 06 Mar 2018



%\cite{Chen:2015jta}
\bibitem{Chen:2015jta} 
  P.~Chen, G.~J.~Ding, A.~D.~Rojas, C.~A.~Vaquera-Araujo and J.~W.~F.~Valle,
  %``Warped flavor symmetry predictions for neutrino physics,''
  JHEP {\bf 1601}, 007 (2016)
  doi:10.1007/JHEP01(2016)007
  [arXiv:1509.06683 [hep-ph]].
  %%CITATION = doi:10.1007/JHEP01(2016)007;%%
  %31 citations counted in INSPIRE as of 06 Mar 2018



%\cite{Vien:2016tmh}
\bibitem{Vien:2016tmh} 
  V.~V.~Vien, A.~E.~Cárcamo Hernández and H.~N.~Long,
  %``The $\Delta(27)$ flavor 3-3-1 model with neutral leptons,''
  Nucl.\ Phys.\ B {\bf 913}, 792 (2016)
  doi:10.1016/j.nuclphysb.2016.10.010
  [arXiv:1601.03300 [hep-ph]].
  %%CITATION = doi:10.1016/j.nuclphysb.2016.10.010;%%
  %24 citations counted in INSPIRE as of 06 Mar 2018



%\cite{Hernandez:2016eod}
\bibitem{Hernandez:2016eod} 
  A.~E.~Cárcamo Hernández, H.~N.~Long and V.~V.~Vien,
  %``A 3-3-1 model with right-handed neutrinos based on the $\varDelta \left( 27\right) $ family symmetry,''
  Eur.\ Phys.\ J.\ C {\bf 76}, no. 5, 242 (2016)
  doi:10.1140/epjc/s10052-016-4074-0
  [arXiv:1601.05062 [hep-ph]].
  %%CITATION = doi:10.1140/epjc/s10052-016-4074-0;%%
  %19 citations counted in INSPIRE as of 06 Mar 2018



%\cite{CarcamoHernandez:2017owh}
\bibitem{CarcamoHernandez:2017owh} 
  A.~E.~Cárcamo Hernández, S.~Kovalenko, J.~W.~F.~Valle and C.~A.~Vaquera-Araujo,
  %``Predictive Pati-Salam theory of fermion masses and mixing,''
  JHEP {\bf 1707}, 118 (2017)
  doi:10.1007/JHEP07(2017)118
  [arXiv:1705.06320 [hep-ph]].
  %%CITATION = doi:10.1007/JHEP07(2017)118;%%
  %9 citations counted in INSPIRE as of 06 Mar 2018



%\cite{deMedeirosVarzielas:2017sdv}
\bibitem{deMedeirosVarzielas:2017sdv} 
  I.~de Medeiros Varzielas, G.~G.~Ross and J.~Talbert,
  %``A Unified Model of Quarks and Leptons with a Universal Texture Zero,''
  arXiv:1710.01741 [hep-ph].
  %%CITATION = ARXIV:1710.01741;%%
  %3 citations counted in INSPIRE as of 06 Mar 2018

%\cite{Bernal:2017xat}
\bibitem{Bernal:2017xat} 
  N.~Bernal, A.~E.~Cárcamo Hernández, I.~de Medeiros Varzielas and S.~Kovalenko,
  %``Fermion masses and mixings and dark matter constraints in a model with radiative seesaw mechanism,''
  JHEP {\bf 1805}, 053 (2018)
  doi:10.1007/JHEP05(2018)053
  [arXiv:1712.02792 [hep-ph]].
  %%CITATION = doi:10.1007/JHEP05(2018)053;%%
  %10 citations counted in INSPIRE as of 16 Nov 2018



%\cite{CarcamoHernandez:2018iel}
\bibitem{CarcamoHernandez:2018iel} 
  A.~E.~Cárcamo Hernández, H.~N.~Long and V.~V.~Vien,
  %``The first $\Delta(27)$ flavor 3-3-1 model with low scale seesaw mechanism,''
  Eur.\ Phys.\ J.\ C {\bf 78}, no. 10, 804 (2018)
  doi:10.1140/epjc/s10052-018-6284-0
  [arXiv:1803.01636 [hep-ph]].
  %%CITATION = doi:10.1140/epjc/s10052-018-6284-0;%%
  %8 citations counted in INSPIRE as of 16 Nov 2018



%\cite{deMedeirosVarzielas:2018vab}
\bibitem{deMedeirosVarzielas:2018vab} 
  I.~De Medeiros Varzielas, M.~L.~López-Ibáñez, A.~Melis and O.~Vives,
  %``Controlled flavor violation in the MSSM from a unified $\Delta(27)$ flavor symmetry,''
  JHEP {\bf 1809}, 047 (2018)
  doi:10.1007/JHEP09(2018)047
  [arXiv:1807.00860 [hep-ph]].
  %%CITATION = doi:10.1007/JHEP09(2018)047;%%
  %3 citations counted in INSPIRE as of 16 Nov 2018



%\cite{CarcamoHernandez:2018djj}
\bibitem{CarcamoHernandez:2018djj} 
  A.~E.~Cárcamo Hernández, J.~C.~Gómez-Izquierdo, S.~Kovalenko and M.~Mondragón,
  %``$\Delta \left( 27\right)$ flavor singlet-triplet Higgs model for fermion masses and mixings,''
  arXiv:1810.01764 [hep-ph].
  %%CITATION = ARXIV:1810.01764;%%
  %1 citations counted in INSPIRE as of 16 Nov 2018


%\cite{CarcamoHernandez:2018hst}
\bibitem{CarcamoHernandez:2018hst} 
  A.~E.~Cárcamo Hernández, S.~Kovalenko, J.~W.~F.~Valle and C.~A.~Vaquera-Araujo,
  %``Neutrino predictions from a left-right symmetric flavored extension of the standard model,''
  arXiv:1811.03018 [hep-ph].
  %%CITATION = ARXIV:1811.03018;%%


%\cite{Carballo-Perez:2016ooy}
\bibitem{Carballo-Perez:2016ooy} 
  B.~Carballo-Perez, E.~Peinado and S.~Ramos-Sanchez,
  %``$\Delta(54)$ flavor phenomenology and strings,''
  JHEP {\bf 1612}, 131 (2016)
  doi:10.1007/JHEP12(2016)131
  [arXiv:1607.06812 [hep-ph]].
  %%CITATION = doi:10.1007/JHEP12(2016)131;%%
  %6 citations counted in INSPIRE as of 06 Mar 2018



%\cite{King:2012in}
\bibitem{King:2012in} 
  S.~F.~King, C.~Luhn and A.~J.~Stuart,
  %``A Grand Delta(96) x SU(5) Flavour Model,''
  Nucl.\ Phys.\ B {\bf 867}, 203 (2013)
  doi:10.1016/j.nuclphysb.2012.09.021
  [arXiv:1207.5741 [hep-ph]].
  %%CITATION = doi:10.1016/j.nuclphysb.2012.09.021;%%
  %74 citations counted in INSPIRE as of 06 Mar 2018



%\cite{King:2013vna}
\bibitem{King:2013vna} 
  S.~F.~King, T.~Neder and A.~J.~Stuart,
  %``Lepton mixing predictions from $\Delta(6n^2)$ family Symmetry,''
  Phys.\ Lett.\ B {\bf 726}, 312 (2013)
  doi:10.1016/j.physletb.2013.08.052
  [arXiv:1305.3200 [hep-ph]].
  %%CITATION = doi:10.1016/j.physletb.2013.08.052;%%
  %80 citations counted in INSPIRE as of 06 Mar 2018



%\cite{Ding:2014ssa}
\bibitem{Ding:2014ssa} 
  G.~J.~Ding and S.~F.~King,
  %``Generalized $CP$ and $\Delta(96)$ family symmetry,''
  Phys.\ Rev.\ D {\bf 89}, no. 9, 093020 (2014)
  doi:10.1103/PhysRevD.89.093020
  [arXiv:1403.5846 [hep-ph]].
  %%CITATION = doi:10.1103/PhysRevD.89.093020;%%
  %41 citations counted in INSPIRE as of 06 Mar 2018



%\cite{Ishimori:2014jwa}
\bibitem{Ishimori:2014jwa} 
  H.~Ishimori and S.~F.~King,
  %``A model of quarks with Δ(6N$^2$) family symmetry,''
  Phys.\ Lett.\ B {\bf 735}, 33 (2014)
  doi:10.1016/j.physletb.2014.06.003
  [arXiv:1403.4395 [hep-ph]].
  %%CITATION = doi:10.1016/j.physletb.2014.06.003;%%
  %13 citations counted in INSPIRE as of 06 Mar 2018



%\cite{King:2014rwa}
\bibitem{King:2014rwa} 
  S.~F.~King and T.~Neder,
  %``Lepton mixing predictions including Majorana phases from Δ(6n$^2$) flavour symmetry and generalised CP,''
  Phys.\ Lett.\ B {\bf 736}, 308 (2014)
  doi:10.1016/j.physletb.2014.07.043
  [arXiv:1403.1758 [hep-ph]].
  %%CITATION = doi:10.1016/j.physletb.2014.07.043;%%
  %46 citations counted in INSPIRE as of 06 Mar 2018



%\cite{Ishimori:2014nxa}
\bibitem{Ishimori:2014nxa} 
  H.~Ishimori, S.~F.~King, H.~Okada and M.~Tanimoto,
  %``Quark mixing from $\Delta$(6N$^2$) family symmetry,''
  Phys.\ Lett.\ B {\bf 743}, 172 (2015)
  doi:10.1016/j.physletb.2015.02.027
  [arXiv:1411.5845 [hep-ph]].
  %%CITATION = doi:10.1016/j.physletb.2015.02.027;%%
  %6 citations counted in INSPIRE as of 06 Mar 2018



%\cite{Everett:2008et}
\bibitem{Everett:2008et} 
  L.~L.~Everett and A.~J.~Stuart,
  %``Icosahedral (A(5)) Family Symmetry and the Golden Ratio Prediction for Solar Neutrino Mixing,''
  Phys.\ Rev.\ D {\bf 79}, 085005 (2009)
  doi:10.1103/PhysRevD.79.085005
  [arXiv:0812.1057 [hep-ph]].
  %%CITATION = doi:10.1103/PhysRevD.79.085005;%%
  %138 citations counted in INSPIRE as of 06 Mar 2018



%\cite{Feruglio:2011qq}
\bibitem{Feruglio:2011qq} 
  F.~Feruglio and A.~Paris,
  %``The Golden Ratio Prediction for the Solar Angle from a Natural Model with A5 Flavour Symmetry,''
  JHEP {\bf 1103}, 101 (2011)
  doi:10.1007/JHEP03(2011)101
  [arXiv:1101.0393 [hep-ph]].
  %%CITATION = doi:10.1007/JHEP03(2011)101;%%
  %79 citations counted in INSPIRE as of 06 Mar 2018



%\cite{Cooper:2012bd}
\bibitem{Cooper:2012bd} 
  I.~K.~Cooper, S.~F.~King and A.~J.~Stuart,
  %``A Golden $A_5$ Model of Leptons with a Minimal NLO Correction,''
  Nucl.\ Phys.\ B {\bf 875}, 650 (2013)
  doi:10.1016/j.nuclphysb.2013.07.027
  [arXiv:1212.1066 [hep-ph]].
  %%CITATION = doi:10.1016/j.nuclphysb.2013.07.027;%%
  %26 citations counted in INSPIRE as of 06 Mar 2018



%\cite{Varzielas:2013hga}
\bibitem{Varzielas:2013hga} 
  I.~de Medeiros Varzielas and L.~Lavoura,
  %``Golden ratio lepton mixing and nonzero reactor angle with $A_5$,''
  J.\ Phys.\ G {\bf 41}, 055005 (2014)
  doi:10.1088/0954-3899/41/5/055005
  [arXiv:1312.0215 [hep-ph]].
  %%CITATION = doi:10.1088/0954-3899/41/5/055005;%%
  %14 citations counted in INSPIRE as of 06 Mar 2018



%\cite{Gehrlein:2014wda}
\bibitem{Gehrlein:2014wda} 
  J.~Gehrlein, J.~P.~Oppermann, D.~Schäfer and M.~Spinrath,
  %``An SU(5) $\times$ A$_5$ golden ratio flavour model,''
  Nucl.\ Phys.\ B {\bf 890}, 539 (2014)
  doi:10.1016/j.nuclphysb.2014.11.023
  [arXiv:1410.2057 [hep-ph]].
  %%CITATION = doi:10.1016/j.nuclphysb.2014.11.023;%%
  %33 citations counted in INSPIRE as of 06 Mar 2018



%\cite{Gehrlein:2015dxa}
\bibitem{Gehrlein:2015dxa} 
  J.~Gehrlein, S.~T.~Petcov, M.~Spinrath and X.~Zhang,
  %``Leptogenesis in an SU(5) $\times$ A$_5$ Golden Ratio Flavour Model,''
  Nucl.\ Phys.\ B {\bf 896}, 311 (2015)
  doi:10.1016/j.nuclphysb.2015.04.019
  [arXiv:1502.00110 [hep-ph]].
  %%CITATION = doi:10.1016/j.nuclphysb.2015.04.019;%%
  %11 citations counted in INSPIRE as of 06 Mar 2018



%\cite{DiIura:2015kfa}
\bibitem{DiIura:2015kfa} 
  A.~Di Iura, C.~Hagedorn and D.~Meloni,
  %``Lepton mixing from the interplay of the alternating group A$_{5}$ and CP,''
  JHEP {\bf 1508}, 037 (2015)
  doi:10.1007/JHEP08(2015)037
  [arXiv:1503.04140 [hep-ph]].
  %%CITATION = doi:10.1007/JHEP08(2015)037;%%
  %47 citations counted in INSPIRE as of 06 Mar 2018



%\cite{Ballett:2015wia}
\bibitem{Ballett:2015wia} 
  P.~Ballett, S.~Pascoli and J.~Turner,
  %``Mixing angle and phase correlations from A5 with generalized CP and their prospects for discovery,''
  Phys.\ Rev.\ D {\bf 92}, no. 9, 093008 (2015)
  doi:10.1103/PhysRevD.92.093008
  [arXiv:1503.07543 [hep-ph]].
  %%CITATION = doi:10.1103/PhysRevD.92.093008;%%
  %42 citations counted in INSPIRE as of 06 Mar 2018



%\cite{Gehrlein:2015dza}
\bibitem{Gehrlein:2015dza} 
  J.~Gehrlein, S.~T.~Petcov, M.~Spinrath and X.~Zhang,
  %``Leptogenesis in an SU(5) x A5 Golden Ratio Flavour Model: Addendum,''
  Nucl.\ Phys.\ B {\bf 899}, 617 (2015)
  doi:10.1016/j.nuclphysb.2015.08.019
  [arXiv:1508.07930 [hep-ph]].
  %%CITATION = doi:10.1016/j.nuclphysb.2015.08.019;%%
  %9 citations counted in INSPIRE as of 06 Mar 2018



%\cite{Turner:2015uta}
\bibitem{Turner:2015uta} 
  J.~Turner,
  %``Predictions for leptonic mixing angle correlations and nontrivial Dirac CP violation from A$_5$ with generalized CP symmetry,''
  Phys.\ Rev.\ D {\bf 92}, no. 11, 116007 (2015)
  doi:10.1103/PhysRevD.92.116007
  [arXiv:1507.06224 [hep-ph]].
  %%CITATION = doi:10.1103/PhysRevD.92.116007;%%
  %25 citations counted in INSPIRE as of 06 Mar 2018



%\cite{Li:2015jxa}
\bibitem{Li:2015jxa} 
  C.~C.~Li and G.~J.~Ding,
  %``Lepton Mixing in $A_5$ Family Symmetry and Generalized CP,''
  JHEP {\bf 1505}, 100 (2015)
  doi:10.1007/JHEP05(2015)100
  [arXiv:1503.03711 [hep-ph]].
  %%CITATION = doi:10.1007/JHEP05(2015)100;%%
  %47 citations counted in INSPIRE as of 06 Mar 2018



%\cite{Ding:2017hdv}
\bibitem{Ding:2017hdv} 
  G.~J.~Ding, S.~F.~King and C.~C.~Li,
  %``Golden Littlest Seesaw,''
  Nucl.\ Phys.\ B {\bf 925}, 470 (2017)
  doi:10.1016/j.nuclphysb.2017.10.019
  [arXiv:1705.05307 [hep-ph]].
  %%CITATION = doi:10.1016/j.nuclphysb.2017.10.019;%%
  %2 citations counted in INSPIRE as of 06 Mar 2018




%\cite{Lindner:2016bgg}
\bibitem{Lindner:2016bgg} 
  M.~Lindner, M.~Platscher and F.~S.~Queiroz,
  %``A Call for New Physics : The Muon Anomalous Magnetic Moment and Lepton Flavor Violation,''
  Phys.\ Rept.\  {\bf 731}, 1 (2018)
  doi:10.1016/j.physrep.2017.12.001
  [arXiv:1610.06587 [hep-ph]].
  %%CITATION = doi:10.1016/j.physrep.2017.12.001;%%
  %71 citations counted in INSPIRE as of 27 Apr 2018



%\cite{Georgi:1979df}
\bibitem{Georgi:1979df} 
  H.~Georgi and C.~Jarlskog,
  %``A New Lepton - Quark Mass Relation in a Unified Theory,''
  Phys.\ Lett.\  {\bf 86B}, 297 (1979).
  doi:10.1016/0370-2693(79)90842-6
  %%CITATION = doi:10.1016/0370-2693(79)90842-6;%%
  %802 citations counted in INSPIRE as of 06 Mar 2018



%\cite{Ross:2007az}
\bibitem{Ross:2007az} 
  G.~Ross and M.~Serna,
  %``Unification and fermion mass structure,''
  Phys.\ Lett.\ B {\bf 664}, 97 (2008)
  doi:10.1016/j.physletb.2008.05.014
  [arXiv:0704.1248 [hep-ph]].
  %%CITATION = doi:10.1016/j.physletb.2008.05.014;%%
  %113 citations counted in INSPIRE as of 06 Mar 2018



%\cite{Bora:2012tx}
\bibitem{Bora:2012tx} 
  K.~Bora,
  %``Updated values of running quark and lepton masses at GUT scale in SM, 2HDM and MSSM,''
  Horizon {\bf 2} (2013)
  [arXiv:1206.5909 [hep-ph]].
  %%CITATION = ARXIV:1206.5909;%%
  %31 citations counted in INSPIRE as of 06 Mar 2018



%\cite{Xing:2007fb}
\bibitem{Xing:2007fb} 
  Z.~z.~Xing, H.~Zhang and S.~Zhou,
  %``Updated Values of Running Quark and Lepton Masses,''
  Phys.\ Rev.\ D {\bf 77}, 113016 (2008)
  doi:10.1103/PhysRevD.77.113016
  [arXiv:0712.1419 [hep-ph]].
  %%CITATION = doi:10.1103/PhysRevD.77.113016;%%
  %314 citations counted in INSPIRE as of 06 Mar 2018



%\cite{Olive:2016xmw}
\bibitem{Olive:2016xmw} 
  C.~Patrignani {\it et al.} [Particle Data Group],
  %``Review of Particle Physics,''
  Chin.\ Phys.\ C {\bf 40}, no. 10, 100001 (2016).
  doi:10.1088/1674-1137/40/10/100001
  %%CITATION = doi:10.1088/1674-1137/40/10/100001;%%
  %2743 citations counted in INSPIRE as of 06 Mar 2018

  
  
  %\cite{deSalas:2017kay}
\bibitem{deSalas:2017kay}
  P.~F.~de Salas, D.~V.~Forero, C.~A.~Ternes, M.~Tortola and J.~W.~F.~Valle,
  %``Status of neutrino oscillations 2018: 3$\sigma$ hint for normal mass ordering and improved CP sensitivity,''
  Phys.\ Lett.\ B {\bf 782} (2018) 633
  doi:10.1016/j.physletb.2018.06.019
  [arXiv:1708.01186 [hep-ph]].
  %%CITATION = doi:10.1016/j.physletb.2018.06.019;%%
  %129 citations counted in INSPIRE as of 06 Jul 2018



%\cite{Esteban:2016qun}
\bibitem{Esteban:2016qun} 
  I.~Esteban, M.~C.~Gonzalez-Garcia, M.~Maltoni, I.~Martinez-Soler and T.~Schwetz,
  %``Updated fit to three neutrino mixing: exploring the accelerator-reactor complementarity,''
  JHEP {\bf 1701}, 087 (2017)
  doi:10.1007/JHEP01(2017)087
  [arXiv:1611.01514 [hep-ph]].
  %%CITATION = doi:10.1007/JHEP01(2017)087;%%
  %241 citations counted in INSPIRE as of 06 Mar 2018



%\cite{KamLAND-Zen:2016pfg}
\bibitem{KamLAND-Zen:2016pfg} 
  A.~Gando {\it et al.} [KamLAND-Zen Collaboration],
  %``Search for Majorana Neutrinos near the Inverted Mass Hierarchy Region with KamLAND-Zen,''
  Phys.\ Rev.\ Lett.\  {\bf 117}, no. 8, 082503 (2016)
  Addendum: [Phys.\ Rev.\ Lett.\  {\bf 117}, no. 10, 109903 (2016)]
  doi:10.1103/PhysRevLett.117.109903, 10.1103/PhysRevLett.117.082503
  [arXiv:1605.02889 [hep-ex]].
  %%CITATION = doi:10.1103/PhysRevLett.117.109903, 10.1103/PhysRevLett.117.082503;%%
  %236 citations counted in INSPIRE as of 06 Mar 2018
  
  
  
  
  %\cite{Chiang:2011cv}
\bibitem{Chiang:2011cv}
  C.~W.~Chiang, Y.~F.~Lin and J.~Tandean,
  %``Probing Leptonic Interactions of a Family-Nonuniversal Z' Boson,''
  JHEP {\bf 1111} (2011) 083
  doi:10.1007/JHEP11(2011)083
  [arXiv:1108.3969 [hep-ph]].
  %%CITATION = doi:10.1007/JHEP11(2011)083;%%
  %35 citations counted in INSPIRE as of 15 May 2018



%\cite{Raby:2017igl}
\bibitem{Raby:2017igl} 
  S.~Raby and A.~Trautner,
  %``Vectorlike chiral fourth family to explain muon anomalies,''
  Phys.\ Rev.\ D {\bf 97}, no. 9, 095006 (2018)
  doi:10.1103/PhysRevD.97.095006
  [arXiv:1712.09360 [hep-ph]].
  %%CITATION = doi:10.1103/PhysRevD.97.095006;%%
  %4 citations counted in INSPIRE as of 08 Jul 2018



% %\cite{Murakami:2001cs}
% \bibitem{Murakami:2001cs} 
%   B.~Murakami,
%   %``The Impact of lepton flavor violating Z-prime bosons on muon g-2 and other muon observables,''
%   Phys.\ Rev.\ D {\bf 65}, 055003 (2002)
%   doi:10.1103/PhysRevD.65.055003
%   [hep-ph/0110095].
%   %%CITATION = doi:10.1103/PhysRevD.65.055003;%%
%   %32 citations counted in INSPIRE as of 27 Apr 2018
  

\end{thebibliography}
\end{document}